\documentstyle[12pt]{article}
\advance \textheight \topskip

\newcommand{\rank}{\mathop{\rm rank}\nolimits}

\begin{document}
\begin{flushright}
Preprint arch-ive/9711192\\
Preprint Siegen University SI-97-23 
\end{flushright}
\centerline{\Large {\bf Grand Unified Theories in Superstrings}}
\centerline{\Large {\bf in the Free World-Sheet Fermion Formulation}%
\footnote{The work of A.A.M., I.A.N. and G.G.V. was supported in part
by the Russian Foundation for Basic Researches, grant No. 95-02-06121a.}}

\vspace{20pt}
\begin{center}
{\Large {\bf H.D. Dahmen $^{a}$, A.A. Maslikov $^{b}$ , I.A. Naumov $^{b}$,\\
  T. Stroh $^{a}$, G.G. Volkov $^{a,b}$. }}\\
\bigskip
$^a$ Department of Physics,
Siegen University,\\
57068 Siegen, Germany.\\
$^b$ Institute for High Energy Physics,\\
142284 Protvino, Moscow Region, Russia.\\
\end{center}

\centerline{ABSTRACT}
In the framework of the four dimensional heterotic superstring with free
world-sheet fermions and using  all constraints due to the conformal and
superconformal
symmetries (Virasoro and super-Virasoro algebras), the two dimensional
current algebra,
the global reparametrization (modular) invariance and the
 autoduality of the fermion charge
lattice we study the rank eight/sixteen
Grand Unified String Theories (GUST) which contain
the non-abelian gauge family symmetry. We develop some methods for
building of corresponding string models.
We explicitly construct GUST on the level one of affine Lie algebra 
with gauge symmetries $ G = SU(5) \times U(1)\times (SU(3) \times U(1))_H$ and
$G = SO(10)\times SU(4)_H $
or $E(6)\times SU(3)_H$ $\subset E(8)$
and consider the full massless spectrum for our string models.
To solve the problem of breaking the $G_{GUT}$ gauge symmetry of the
level one current algebra we extend our gauge group to $G \times G$ 
and, alternatively, we can consider as an observable gauge
symmetry the diagonal subgroups  $G^{\rm symm}$ of the rank 16 group
$G \times G$ $\subset SO(16) \times SO(16)$ or $\subset E(8) \times E(8)$.
This trick gives us the wide possibilities to construct the GUSTs and
investigate them comparing their predictions with the low energy
phenomenology.  Using the autoduality of the fermionic charge lattice we
studied the possibilities of constructing of some GUST gauge groups. For
example, we explain why all attempts to construct the  realistic GUTs with 
$SO(10) \times SU(3) \times U(1)$ gauge symmetry fail giving instead
$SU(4)_H$ as the horizontal gauge symmetry.
We discuss the possible
fermion matter and Higgs sectors in some GUST theories.
We study string amplitudes and for one realistic model we 
explicitly calculate 
the renormalizable (N=3) and nonrenormalizable (N=4,5,6)
contributions to the superpotential.
We observe that the existence of realistic quark and lepton mass spectra
depends on the form of the horizontal family gauge groups and there is a
possibility for the existence of this symmetry with low mass breaking scale.
Due to consideration of the $SU(3)_H$ horizontal factor  there can
exist "superweak" light chiral matter ($m_H^f
< M_W$) in GUST under consideration.

\tableofcontents

\section{Introduction}
\subsection{The Origin of Grand Unified Theories in  Superstrings.}

 The idea of the common origin of all known (and presently unknown)
types of interaction (theory of everything) is one of the dominating ones
in the contemporary high energy physics and cosmology because is allows
to understand the unified substance of all interactions and the
dynamics of the Universe evolution in the framework of the Big Bang
concept.

Grand Unified Theories (GUTs) based on the principle of   gauge invariance
in the framework of  simple Lie groups allow to  describe known interactions
related to the internal symmetries (weak, electromagnetic, and strong,
except gravity) in a universal way using common coupling constant.
As a result, GUTs allow  to predict the low-energy value of the
Weinberg  angle, which corresponds to the experiment within the limits
of accuracy. Besides multitude of various symmetries between interactions,
GUTs also predict the existence of new interactions, for example, those
responsible for nuclear matter instability and baryon quantum number
violation.
(The latter gives a great impact to the  experimental searches of the proton
and neutron decay with $\Delta B \neq 0$).
However the gauge symmetry dynamics, which  GUTs evolution  is based upon,
does not allow to obtain clear and definite predictions for the choice of
the GUT gauge group itself, its representations needed for breaking the
primordial GUT symmetry and superpotential construction, which would
determine the mass origin. The evolution of GUTs reveals that the
principle of gauge invariance by itself is not sufficient for further
progress of GUTs.

In the strings the  local gauge covariance  
at first is obtained
as a consequence of the 2-dimensional string dynamics (the same is true
for gravity), i.e. in strings we come to a deeper
understanding of the origin of local gauge
symmetries. As a result of this approach we can get the information about
the form of the Grand Unified Group, rank or/and dimension of Lie group,
kinds of the possible representations of these groups.
Further, the hypothesis that the particles construction based on the
unified mechanism in the framework of the string theory allows to make
a number of new clear predictions about Grand Unified Theories involving
gravity, thus it establishes
 the ground for creation "Theory of everything" (TOE).
The two-dimensional strings dynamics is based on the infinite dimensional
symmetries (conformal Virasoro, affine Lie algebras) and on the modular
invariance. It also includes Kaluza's ideas about relations between
the internal symmetries and the gravity in the $D>4$ dimensional space.
The string dynamics allows to determine quite definitely both the
dimension of our space-time and the structure of TOE as a union of
the $N=1$ supergravity and grand unified gauge group of all known
interactions. The choice of the unifying group also appear to be
restricted (for example $E_8\times E_8$ for $D=10$). Moreover
the possible representations are determined quite definitely correspondingly
to the level number of affine Lie algebra.

The evolution of the 4-dimensional Grand Unified String Theories (GUSTs)
has very short history and seems to be very appealing concerning the
unique possibility to find the very principle which would allow to
find the unique GUST gauge group like $E_8\times E_8$ for the $D=10$ case.

For a couple of years superstring theories, and particularly the heterotic
string theory, have provided an efficient way to construct the Grand Unified
Superstring Theories ($GUST$) of all known interactions, despite the
fact that it is
still difficult to construct unique and fully realistic low energy models
resulting after decoupling of massive string modes. This is because
of the fact that only
10-dimensional space-time  allows existence of two consistent (invariant
under reparametrization, superconformal, modular, Lorentz and SUSY
transformations) theories with the gauge symmetries $E(8)\times E(8)$ or
${spin(32)}/{Z_2}$ \cite{13i,14i} on the level one of affine Lie algebra
and which after compactification of
 the six extra space coordinates (into the Calabi-Yau \cite{15i,16i} manifolds
or into the
orbifolds) can be used for constructing GUSTs. Unfortunately, the process of
compactification to four dimensions is not unique.
 On the other hand, constructing
the theory directly in $4$-dimensional space-time requires including a
considerable number of free bosons or fermions into the internal string sector
of the heterotic superstring \cite{17i,18i,19i,20i}.
The freedom of the choice for the spin-boundary conditions of the right-moving
world-sheet fermions (bosons) can lead to as large internal symmetry group
such as e.g. rank  $22$ group, the left-moving fermions (bosons) give
the $N=1$ ($N=2$) local (global) supersymmetry and and some global internal
symmetries, the former are important in string amplitude constructions.

The way of breaking this primordial symmetry
is again not unique and leads to a huge number of possible models,
each of them
giving different low energy predictions.

Because of the presence of the affine Lie algebra 
$\hat g$ (which is a 2-dimensional manifestation of gauge symmetries of the
string itself) on the world sheet, string constructions yield definite
predictions concerning the representation of the symmetry group 
that can be used for
low energy models building\cite {21i,22i}.

Here we bring into play the multitude of experimental observations
made around Standard Model in the recent 10-15 years. The most important
questions are related to the progress in the understanding of
the chirality generations and their number, the quark and lepton masses
hierarchy within the three known families, the families mixing  mechanism,
the CP violation in the neutral K-meson decays,
the mixing amplitudes of the neutral $B_{d,s}$ mesons,
the baryon number violations, the neutrino mass and
the mixing problem, the neutrino mass origin --- Majorana or Dirac mass, 
the problem of compositeness of the Higgs and fermion matter.
Another question is  about supersymmetry, namely the SM extension in the
framework of SUSY $N=1$, SUSY breaking problem, ($N=1$ or $2$, since
string theories usually predict supersymmetric extensions  of the
GUSTs  and $N=1$ SUGRA).

There are some important directions in the
further GUST development which could be described as follows:

\begin{itemize}
\item The development of the 4-dimensional superstring theory, search
for the only unified group for GUST.
\item The  development of the GUST with current algebras on the level
$1$ or $2$, $3$. Searches of the GUST using charge lattices for
the solving of the problem of breaking GUST    down to  SM.
\item Obtaining the $N=1$ SUGRA${}\times{}$GUST from the string theory.
\item Solving the generation problem in the framework of GUST with the
light and massive generations number $N=3$ or $4$ and the horizontal symmetry
group.
\item The quark and lepton mass hierarchy problem in the framework of the
GUST.
\end{itemize}

\subsection{The Possible Ways of  E(8)-GUST Breaking
Leading to the $N_G=3$ or $N_G=3+1$ Families}\label{sbsec32}

There are not so many GUSTs trying to describe correctly the observable sector
of the Standard Model. They are well known: the SM gauge group, the Pati-Salam
($SU(4^c)\times SU(2)_L\times SU(2)_R$) gauge group,
the flipped $SU(5)\times U(1)$ gauge group
and SO(10) gauge group, which includes flipped $SU(5)\times U(1)$
\cite{20i,Leont,20ii}.
Also  some attempts to construct Grand Unified  Models
with $SU(5) \times U(1)$ and $SO(10)$ gauge groups on the level two \cite{33i}
and on the level 3 with effectively 3 generations \cite{Kakush}
were made.
Also the method of constructing the GUSTs with $G \times G$-
gauge groups was suggested in the papers \cite{25i}.

Low energy phenomenology gives additional constraints for
the search of the unique vacuum of string theory.
Really, any additional possible extension of the SM at high energies due
to new exotic gauge symmetries
($SU(2)_L \times SU(2)_R$-gauge group with  right-handed currents;
$SU(4^c)$-color group, in which leptons are considered like quarks of the
fourth color;
Horizontal symmetries with non-abelian gauge groups, $SU(3)_H$,$SU(4)_H$,
$SU(3)_H \times U(1)_H$, \ldots; the color-family symmetry, in which the
number of colors equal to the number of generations, $N^c = N_g=3$ or $4$,
and the form for color gauge group and family gauge group are equal to each
other, $SU(3^c)$ or $SU(4^c)$ and $SU(3)_H$ or $SU(4)_H$, respectively)
could give us, may be, very strong constraints for constructing
this GUT in heterotic strings.
There are good physical reasons for including the horizontal $SU(3)_H$ group
into the unification scheme. Firstly, this group naturally accommodates three
fermion families presently observed (explaining their origin) and, secondly,
it can provide correct and economical description of the fermion mass spectrum
and mixing  without invoking the high dimensional representation of
conventional $SU(5)$, $SO(10)$ or $E(6)$ gauge groups. The construction of a
string model (GUST)
containing the horizontal gauge symmetry provides additional strong
motivation
to this idea. Moreover, the fact that in GUSTs high dimensional
representations
are forbidden by the current algebra is a very welcome feature in this context.

That is why we consider this approach to be
useful in the investigations of GUTs that include horizontal gauge
symmetry. For the heterotic string the groups $E_8\times E_8$ and
$spin(32)/Z_2$ on the level one of the affine Lie algebra are characteristic.
Hence it is interesting to consider GUT
based on its various rank 8 and 16 subgroup.
String theories possess infinite dimensional symmetries that place many
specific constraints on the theory spectrum. These symmetries origin from
the conformal invariance, the modular invariance  and the two dimensional
current algebras.

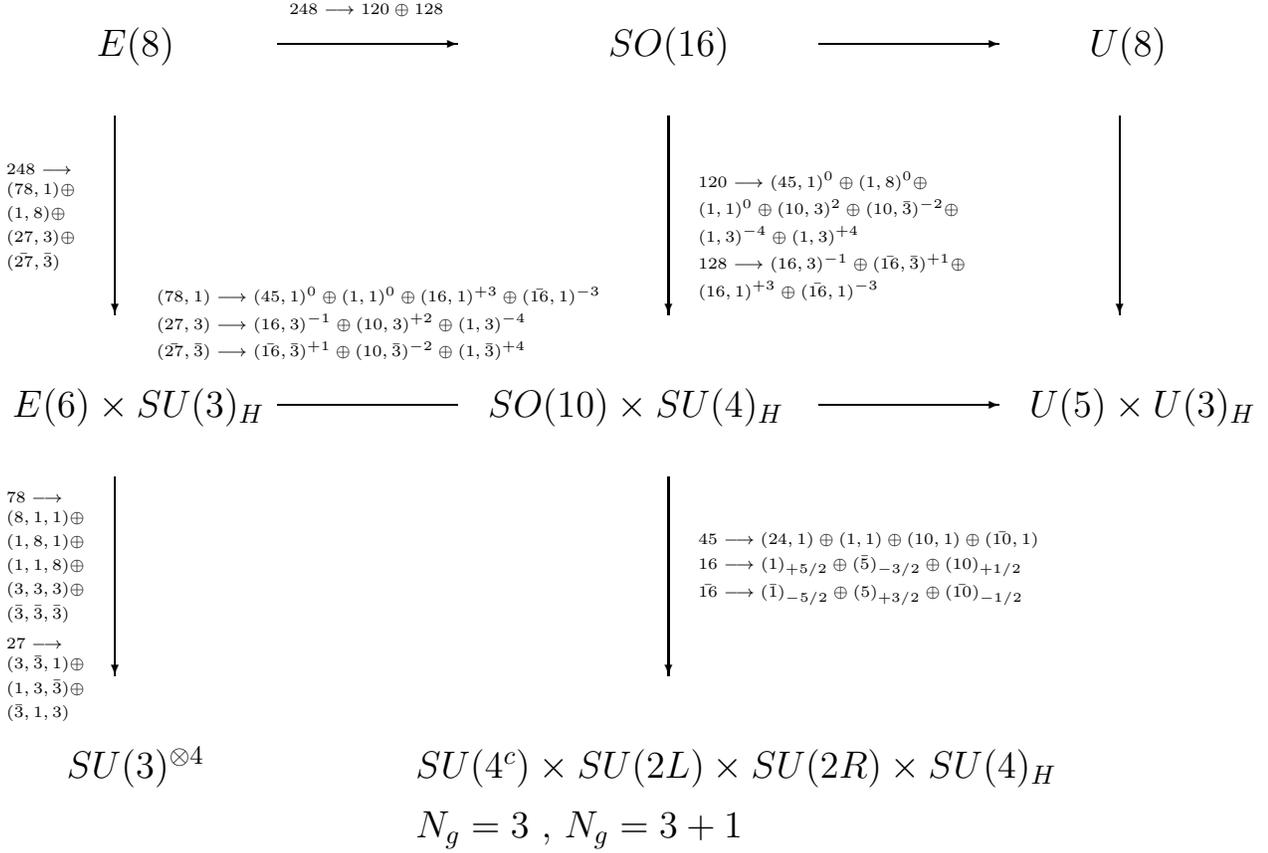
\begin{figure}
\caption{The possible ways of E(8) gauge symmetry breaking leading to the
3+1 or 3 generations.}
\label{fig1}
\setlength{\unitlength}{1mm}
\begin{picture}(165,135)(10,0)
\setlength{\unitlength}{0.8mm}
{\tiny
\put(20,140){{\large $E(8)$ } }
\put(105,140){{\large $SO(16)$ }}
\put(185,140){{\large $U(8)$ }}
\put(6,80){{\large $ E(6)\times SU(3)_H $ }}
\put(85,80){{\large $ SO(10)\times SU(4)_H $ }}
\put(175,80){{\large $ U(5)\times U(3)_H$ }}
\put(15,20){{\large $ SU(3)^{\otimes 4} $ }}
\put(73,20){{\large $ SU(4^c)\times SU(2L)\times SU(2R)\times SU(4)_H$ }}
\put(73,10){{\large $N_g = 3 $ , $N_g = 3 + 1 $ }}

\put(23,130){\vector(0,-1){33}}
\put(115,130){\vector(0,-1){33}}
\put(50,142){\vector(1,0){30}}
\put(23,70){\vector(0,-1){33}}
\put(115,70){\vector(0,-1){33}}
\put(50,82){\line(1,0){30}}
\put(140,142){\vector(1,0){30}}
\put(190,130){\vector(0,-1){33}}
\put(140,82){\vector(1,0){30}}

\put(5,105){\shortstack[l]{ {$ 248 \longrightarrow $} \\ {$(78,1)\oplus $} \\
 {$ (1,8)\oplus $} \\ {$ (27,3)\oplus $} \\ {$ (\bar{27},\bar 3) $}  }}

\put(5,30){\shortstack[l]{ {$ 78 \longrightarrow $} \\ {$(8,1,1)\oplus $} \\
      {$ (1,8,1)\oplus $} \\ {$ (1,1,8)\oplus $} \\
          {$ (3,3,3)\oplus $} \\{$ (\bar 3,\bar 3,\bar 3) $} \\ \\
   {$ 27 \longrightarrow $} \\ {$(3,\bar 3,1)\oplus $} \\
      {$ (1,3,\bar 3)\oplus $} \\ {$ (\bar 3,1,3) $}    }}

\put(52,147){\shortstack[l]{ {$  248 \longrightarrow 120 \oplus 128 $} }}

\put(120,100){\shortstack[l]{ {$ 120 \longrightarrow (45,1)^0 \oplus (1,8)^0
 \oplus $} \\ {$ (1,1)^0 \oplus  (10,3)^2 \oplus (10,\bar 3)^{-2} \oplus $} \\
              {$ (1,3)^{-4} \oplus (1,3)^{+4} $} \\
 {$ 128 \longrightarrow (16,3)^{-1} \oplus (\bar{16},\bar 3)^{+1} \oplus $} \\
   {$ (16,1)^{+3} \oplus  (\bar{16},1)^{-3} $}    }}

\put(30,90){\shortstack[l]{
 {$ (78,1) \longrightarrow (45,1)^0 \oplus (1,1)^0
 \oplus  (16,1)^{+3} \oplus (\bar{16},1)^{-3} $} \\
 {$ (27,3) \longrightarrow (16,3)^{-1} \oplus (10,3)^{+2} \oplus (1,3)^{-4} $}
\\
 {$ (\bar{27},\bar{3}) \longrightarrow (\bar{16},\bar 3)^{+1} \oplus (10,\bar
3)^{-2}
            \oplus (1,\bar 3)^{+4} $} }}

\put(120,50){\shortstack[l]{
 {$ 45 \longrightarrow (24,1) \oplus (1,1) \oplus  (10,1)
                \oplus (\bar{10},1) $} \\
 {$ 16 \longrightarrow (1)_{+5/2} \oplus (\bar 5)_{-3/2}
                        \oplus (10)_{+1/2} $} \\
 {$ \bar{16} \longrightarrow (\bar 1)_{-5/2} \oplus (5)_{+3/2}
                        \oplus (\bar{10})_{-1/2} $} }}

}
\end{picture}

\end{figure}

All this leads us naturally to considering  possible forms for horizontal
symmetry
$G_H$, and $G_H$ quantum number assignments for quarks (anti-quarks) and
leptons (anti-leptons) which can be realized within GUST's framework.
To include the
horizontal interactions with three known generations in the ordinary
GUST it is
natural to consider rank eight gauge symmetry.
 We can consider
$SO(16)$ or $E(6) \times SU(3)$ which are the maximal subgroups of $E(8)$
and which contains the rank eight
subgroup $SO(10)\times (U(1)\times SU(3))_H$ \cite {24i}. We will be,
therefore,
concerned with
the following chains (see Fig. \ref{fig1}):
\vskip 0.3cm
\small
$$
\begin{array}{cccc}
\:\:E(8) \longrightarrow SO(16) \longrightarrow U(8)\\
 \longrightarrow SU(5) \times U(1)_{Y_5}
\times (SU(3) \times U(1))_H  \\
\end{array}
$$
or
\vskip 0.3cm
\small
$$
\begin{array}{cccc}
\:\:E(8) \longrightarrow SO(16) \longrightarrow
\underline {SO(10) \times SU(4)_H} \\
\longrightarrow SU(4^c) \times SU(2)_L \times SU(2)_R \times SU(4)_H
\end{array}
$$
 \vskip 0.3cm
\small
or
$$
\begin{array}{cccc}
\:\:E(8) \longrightarrow E(6) \times SU(3) \longrightarrow
 (SU(3))^{ \times 4}.  \\
\end{array}
$$
 \vskip 0.3cm
\normalsize

According to this scheme one can get $SU(3)_H\times U(1)_H$ gauge family
symmetry with $N_g = 3 + 1 $ (there are also other possibilities as e.g.
 $E(6)\times SU(3)_H\subset E(8)$
 $N_g = 3 $ generations can be obtained due to the third way of $E(8)$
gauge symmetry breaking via $E(6)\times SU(3)_H$, see Fig.\ref{fig1}),

In this paper  starting from the rank 16 grand unified gauge group (which is
the minimal rank allowed in strings) of the form $G\times G$
\cite{25i,26i} and making use of the current algebra which select
the possible gauge group
representations we construct the string models based on the diagonal subgroup
$G^{\rm symm}\subset G\times G\subset SO(16)\times SO(16)~(\subset
E_8\times E_8)$ \cite {25i}. We discuss and consider
$G^{\rm symm}=SU(5)\times U(1)\times (SU(3)\times U(1))_H$
$\subset SO(16)$ where the factor $(SU(3)\times U(1))_H$ is interpreted as the
horizontal gauge family symmetry. We explain how the unifying gauge symmetry
can be broken down to the Standard Model group. Furthermore, the horizontal
interaction predicted in our model can give an alternative description of the
fermion mass matrices without invoking high dimensional Higgs representations.
For this we will use some ways of breaking the non-abelian gauge horizontal
symmetries and  will also take into account the nonrenormalizable
contributions to the superpotential.
In contrast with other GUST constructions (flipped $SU(5)$ gauge group),
our model does not contain particles with exotic fractional electric charges
\cite {27i,25i}.
This important virtue of the model is due to the symmetric construction
of the electromagnetic charge  $Q_{em}$ from $ Q^I$ and $Q^{II}$ -- the
two electric charges of each of the $U(5)$ groups\cite {25i}:

\begin{eqnarray}\label{eq102}
Q_{em} &=& Q^{II} \oplus  Q^I.
\end{eqnarray}

We consider the possible forms of the
$G_H= SU(3)_H$ ,$SU(3)_H \times U(1)$,$SU(4)_H$,
 $G_{I} \times G_{II}$ \ldots - gauge family
symmetries  in the framework of Grand Unification Superstring Approach.
Also we will study the
 matter spectrum of these GUST and the possible Higgs sectors.
The form of the Higgs sector is very important for GUST-, $G_H$-, and
SM-gauge symmetries breaking
and for constructing the Yukawa couplings.

In this paper we analyze the general constructions of the GUTs
based on 4
dimensional heterotic string in free fermions approach.
We consider possible ways of breaking the "string" gauge group
$E_8\times E_8$ on the level one of affine Lie algebra down to low energy
supersymmetric model that includes the
Standard Model group and the horizontal factors $SU(3)_H$ or $SU(4)_H$, \ldots
and naturally
describes $3$ presently observable generations
with one possible massive generation.

The paper is organized as follows.

In Chapter 2 we discuss in the framework of the heterotic superstring
the conformal and superconformal
symmetries which are the basic fundament  for building gauge symmetries
of Grand Unified Theories and N=1 supergravity. The external string sector
is responsible for the construction of the N=1 supergravity.
The cancelation of the conformal
anomaly gives us the useful information about the form and rank of the
gauge group of the GUT, which is constructed in the internal string sector.
 In the heterotic superstring the internal sector consists of the 18 left
real world-sheet fermions with N=1 hidden local supersymmetry and 44 right
real world-sheet fermions (N=0 SUSY). The latter define the gauge group for
the Grand Unified Theory of maximal rank, $r = 22$.
In principle, using Ising models for 12 left and 12 right real world-sheet
fermions the rank of this group can be decreased down to the
$r=16$ \cite{20ii} or we can get the hidden local group of rank less or equal
to 6 and some global symmetries which are defined by in the left sector
and which will be important for calculation of the string amplitudes.
We consider here the string scattering amplitudes and ghost pictures,
what will be very important for calculating the renormalizable and
nonrenormalizable contributions to the superpotential.

In Chapter 3 we analyze several
 restrictions on various unitary representations in the  superstring GUTs
 from the point of the current algebra dynamics-level ,
Sugawara-Sommerfeld
constructions of the stress tensor and conformal weight of the primary
physical states. It is emphasized that the breaking of gauge group down
to symmetric diagonal subgroup effectively increases affine Lie algebra 
level to 2 and
makes the existence of the representations for Higgs fields needed for GUT
symmetry breaking possible.
For this we prefer to consider  the GUT of rank 16 gauge of
group $E_8\times E_8$ and its symmetric diagonal subgroup of rank 8. We
outline the perspective way including the symmetric subgroup on the
intermediate stage that does not involve higher level of affine Lie
algebra representations.

Chapter 4 presents various string models including Grand Unification group
$SU(5) \times U(1)$, $SO(10)$ and $E(6)$ along with horizontal gauge symmetry
$G_{HI} \times G_{HII}$, where $G_H= SU(3) \times U(1)$, $SU(4)$ or
$SU(2) \times U(1) \times U(1) $,
describing three light and one possible massive generations.
For all these models  it is essential that gauge group of GUT is breaking down
to the symmetric diagonal subgroups. The rules of building the basis of spin
boundary conditions and GSO projections based on modular invariance are given
in Appendix A. There are two possibilities of the anomaly cancelations
of the $SU(3)_H$ horizontal groups are considered, "axial-vector" like and
"vector" like horizontal models. In the first case the spectrum of the models 
includes "sterile", the Standard Model  singlet-particles. The second
case corresponds to the existing in the spectrum the additional $N_g$
multiplets with the opposite to the ordinary particles $G_H$ quantum numbers.
These multiplets cancel the $SU(3)_H$- or $SU(4)_H$-anomalies and are charged
under SM (see the model 4). We use
a computer program for calculating the full massless spectrum of states of
the considered models. The full spectra for the presented models are given
in corresponding tables.

The further analysis of autodual  charge lattices with complex fermions 
(level one of current algebra) and possible gauge groups
is given in Chapter 5. Based on this method we showed that it is impossible
to construct the GUT with gauge group $E(6)\times SU(3)$ and $N=1$
supersymmetry.
Also in constructing GUTs with $SO(10) \times SU(3)_H \times U(1)_H$
the autoduality of
the charge lattice does not allow to obtain the desired group 
resulting in extending the horizontal group up the $SU(4)_H$-gauge
symmetry.
Based on the method proposed
there we build 4D superstring GUT model with
${\left[E_6\times SU(3)\right]}^2$ and $SU(3)^{\otimes 4}$ gauge groups
with $N=2$ SUSY.
Also for completeness we construct the GUT with
$[E(6) \times SU(2) \times U(1)]^{\otimes 2}$ gauge group.

In the Chapter 6
the string amplitudes , superpotential and nonrenormalizable
contribution permitted by string dynamics are presented.
We study the renormalizable and nonrenormalizable vertex operators and
for the Model~1 we explicitly calculate
 the Yukawa coupling constants of the three-, four-, five-, and six-
point vertex operators (F-type).
The last form of obtained superpotential with taking into account the
five -point non-renormalizable vertex operators implies that two
generations remain massless comparing to the $m_W$ scale.
Using the condition of  $SU(3)_H$ anomaly cancelation  the theory
predicts the existence of the Standard Model singlet
"neutrino-like" particles that participate only in horizontal
interactions.  As following from the form of the superpotential some of
them   could be light (less than $m_W$) that will be very interesting in
sense of experimental searches.

In chapter 7 we analyzed the paths of unification with the $G\times G$ groups.
On the $[U(5) \times U(3)]^{\otimes 2}$
example we consider the ways of gauge group symmetry breaking and we show
that in the case of non-flipped $U(5)$ groups it is possible to solve
the problem of the GUT and string unification scale splitting. Also we
consider here the string thresholds corrections.

\section{Superconformal and Conformal Symmetries in Constructing the GUTs
in Heterotic String and String Scattering Amplitudes.}

\subsection{Conformal Symmetry and their Representations.}

The action for bosonic string has the following usual form
\begin{equation}
S=\int{\cal L}= -\frac{1}{8\pi } \int d\tau d\sigma
\sqrt{-\mbox{det}(h_{\alpha\beta})} h^{\alpha\beta}
\eta_{\mu\nu} \partial_\alpha X^{\mu} \partial_\beta X^{\nu}
\sim -\frac{1}{8\pi } \int d\tau d\sigma
[(\partial_\sigma X)^2 - (\partial_\tau X)^2 ]
\end{equation}
where we used the standard metrics signature:
$$
\eta_{\mu\nu}\sim (- , +,\ldots , +) \quad ,\qquad
\eta_{\alpha\beta}\sim (- , +)\ .
$$

The corresponding partition function  is as follows:
\begin{equation}
Z=\int\ DX\ Dh\ e^{iS}
\end{equation}
After transition to Euclidean 2D space (complex plane):
\begin{equation}
\label{perehod}
t=+i\tau \quad , \qquad z=\exp (t +i\sigma )
\end{equation}
we have
\begin{equation}
\label{act}
S_E =+\frac{1}{2\pi } \int d{\bar z} dz h^{z \bar z}
\eta_{\mu\nu} \partial_{\bar z} X^{\mu} \partial_z X^{\nu} \quad ,
\qquad Z\sim\int DX\ e^{-S_E}\ ,
\end{equation}
where $h^{z \bar z}=1/2$, and now the decomposition of a string state
into running left or right moving waves becomes a decomposition into
analytic and anti-analytic functions on the 2D Euclidean surface.
The points $0$ and $\infty $ correspond to the $in$ and $out$ states.
The  lines of constant time are mapped into circles on the $z$ plane;
the operation of time translation, $t\rightarrow t+a$, becomes
the dilatation, $z\rightarrow e^a z$, and we should identify the
generator of dilatation with the Hamiltonian of the original string
theory.

The propagator of $X$-field is the Green's function for the Laplace
equation in 2D: $\langle X_\mu (z_1) X_\nu (z_2)\rangle =
-2\eta_{\mu\nu} \ln |z_1 -z_2 |$.
The propagator of the analytic field is then
\begin{equation}
\label{corr1}
\langle X_\mu (z_1) X_\nu (z_2)\rangle =
-\eta_{\mu\nu} \ln (z_1 -z_2 )
\end{equation}
The analytic part of the energy-momentum tensor for this field
is given by normal ordering:
\begin{equation}
\label{tensor1}
T_{zz}=-1/2\ :(\partial_z X)^2 : \ .
\end{equation}
The energy-momentum tensor defines the infinitesimal conformal
mapping by the generator:
\begin{equation}
\label{generator}
T_\epsilon=\oint\frac{dz}{2\pi i} \epsilon (z) T(z)
\end{equation}
(a priori the integral is taken over an equal-time surface
$|z|=const$, but since the integrand is an analytic function,
we may arbitrary deform the contour).

When we have equations like (\ref{corr1}, \ref{tensor1})
we can compute the commutator of the generator (\ref{generator})
with some local field operator.
The commutators  are interesting
for us in sense of their vacuum expectations (v.e.). 
Recall that the functional integral with $\exp (-S)$ weight (v.e.)
defines the operator product by setting the operators in time order
(this is radial order in our case). So for the equal-time commutator
we can write:
\begin{equation}
\langle [ T_\epsilon , \phi (z)]\rangle =\frac{1}{Z}\int DX\ e^{-S}\
\oint_{\mbox{around }z}dz'T(z')\phi (z')
\end{equation}
This means that equations for commutators which describes the conformal
transformation properties of fields have encoded in convenient formulae of
operator product expansion (OPE).

The OPE's with participation of energy-momentum tensor $T$
are very important for conformal theory. For example such OPE
with primary fields (operator which transforms under reparametrizations
as tensor)

\begin{eqnarray}
 \phi(z) \rightarrow
 (\frac {dz^{'}}{dz})^h \phi(z^{'})
\end{eqnarray}

of conformal dimension $h$ are
\begin{equation}
T(z') \phi (z) \sim \frac{h}{(z'-z)^2} \phi (z)
+\frac{1}{(z'-z)} \partial_z \phi (z) + \ldots \ .
\end{equation}

The operator of free bosonic field like $X_\mu$ actually does not have
well defined conformal dimension, since this field can have logarithmic
branch cuts (see (\ref{corr1})). The "good" operators are for
example $\partial_z X$  (dim$=1$) and exponential operator
$e^{ik\cdot X(z)}$  (dim$=k^2 /2$).

Also the OPE $\ T(z)\cdot T(w)$ is very important since it defines
the conformal anomaly $c$ (central charge of Virasoro algebra)
\begin{equation}
T(z)T(w)\sim \frac{c/2}{(z-w)^4} + \frac{2}{(z-w)^2} T(w)
+\frac{1}{(z-w)} \partial_w T + \ldots
\end{equation}
For above-mentioned $D$-dimensional system of fields $X_{\mu}$ ,
$\mu = 1,\ldots ,D$ the central charge $c=c_X=D$.
where c is a central charge or conformal anomaly.

If we take the moments of the energy-momentum operator T(z) we will get the
conformal generators with the following Virasoro algebra:

\begin{eqnarray}
[L_n,\,L_m]\,= (n-m)\,L_{n+m}\,+\,\frac{c}{12}\,n\,(n^2-1)\, {\delta}_{n,-m}.
\end{eqnarray}

Using the Virasoro algebra we can construct the representations of the 
conformal group
where highest weight state is specified by two quantum numbers,
conformal weight $h$ and central charge $c$, such that:

\begin{eqnarray}
L_0 |h,c\rangle =&& h |h,c\rangle \nonumber\\
L_n |h,c\rangle =&& 0,\,\,\,n=1,2,3, \ldots.
\end{eqnarray}
For physical states   the conformal weight $h=1$.

It is interesting, that 2D metric field $h$ is excluded from
action (\ref{act}) in conformal gauge, which corresponds to choice:
\begin{equation}
\label{gaug}
ds^2=e^{\varphi } (dt^2 +d\sigma^2) \qquad \mbox{or} \qquad
h^{\alpha\beta } =e^{\varphi } \eta^{\alpha\beta }
\end{equation}
In terms of complex plane this means
$h_{zz}=h_{\bar z\bar z}=0$.

However, now we should count the determinant
of gauge transformation into partition function.

Under reparametrization of world sheet the gauge conditions
transform as:
\begin{equation}
\delta h_{zz} = \nabla_z \xi_z \qquad ,
\qquad \delta h_{\bar z \bar z} = \nabla_{\bar z} \xi_{\bar z} \ .
\end{equation}
Setting these components equal to zero as a gauge condition leads to
a nontrivial determinants of $\nabla $'s.
In correspondence with standard
Faddeev-Popov techniques this determinant may be represented
by functional integral of exponent of ghosts-antighosts action.
Ghost $c$ ($\gamma$ see below) has quantum numbers of gauge parameter
(transforms as dual adjoint representation),
antighost $b$  ($\beta$ see below) transforms as adjoint representation
of gauge algebra and both have statistic which opposite to
statistic of gauge generators.

So, the addition contribution to partition function arise :
\begin{equation}
\label{bc-act}
\int\ Db\ Dc\ \exp -\frac{1}{2\pi}\int dz d\bar z
(b_{zz}\nabla_{\bar z} c^z +{\bar b}_{\bar z \bar z}\nabla_z {\bar c}^{\bar z})
\end{equation}
We write down the ghost action in conformal gauge.

The anticommuting fields $c^z$ and $b_{zz}$ have conformal dimension
(-1) and (+2) correspondingly and in conformal gauge they 
obey the classical equation of motion
\begin{equation}
\partial_{\bar z} c^z =\partial_{\bar z} b_{zz} =0 \ .
\end{equation}
Meantime their quantum propagator is
\begin{equation}
\label{pr}
\langle b_{zz}(z) c^w(w) \rangle = \frac{1}{z-w} \ .
\end{equation}

Now we can rewrite ghost action (see (\ref{bc-act})) in curve space
and calculate energy-momentum tensor. Otherwise this tensor can be
reconstructed from the requirement that it reproduce the right OPE
for $b$ and $c$ with right conformal dimension. In both cases the
final result in conformal gauge is
\begin{equation}
T_{bc}(z) = -2b\partial_z c -(\partial_z b) c
\end{equation}

Now if we consider OPE $T_{bc}(z)\cdot T_{bc}(w)$
we can calculate the contribution of this ghost system into
central charge of Virasoro algebra. It is useful to investigate
more general system with propagator (\ref{pr}) and "antighost"-"ghost"
conformal dimensions $j$ , $(1-j)$. Such system gives contribution
to central charge
\begin{equation}
\label{cbc}
c_{bc} = (-)(1-3k^2)
\end{equation}
where $k=2j-1$ and minus sign corresponds to commuting fields.
In our case $j=2$ and $c_{bc}=-26$. This means that in bosonic string 
only in the  case of $D=26$ the sum $c_{bc}+c_X$ vanishes 
and conformal anomaly is absent.

Note, that $b-c$ ghosts system contains the conserved in flat case
anomalous current of ghost number
$$J^{bc}_z =c^z b_{zz} \quad ,\qquad \partial_{\bar z} J^{bc}_z
=1/8\ k\sqrt{h} R^{(2)} \ ,$$
where $k=3$ and $R^{(2)}$ is the intrinsic curvature of the world
sheet described by $h_{\alpha\beta}$.
We can describe this current by one bosonic field $\varphi $:
$J^{bc}_z=J^{\varphi }_z =\partial_z \varphi^z$  requiring that
OPE's of theory remain the same.
For example, from conservation of $\ T\cdot J\ $ OPE we can 
construct the following
energy-momentum tensor (with changing sign of kinetic term)
\begin{equation}
T_\varphi = +\frac{1}{2}(\partial_z\varphi \partial_z\varphi
+k\partial_z^2 \varphi ) \ ,\qquad 
\langle\varphi (z_1) \varphi (z_2)\rangle =\ln (z_1 -z_2 )
\end{equation}
where $k=3$ (cf.(\ref{tensor1})). Theory with such energy-momentum
tensor corresponds to theory of free bosonic field with incorrect
sign of kinematic term and with non-zero $(-3)$ vacuum charge on infinity
$\langle 0| e^{+3\varphi }|0\rangle =1 $.
The conformal anomaly for such system is $c=1-3k^2 |_{k=3}=-26$
(cf.(\ref{cbc}))
In this theory the conformal dimension of exponential operators is 
defined by formula
\begin{equation}
\label{dim1}
\dim(:e^{\alpha\varphi }:)=\frac{1}{2}\alpha (\alpha -k)
\end{equation}
So in our case $k=3$ we can identify ghosts with exponentials
\begin{equation}
b\sim :e^{-\varphi }: \quad \mbox{dim}=2 \quad , \qquad
c\sim :e^{\varphi }: \quad \mbox{dim}=-1 \ .
\end{equation}
Finally, it can be showed that exponentials of free bosonic fields
are anticommuting operators.
So we can say that we have completed the bosonization of the
reparametrizations ghosts.

\bigskip

\subsection{Superconformal Symmetry in Superstrings.}

The action:
\begin{equation}
S=-\frac{1}{4\pi }\int d^2\sigma\sqrt{h}[ 1/2\ h^{\alpha\beta }\partial_\alpha
X^\mu \partial_\beta X_\mu +i {\bar\psi}^\mu \gamma^\alpha \nabla_\alpha
\psi_\mu - i/2\ ({\bar\chi}_\alpha \gamma^\beta \gamma^\alpha \psi^\mu )
(\partial_\beta X_\mu -i/4\ {\bar\chi}_\beta \psi_\mu )]
\end{equation}
is invariant under the local supersymmetry transformations
\begin{eqnarray}
\label{transform}
&&\delta h_{\alpha\beta } =2i{\bar\varepsilon}
\gamma_{(\alpha}\chi_{\beta )}\quad ,
\qquad \delta X^\mu = i{\bar\varepsilon}\psi^\mu \nonumber\\
&&\delta\psi^\mu =1/2\ \gamma^\alpha (\partial_\alpha X^\mu 
-i/2\ {\bar\chi}_\alpha \psi^\mu )\varepsilon \quad ,\qquad \delta\chi_\alpha
=2\nabla_\alpha \varepsilon \ .
\end{eqnarray}
If we choose the so called covariant superconformal gauge
which is the extension of conformal gauge (\ref{gaug})
\begin{equation}
\label{gaug2}
h^{\alpha\beta } =e^{\varphi } \eta^{\alpha\beta }\quad ,\qquad
\chi_\alpha =\gamma_\alpha \zeta
\end{equation}
then the fields $\varphi $ and $\zeta $ decouple due to super-Weyl
invariance and we get after going to Euclidean coordinates
the free fields action.
\begin{equation}
S = \frac{1}{4\pi } \int d{\bar z} dz 
(\partial_{\bar z} X^{\mu} \partial_z X_{\mu}
-\psi^\mu \partial_{\bar z} \psi_\mu
- {\bar \psi}^\mu \partial_z {\bar \psi}_\mu )
\end{equation}
$\psi^\mu$ is analytic field and ${\bar\psi}^\mu$ is anti-analytic.
The energy-momentum tensor of the $(X^\mu ,\psi^\mu )$ supermultiplet
is
\begin{equation}
T_B = -1/2\ (\partial_z X^\mu )^2 +1/2 \psi^\mu \partial_z \psi_\mu
\end{equation}
its superpartner (supercurrent) is
\begin{equation}
\label{TF}
T_F = -1/2\ \psi^\mu \partial_z X_\mu
\end{equation}
this is the generator of the local supersymmetry transformations.
The operator product expansions
 $T_B(z)$ and $T_F(z)$ give the N=1, (1,0), superconformal algebra:

\begin{eqnarray}
T_B(w)T_B(z)&&\sim \frac{3{\hat c}/4}{(w-z)^4} +
 \frac{2}{(w-z)^2}T_B(z)+ \frac{1}{w-z} \partial_z T_B(z),\nonumber\\
T_B(w)T_F(z) &&\sim \frac{3/2}{(w-z)^2}T_F +
  \frac{1}{w-z} \partial_z T_F(z),\nonumber\\
T_F(w)T_F(z) &&\sim \frac{{\hat c}/4}{(w-z)^3} +
 \frac{1/2}{(w-z)}T_B(z),
\end{eqnarray}
where $\hat c = 2/3c= 6$.

It can be show that contribution into conformal anomaly from
$\psi$-fields is $c_\psi =D/2$ , where $D$ is dimension of space-time.

Since we have added new gauge-fixing condition (\ref{gaug2}),
which transforms according to
$\delta\chi_z =\nabla_z \varepsilon $ (\ref{transform})
we should count the Jacobian for this variables.
The Jacobian may be represented by a path integral over a
pair of free {\it commuting} antighost-ghost fields $\beta$ and $\gamma$
with conformal dimensions $(3/2)$ and $(-1/2)$ correspondingly,
with the action
\begin{equation}
\label{first.ord.}
S=\frac{1}{2\pi}\int d^2 z (\beta\nabla_z \gamma + c.c.)
\end{equation}

Now for system $\beta ,\ \gamma$ we can apply the same arguments as
 we  have used for $b,\ c$ ghosts.
The system of commuting superconformal ghosts $\beta ,\ \gamma$
have $k=2$ and give contribution to the central charge
$c_{\beta\gamma}=-(1-3k^2)|_{k=2}=+11$.
Eventually  we have 
\begin{equation}
c_X=D\ ,\qquad c_{bc}=-26\ ,\qquad c_\psi =D/2\ ,\qquad
c_{\beta\gamma}=+11\ .
\end{equation}
In $D=10$ spacetime dimensions $c=0$ so there is no conformal
anomaly in this case.

By analogy with $b,\ c$ case we can define the anomalous ghost
number current
$$J_z =\beta\gamma \quad ,\qquad \partial_{\bar z} J_z
=1/8\ k\sqrt{h} R^{(2)} \ ,\ k=2\ .$$
And now we can do "the bosonization of bosons".
For this let us put $J_z =-\partial_z c$ (please do not confuse with
anticommuting ghost $c$) and introduce the energy-momentum tensor
(now with correct sign of kinematic terms)
\begin{equation}
\label{tensor2}
T^{(c)}_{zz}=-1/2\ (\partial_z c \partial_z c -k \partial_z^2 c)
=-1/2\ (J_z J_z -k \partial_z J_z )\ ,\ k=2\ .
\end{equation}
We should use different signs for kinematic terms of $b-c$
and $\beta -\gamma$ systems since OPE's $\ J^{(bc)}J^{(bc)}$
and$\ J^{(\beta\gamma)}J^{(\beta\gamma)}$ have anomalous terms
of opposite signs.
The conformal dimension of exponential operators for systems
with tensor (\ref{tensor2}) is defined by formula
\begin{equation}
\label{cghost}
\dim(e^{\alpha c }) =-1/2\ \alpha (\alpha +k)|_{k=2}
=-\frac{\alpha^2}{2}-\alpha
\end{equation}

However, now the process of bosonization is not completed on this stage
since
1) the tensor (\ref{tensor2}) gives contribution into central
charge $c_c =1\oplus 3k^2 |_{k=2} =+13$ not $+11$ (correct sign
of kinematic terms plays its role);
2) dim$(e^{+c})=-3/2$ and dim$(e^{-c})=+1/2$ , but
dim$(\beta )=+3/2$ and dim$(\gamma )=-1/2$ (opposite signs);
3) exponentials are anticommuting operators but
ghosts $\beta , \ \gamma$ are commuting.

All of these problems are solved by adding a system of fermions
$\xi$ and $\eta$ with dimensions 0 and 1 , and energy-momentum
tensor
\begin{equation}
T^{(\xi\eta )}=\partial_z \xi\cdot \eta
\end{equation}
This system has central charge $-2$, so it is exactly what we need to combine
with the $c$-field system to give $11$.
Finally, the combinations
\begin{equation}
\label{xi}
\beta\sim\partial_z \xi e^{-c} \sim e^{-c+\chi }\partial_z \chi \ ,
\quad \gamma\sim \eta e^{+c} \sim e^{c-\chi }
\ ,\quad\langle\chi(z_1)\chi(z_2)\rangle=-\langle c(z_1) c(z_2)\rangle
=\ln (z_1 -z_2 )
\end{equation}
have the correct dimensions and the correct operator products.

Note that system with energy-momentum tensor (\ref{tensor2}) and $k=2$
corresponds to free massless bosonic fields $c$ with vacuum
charge $+2$ on infinity. So for such a system only correlation functions
of operators with total charge $(-2)$ do not vanish.

\bigskip

The string theory that holds the most promise of describing the
physical world is that of the heterotic string.
The heterotic string uses the 16 dimensions left from the
compactification of 26D down to 10D to introduce a rank 16 Lie
group ($E_8\times E_8$ for example). Closed string has right-
and left-moving modes, which do not interact. The heterotic string
splits these modes apart. The right-moving modes are purely bosonic and live
in a 26D space which, has been compactified to 10D, leaving us with
a rank 16 gauge symmetry. However the left-moving modes only
live in a 10D space and contain the supersymmetric theory.
When left and right halves are put together, they produce a self-consistent,
ghost-free, anomaly-free, one-loop finite theory, the heterotic
string in D=10. Of course it is needed  to investigate the further 
compactification from D=10 to D=4.

However, using the Fermi-Bose equivalence in two dimensions we can
construct the heterotic string theory in D=4 in free fermionic approach.
Then the compactification will be encoded in boundary conditions
for free world sheet fermions.

\bigskip

 In the heterotic string theory in the left-moving
  (supersymmetric) sector there are $d-2$
 (in the light-cone gauge) real fermions $\psi^{\mu}$,
 their bosonic superpartners $X^{\mu}$, and $3(10-d)$ real
 fermions $\chi^I$. In the right-moving sector there are $d-2$
 bosons ${\bar X}^{\mu}$ and $2(26-d)$ real fermions.
In the heterotic string theories \cite{13i,14i} ${(N=1\: SUSY)}_{LEFT}$
${(N=0\: SUSY)}_{RIGHT}$ $\oplus$ ${\cal M}_{c_{L};c_{R}}$ with $d \leq 10$,
the conformal anomalies of the space-time sector are cancelled by the
conformal anomalies of the internal sector ${\cal M}_{c_L;c_R}$,
where $c_L=15-3d/2$ and $c_R=26-d$ are the conformal anomalies in the
left- and right-sector of the string. In the fermionic formulation of the
four-dimensional heterotic string theory in
addition to the two transverse bosonic coordinates $X_{\mu}$ ,$\bar{X}_{\mu}$
and their left-moving superpartners ${\psi}_{\mu}$, the internal sector
${\cal M}_{c_L;c_R}$ contains 44 right-moving ($c_R=22$) and 18 left-moving
($c_L=9$) real fermions (each real world-sheet fermion has
$c_f=1/2$).

 In the left supersymmetric sector world-sheet supersymmetry is non-linearly
 realized via the supercharge
 \begin{equation}
 T_F=\psi^{\mu}\partial X_{\mu} +f_{IJK}\chi^I\chi^J\chi^K\ ,\label{sch}
 \end{equation}
 where $f_{IJK}$ are the structure constants of a semi-simple
 Lie group $G$ of dimension $3(10-d)$.

 The possible Lie algebras of dimension 18 for $d=4$
 are $SU(2)^6$, $SU(3)\times SO(5)$, and $SU(2)\times SU(4)$.

 It has been shown \cite {bank'} that the N=1  space-time SUSY
vacuum
of heterotic string with local (1,0) worldsheet superconformal invariance
extends to a global N=2, (2,0), superconformal invariance:
\begin{eqnarray} \label {equation 2.0}
{T_F^+}(w) {T_F^{-}}(z) &&\sim \frac{\hat c}{(w-z)^3} + \frac{2\partial J}
{(w-z)^2} +
 \frac{2T + \partial J}{w-z}\nonumber\\
J(w) {T_F^{\pm}}(z) &&\sim \pm \frac{T_F^{\pm}}{w-z}\nonumber\\
J(w) J(z) &&\sim \frac{\hat c/2}{(w-z)^2}.
\end{eqnarray}

\bigskip

\subsection{Vertex Operators and String Scattering Amplitudes.}

To define asymptotic states of the string we make use of
the conformal
invariance of the theory. Using the conformal mapping
$z-z_0 =e^w$ we can map the asymptotic region of an infinitely
extended closed-string world surface into the neighborhood of the point
$z=z_0$. We introduce a vertex operator corresponding to this point.
A vertex operator for a given physical state is a collection
of two-dimensional conformal fields that represent the quantum number of
the state under all the symmetries of the model.

The integrated out vertex operator for state with set of quantum numbers
$\Lambda $ and momentum $K^\mu$ has the form
\begin{equation}
\label{vertex}
V_\Lambda (K)=\int d^2 z\sqrt{h}\ W_\Lambda (z,\bar z)
e^{i/2\ K\cdot X}e^{i/2\ K\cdot \bar X}\ .
\end{equation}
The integration is necessary since the vertex may be placed
in any arbitrary point of string worldsheet.
The operator $W_\Lambda (z,\bar z) e^{i/2\ K\cdot X}e^{i/2\ K\cdot \bar X}$
is a primary field of dimension $(1,1)$ (see explanation below).

Scattering amplitude should be the sum of
the functional integrals for 2D
quantum field theory with insertion of corresponding vertex operators
into corresponding genus Riemann surfaces (string worldsheets).
After fixing the gauge (\ref{gaug2}) the string action obtains ghosts
additions. For tree amplitude the corresponding Riemann surface is sphere
and the $N$ point amplitude can be written as
\begin{eqnarray}
A(\Lambda_1 , K_1 ; \ldots ;\Lambda_N , K_N )&&
=g^{(N-2)}\int DX\ D\psi\ D\phi_{int.}\ Db\ Dc\ D\beta\ D\gamma \\
&&\exp -\biggl[ S_{str.}(X_\mu ,\psi_\mu , \phi_{int.})
+S_{gh.}(b,c,\beta ,\gamma )\biggr] \cdot
\prod^N_{i=1} V_{\Lambda_i}(K_i)\ , \nonumber
\end{eqnarray}
where $g$ is three-string interactions constant and
fields $\phi_{int.}$ describe internal string degrees of freedom,
which can arise in non-critical dimensions.
Of course this amplitude can be written in other equivalent forms
if we use the bosonization procedure.

Now the amplitude can be rewritten as amplitude of free theory
in 2D flat space
\begin{equation}
\label{A}
A(\Lambda_1 , K_1 ; \ldots ;\Lambda_N , K_N )
=g^{(N-2)}\int\prod_{i=1}^N d^2 z_i
\biggl\langle\prod_{j=1}^N e^{i/2\ K_j\cdot X} W_\Lambda (z_j,{\bar z}_j)
e^{i/2\ K_j\cdot \bar X} \biggl\rangle \ .
\end{equation}

Note, that expression (\ref{A}) is not yet well defined since
gauge fixing (\ref{gaug}, \ref{gaug2}) do not completely
alienate the reparametrization invariance.
The remaining symmetry is the unbounded three-parameter group
of fractional linear transformations with complex coefficients SL(2,~C)
\begin{equation}
\label{sl}
z\longrightarrow \frac{az+b}{cz+d}
\end{equation}
where $ad-bc=1$.

A natural way to cure this problem is to divide (\ref{A}) by the
group volume of SL(2,~C). Using the infinitesimal form of an SL(2,~C)
transformation (\ref{sl}) given by
$$
\delta z = \lambda_{-1} +\lambda_0 z +\lambda_1 z^2
$$
we can find the Jacobian for the change of variables from any three
of the complex $z_i$ to the three complex parameters $\lambda_i$:
\begin{equation}
\label{det}
\biggl| \frac{\partial (z_i , z_j , z_k )}
{\partial (\lambda_{-1},\lambda_0 ,\lambda_1 )}\biggr|^2
=|z_i -z_j |^2 |z_i -z_k |^2 |z_j -z_k |^2
\end{equation}
Now we can cancel the integral over group parameters
and fix arbitrary values of three $z_i$.
The standard convenient choice is $z_i\rightarrow\infty$, $z_j =1$, $z_k =0$.

Now, let us consider the construction of vertices operators
$V_\Lambda $ (\ref{vertex}).
These operators must be the reparametrization invariant and hence
the SL(2,~C) invariant. The special case of SL(2,~C) is
$z\rightarrow az$, then $dz\ d{\bar z}\rightarrow |a|^2 dz\ d{\bar z}$.
So, for invariance of $V_\Lambda$ we find
$$
W_\Lambda (z,\bar z) e^{i/2\ K\cdot X}e^{i/2\ K\cdot \bar X}
\longrightarrow |a|^{-2}\cdot
W_\Lambda (z,\bar z) e^{i/2\ K\cdot X}e^{i/2\ K\cdot \bar X}
$$
This means that this operator is a primary field of dimension $(1,1)$.
Moreover for massless case $K^2 =0$ and hence
the operator $W_\Lambda (z,\bar z) $
is a primary field of dimension $(1,1)$ too.
Also this result can be found from the requirement of BRST invariance
of physical states.

For bosonic string the form of vertex operators is completely defined
by the requirement of the correct conformal dimension and by the 
requirement of representation the quantum number of
the state under all the symmetries of the model.

For superstring (critical dimension is D=10)
the situation is more complex. In this case,
firstly,  ghosts degrees of freedom
begin to play important role
in particular $c$-ghosts (see (\ref{cghost})) (bosonized bosonic ghosts)
which correspond to free bosonic system with non-zero vacuum charge
$+2$ on infinity; and secondly, space-time
fermions states with very non-trivial construction of vertex operators
arise in spectrum of theory.

It is known that in superstring space-time bosonic states
correspond to the Neveu-Schwarz sector of the string with
anti-periodic boundary conditions for worldsheet fermionic
field $\psi^{\mu} (\tau ,\sigma +2\pi) = -\psi^{\mu} (\tau ,\sigma )$;
while space-time fermionic states correspond to the Ramond sector
with periodic boundary conditions
$\psi^{\mu} (\tau ,\sigma +2\pi) = \psi^{\mu} (\tau ,\sigma )$.

If we make the replacement $z=e^w =\exp (t+i\sigma )$ (\ref{perehod})
and carry out this conformal mapping of the tensor $\psi_{\mu}$ of
dimension $1/2$, we will find
\begin{equation}
\label{aaa}
\psi_{\mu} (w)\rightarrow {\psi '}_{\mu} (z)=
\biggl( \frac{dz}{dw} {\biggr)}^{\frac{1}{2}} \psi_{\mu} (z(w))=
e^{w/2}\psi_{\mu} (z(w)) =\sqrt z \psi_{\mu} (z) \ .
\end{equation}
So, after going once around the string $w\rightarrow w+2i\pi $
we find a factor $e^{i\pi}=-1$ on the right-hand side of (\ref{aaa}).
Hence, the periodic (Ramond) $\psi_\mu$ is double-valued field
on the $z$ plane.
So the analytic function $\psi^\mu (z)$ has a square-root branch point
at $z_0$ (point to which the asymptotic string is mapped).
The vertex operators which create the Ramond states
(so called {\it spin operators}), then would  be the operators which create
this branch cut structure.

To construct such an operator let us replace the world-sheet
fermions $\psi^\mu$ with pairs and bosonize them
\begin{equation}
\frac{1}{\sqrt 2} (\psi^1 \pm i\psi^2 ) =\psi^{1\pm i2}
\sim e^{\pm i\phi_{12} }\ \mbox{ etc.}
\end{equation}
(Note, that the form of fermionic action in terms of the 
$\psi^{1\pm i2}$ fields is
$\int d^2 z \psi^{1+ i2} \partial_+ \psi^{1- i2} +c.c.$
 and is similar to ghosts action.)
Now if we construct (spin) operator
\begin{equation}
S_{\beta }=\prod_{(k,l)=(1,2)}^{(9,10)} e^{\pm i/2\ \phi_{kl}}
= e^{i\alpha_\beta \cdot \phi }
\end{equation}
it will create branch cuts for the $\psi^\mu (z)$.
Here index $\beta$ defines the choice of signs $\pm$ for all (five for D=10)
exponentials. So we have $2^5 =32$ components of spin operator.
But $32=16 +\bar{16}$ is dimension of spinor representation
for SO(10) group.
Moreover, the coefficients in exponentials define the weight
vectors of spinor representation
$\alpha_\beta = (\pm 1/2 , \pm 1/2 , \ldots , \pm 1/2)$
and we can construct the generators of SO(2N) in their bosonized form.
\begin{equation}
M^{\mu\nu} =\oint\frac{dz}{2i\pi} j^{\mu\nu}(z)\ ,\quad
j^{\mu\nu}(z) = -:\psi^\mu \psi^\nu :
\end{equation}
After transition from the indices $\mu ,\nu$ to $\mu\pm i\nu\sim a(\bar a)$
and bosonization we can rewrite currents
\begin{equation}
\label{curr}
j^{ab} =:e^{i\alpha_{ab}\cdot\phi}:\ ,\quad
\alpha_{ab}^i=\delta_a^i +\delta_b^i \ ,\qquad
j^{a\bar b} =:e^{i\alpha_{a\bar b}\cdot\phi}:\ ,\quad
\alpha_{a\bar b}^i=\delta_a^i -\delta_b^i \ ,\qquad
j^{a\bar a} =i\partial_z \phi^a\ .
\end{equation}
The N generators $M^{a\bar a}$ provide a Cartan subalgebra of SO(2N).
The formulae (\ref{curr}) allows us to compute the weight vector of
any operator which is the exponential of a boson field.
From the operator product
\begin{equation}
j^{a\bar a}(z)\ e^{i\alpha_\beta \cdot\phi }(w)\sim
\frac{1}{(z-w)}\ \alpha_\beta^a\ e^{i\alpha_\beta \cdot\phi }(w)
\end{equation}
we find the commutator
\begin{equation}
[M^{a\bar a} , e^{i\alpha_\beta \cdot\phi} (w)]=\alpha_\beta^a \
e^{i\alpha_\beta \cdot\phi} (w) \ .
\end{equation}
Thus, the weight vector for this exponential operator is exactly
$\alpha_\beta^i$.

Besides, we have the operator product
\begin{equation}
\psi^a (z) e^{i\alpha_\beta \cdot\phi } (w)\sim
(z-w)^{\alpha_\beta^a } :e^{i(\delta_a^i +\alpha_\beta^i )\cdot\phi_i}(w):\ .
\end{equation}
The degree of the singular terms is $\alpha_\beta^a =\pm 1/2$,
so the spin operator $S_\beta $ with weight $\alpha_\beta^i $
induces the square-root singularity in $\psi^a (z)$.

However, the operator $S_\beta (z)$ has dimension
$|\alpha_\beta |^2 /2 =N/8 $; for a $D=10$ superstring theory, this
equals to $5/8$; but the vertex operator should has dimension equal to $1$.

The solution of the problem is given by taking into account the superconformal
ghosts degrees of freedom.
Really, $\delta X^\mu = i\varepsilon\psi^\mu $ (\ref{transform})
and ghost $\gamma $ has a quantum numbers of parameter $\varepsilon $.
So, if the Ramond vertex operator creates the branch cut for
the field $\psi^\mu$, then the ghosts $\gamma ,\beta$ should have
a square-root branch point too.
As result we must add to the spin operator $S_\beta $ a spin
field for the commuting spinor ghosts
$e^{-1/2\ c(z)}$ with conformal dimension $3/8$ (see (\ref{cghost})).
So, our first candidate for the covariant fermion vertex is
\begin{equation}
V^f_{(-1/2)} \sim e^{-1/2\ c(z)} S_\beta (z) e^{iK\cdot X(z)} \ .
\end{equation}
This is, however, not sufficient to describe fermion scattering because
$V^f_{(-1/2)}$ has fermion ghost charge of $-1/2$, and only four-fermion
amplitude would cancel the ghost background charge $+2$.
Thus we need a second version of the fermion vertex, $V^f_{(+1/2)}$,
with positive ghost charge.

In the BRST formulation the physical states are constructed as an 
invariant under the BRST-transformation:
$$
Q_{BRST}|{\rm phys}\rangle =0 \quad ,\qquad Q^2_{BRST} =0 \ .
$$
Due to the latter condition the state of the form
$|{\rm null}\rangle =Q_{BRST}|\psi\rangle$ satisfies to former condition,
however such states have zero norm.
Thus it is only the operators with nontrivial "BRST cohomology"
that are of concern, and we may choose the states to be the characteristic
representatives of the cohomology classes.

In our case of superstring
\begin{equation}
Q_{BRST} =\oint\frac{dz}{2i\pi }\biggl\{
[cT_B(X,\psi ;\beta ,\gamma ) -bc\partial c]
+1/2\ \gamma\psi^\mu \partial X_\mu +1/4\ \gamma^2 b\biggr\} \ .
\end{equation}
Here $T_B(X,\psi ;\beta ,\gamma )$ is the combined stress tensor of all
fields mentioned.
In fact, the only operators which commute with $Q_{BRST}$
(corresponding to BRST-invariant states) with positive ghost charge are
of the form $[Q_{BRST},V]$. These are all null-vectors (spurious),
with the exception of
\begin{equation}
\label{pic.ch.1}
{V'}_{\rm phys} =[Q_{BRST},2\xi V_{\rm phys}]\ ,
\end{equation}
which is not spurious because $\xi$ is not part of the irreducible algebra
of the fermionic ghosts $\beta $ and $\gamma $; since
the irreducible current algebra representation involves
boson $c$, fermions $\eta $ and $\partial\xi $ but not $\xi $
itself (\ref{xi}).

So, if we take $V_{\rm phys}=V^f_{(-1/2)}$, we will get
the second version of fermionic vertex operator, 
with ghost number 1/2 \cite{FMS}.
We will call this the "$+1/2$ picture".
\begin{equation}
V^f_{(+1/2)}\sim \biggl\{ e^{c/2} [\partial X_\mu +i/4\ (K\cdot\psi )
\psi_\mu ]\gamma^\mu_{\alpha\beta }S^\beta +1/2\ e^{3/2\ c}\eta bS_\alpha
\biggr\} e^{iK\cdot X}\ .
\end{equation}
where a total derivative ignored.
The second term will not contribute to expectation values due to
ghost-charge conservation. So, all contribution comes from part
$j_{BRST}\sim 1/2 (e^c  e^{-\chi }) \psi^\mu \partial_\mu
\sim1/2\ \gamma T_F$ (see (\ref{TF})).
Since $e^{-\chi }(z_1)\xi (z_2)\sim\frac{1}{z_1 -z_2}$
we can rewrite the formula (\ref{pic.ch.1}) as
\begin{equation}
\label{pic.ch.2}
V_{q+1}=:e^c T_F V_q :\ .
\end{equation}

Really, there are infinite number of pictures (for bosonic vertex
of superstring too). This is due to, that system of {\it commuting}
ghosts $\beta $, $\gamma $ have the first-order action (\ref{first.ord.}).
So, for $\beta $, $\gamma $ -- system exists an infinite number
of linearly independent vacuums with different ghost numbers.
The spectrum is unbounded above and below, so the choice of vacuum
is somewhat arbitrary. We call these different sectors of the
theory different "pictures", and it can be shown that, on-shell
all pictures are equivalent \cite{FMS}.

Since we are dealing with a free field theory, there are not transitions
between vacuums and this situation is not catastrophic.

The formula (\ref{pic.ch.1}) or (\ref{pic.ch.2}) describes the picture
changing operation with increasing of ghost charge by 1.

\section{World-Sheet Current Algebra (WSCA)
         and Main Features of Rank Eight GUST}\label{sec4}

\subsection{The Representations of Affine Lie Algebra}\label{sbsec41}

Let us begin with a short review of the WSCA results \cite{21i,22i}.
In heterotic string the WSCA is constructed by the operator product
expansion (OPE) of the fields $J^a$ of the conformal dimension $(0,1)$:

\begin{equation}
{J^a}(w) {J^b}(z)\sim {\frac{1}{{(w-z)}^2}} k {\delta}^{ab} +
{\frac{1}{w-z}} i f^{abc} J^c +  \ldots
\end{equation}
or
\begin{equation}
[J^a_m, J^b_n] = i f_{abc} J^c_{m+n} + k m{\delta}^{ab}{\delta}_{m+n}
\end{equation}

 The structure constants $f^{abc}$ for the group $g$ are normalized so that
\begin{equation}
 f^{acd}f^{bcd} = Q_{\psi} {\delta}^{ab} =\tilde h {\psi}^2
{\delta}^{ab}
\end{equation}
 where $Q_{\psi}$ and $\psi$ are  the
quadratic Casimir and the highest weight of the adjoint representation and
$\tilde h$ is the dual Coxeter number.
The $\frac{\psi}{{\psi}^2}$ can be
expanded as in integer linear combination of the simple roots of $g$:
\begin{equation}
 \frac{\psi}{{\psi}^2} = \sum_{i=1}^{\rank g} m_i {\alpha}_i.
\end{equation}
The dual Coxeter number can be expressed through the integers numbers $m_i$
\begin{equation}
\tilde h = 1 + \sum_{i=1}^{\rank g} m_{i}
\end{equation}
and for the simply laced groups (all roots are equal and ${\psi}^2 =2$):
$A_n$, $D_n$, $E_6$, $E_7$, $E_8$
they are equal $n+1$, $2n-2$, $12$, $18$ and $30$, respectively.

A Sugawara-Sommerfeld construction of the energy-momentum tensor $T(z)$
algebra in terms of bilinears in the affine Lie generators
$J^a_n(z)$ \cite {21i,21ii},\cite {22i},
\begin{eqnarray}
T(z)=\sum_n L_n z^{-n-2}=
\frac{1}{2k + Q_{\psi}}\sum_{n,m}:J^a_{n-m}J^a_m: z^{-n-2},
\end{eqnarray}
allows to get by calculating the commutator between two
 generators of the Virasoro algebra
 the following expression for the central Virasoro
"charge":
\begin{eqnarray}\label{eq 101}
c_g = \frac {2 k \dim [g]}{2 k + Q_{\psi}}=
\frac{x \dim[g]}{ x + \tilde h}.
\end{eqnarray}

The WSCA $\hat g$ allows to grade the representations $R$ of the gauge group by
a level number $x$ (a non negative integer) and by a conformal weight $h(R)$.
An irreducible representation of the affine algebra $\hat g$ is characterized
by the vacuum representation of the algebra $g$ and the value of the central
term $k$, which is connected to  the level number by the relation
$x=2 k/{\psi}^2$.
The value of the level
number of the WSCA determines the possible highest weight unitary
representations
which are present in the spectrum in the following way

\begin{equation}\label{eq1}
x=\frac{2 k}{{\psi}^2} \geq \sum_{i=1}^{\rank g}n_{i} m_{i},
\end{equation}

\noindent where the sets of non-negative integers $\{m_i = m_1,\ldots, m_r\}$
and $\{ n_i = n_1,\ldots,n_r \}$ define the highest root and the highest weight
in terms of fundamental weights
of a representation $R$ respectively \cite{21i,22i}:
\begin{equation}
{\mu}_0 = \sum_{i=1}^{\rank g} n_{i} {\lambda}_{i}
\end{equation}

In fact, the WSCA on the level one is realized in the 4-dimensional heterotic
superstring theories with free world sheet fermions which allow a
complex fermion description \cite{18i,19i,20i}. One can obtain WSCA on a higher
level working with real fermions  and using some tricks \cite{33i}.
For these models the level of WSCA coincides with the Dynkin index of
representation $M$ to which free fermions are assigned,

\begin{equation}\label{eq2}
x = x_M = \frac{Q_M}{{\psi}^2}\frac{\dim M}{\dim g}
\end{equation}
($Q_M$ is a quadratic Casimir eigenvalue of representation $M$) and equals
one in cases when real fermions form vector representation $M$ of $SO(2N)$,
or when the world sheet fermions are complex and $M$ is the fundamental
representation of $U(N)$ \cite{21i,22i}.

Thus, in strings with WSCA on the level one realized on the world-sheet, only
very restricted set of unitary representations can arise in the spectrum:

\begin{enumerate}
\item  singlet and totally antisymmetric tensor representations of $SU(N)$
groups, for which $m_i = (1,\ldots,1) $;
\item singlet, vector, and spinor representations of $SO(2N)$ groups with\\
$m_i = (1,2,2,\ldots,2,1,1)$;
\item singlet, $\underline{27}$, and $\bar{\underline{27}}$-plets of $E(6)$
corresponding to $m_i = (1,2,2,3,2,1)$;
\item singlet  of $E(8)$ with $m_i = (2,3,4,6,5,4,3,2)$.
\end{enumerate}
Therefore only these representations can be used to incorporate matter
and Higgs fields in GUSTs with WSCA on the level 1.

In principle it might be possible to construct explicitly an example of
a level 1 WSCA-representation of the simply laced algebra $\hat g$
(A-, D-, E-types)
from the level one representations of the Cartan subalgebra of $g$.
This construction is achieved using the vertex operator of string, where
these operators are assigned to a set of lattice point corresponding to
the roots of a simply-laced Lie algebra $g$.

\subsection{The Features of the Level one WSCA and Conformal Weights 
in Matter and Higgs Representations in Rank 8 and 16 GUST 
Constructions}\label{sbsec42}
For example, to describe chiral matter fermions in GUST with the gauge
symmetry group $SU(5)\times U(1)\subset SO(10)$ one cane use
the following sum of the level-one
complex representations: $\underline {1}(-5/2) + \underline {\bar{5}}(+3/2) +
\underline {10}(-1/2) = \underline{16}$. On the other hand, as a real
representations of $SU(5)\times U(1)\subset SO(10)$, from which Higgs fields
can arise, one can take for example $\underline{5} + \underline{\bar 5}$
representations arising from real representation $\underline {10}$
of $SO(10)$. Also, real Higgs representations like $\underline {10}$(-1/2)
+ $\underline{\bar{10}}$(+1/2) of $SU(5)\times U(1)$ originating
from $\underline{16}$+$\underline{\bar{16}}$ of $SO(10)$, which has been used
in ref. \cite{10i} for further symmetry breaking, are allowed.

Another example is provided by the decomposition of $SO(16)$ representations
under $SU(8)\times U(1)\subset SO(16)$. Here, only singlet,
$v=\underline{16}$,
$s = \underline{128}$, and $s^{\prime} = \underline{128}^{\prime}$
representations of $SO(16)$
are allowed by the WSCA ($s = \underline{128}$ and
$s^{\prime}= \underline{128}^{\prime}$ are
the two nonequivalent, real spinor representations with the highest
weights
${\pi}_{7,8} =1/2 ({\epsilon}_1+{\epsilon}_2, + \dots +
{\epsilon}_7 \mp {\epsilon}_8)$,
${\epsilon}_{i}{\epsilon}_{j}= {\delta}_{ij}$).
 From the item 2. we can
obtain the following $SU(8)\times U(1)$ representations: singlet,
$\underline{8}$+$\underline{\bar 8}$ $(=\underline{16})$,
$\underline{8}+\underline{56}+\underline{\bar {56}}+\underline{\bar 8}$
$(=\underline{128})$, and $ \underline{1}+\underline{28}+\underline{70} +
+\underline{\bar{28}} + \underline{\bar {1}}$ $(=\underline{128}^{'})$.
The highest weights of $SU(8)$ representations
 ${\pi}_1 ={\pi}( \underline{8})$, ${\pi}_7 = ={\pi}(\underline{\bar 8})$
and
${\pi}_3 =={\pi}(\underline{56})$,
${\pi}_5 =={\pi}(\underline{\bar {56}})$  are:
\begin{eqnarray}
{\pi}_{1}& = &1/8 ( \underbrace{7{\epsilon}_1 - {\epsilon}_2
- {\epsilon}_3 - {\epsilon}_4 - {\epsilon}_5}  \underbrace{- {\epsilon}_6
-{\epsilon}_7 -  {\epsilon}_8}),\nonumber \\
{\pi}_{7}& = & 1/8(  \underbrace{{\epsilon}_1 + {\epsilon}_2
+ {\epsilon}_3 + {\epsilon}_4 + {\epsilon}_5}  \underbrace{+ {\epsilon}_6
+{\epsilon}_7 - 7 {\epsilon}_8}),\nonumber \\
{\pi}_{3}& = &1/8( \underbrace{{5\epsilon}_1+ 5{\epsilon}_2 + 5{\epsilon}_3
- 3{\epsilon}_4 - 3{\epsilon}_5}  \underbrace{- 3{\epsilon}_6 -
3{\epsilon}_7 - 3{\epsilon}_8}),\nonumber\\
 {\pi}_{5}& = &1/8( \underbrace{{3\epsilon}_1+ 3{\epsilon}_2 + 3 {\epsilon}_3
+ 3{\epsilon}_4 + 3{\epsilon}_5}  \underbrace{ - 5{\epsilon}_6 -
5{\epsilon}_7  - 5{\epsilon}_8}).
\end{eqnarray}
Similarly,the highest weights of $SU(8)$ representations
 ${\pi}_2 =={\pi}( \underline{28})$,  ${\pi}_6 =={\pi}(\underline{\bar {28}})$
and
${\pi}_4 =={\pi}(\underline{70})$ are:
\begin{eqnarray}
{\pi}_{2}& = &1/4 ( \underbrace{3{\epsilon}_1 + 3{\epsilon}_2 -{\epsilon}_3
-{\epsilon}_4 -{\epsilon}_5}  \underbrace{ -{\epsilon}_6
-{\epsilon}_7 -  {\epsilon}_8}),\nonumber \\
{\pi}_{6}& = &1/4 ( \underbrace{{\epsilon}_1 + {\epsilon}_2
+ {\epsilon}_3 + {\epsilon}_4 + {\epsilon}_5}  \underbrace{
+  {\epsilon}_6 -
3{\epsilon}_7 - 3 {\epsilon}_8 }),\nonumber \\
{\pi}_{4}& = &1/2( \underbrace{{\epsilon}_1+ {\epsilon}_2 + {\epsilon}_3
+ {\epsilon}_4 - {\epsilon}_5}  \underbrace{ - {\epsilon}_6 -
{\epsilon}_7 - {\epsilon}_8}).
\end{eqnarray}

However, as we will demonstrate, in each of the string sectors
the generalized Gliozzi--Scherk--Olive projection (the $GSO$ projection
in particular guarantees the modular invariance and supersymmetry of the
theory and also give some nontrivial restrictions on gauge
groups and its representations) necessarily eliminates either
$\underline{128}$ or $\underline{128}^{\prime}$. It is therefore important
that, in order to incorporate chiral matter in the model, only one spinor
representation is sufficient. Moreover, if one wants to solve the
chirality problem applying further $GSO$ projections (which break the
gauge symmetry) the representation $\underline {\bar{10}}$ which
otherwise, together with $\underline{10}$, could form real Higgs
representation, also disappears from this sector. Therefore, the existence of
$\underline{\bar{10}}_{-1/2}$+$\underline{10}_{1/2}$, needed for breaking
$SU(5)\times U(1)$ is incompatible (by our opinion) with the possible
solution of the chirality problem for the family matter fields.

Thus, in the rank eight group $SU(8) \times U(1) \subset SO(16)$ with Higgs
representations from the level-one WSCA only, one cannot arrange for further
symmetry breaking. Moreover, construction of the realistic fermion mass
matrices
seems to be impossible. In old-fashioned GUTs (see e.g.\cite{23i}), not
originating from strings, the representations of the level two were commonly
used to solve these problems.

The way out of this difficulty is based on the following important
observations. Firstly, all higher-dimensional 
representations of (simply laced)
groups like $SU(N)$, $SO(2N)$ or $E(6)$, which belong to the level two
representation of the WSCA (according to equation \ref{eq1}),
appear in the direct product of the level one representations:

\begin{eqnarray}\label{eq3}
R_G(x=2) \subset R_G(x=1) \times {R_G}^{'}(x=1).
\end{eqnarray}

For example, the level-two representations of $SU(5)$
 will appear in the corresponding direct products of
\begin{eqnarray}
 \underline{15},
\underline{24}, \underline{40}, \underline{45}, \underline{50},
\underline{75}
\subset \underline{5}\times
\underline{5}, \underline{5}\times\underline{\bar 5}, \underline{5}\times
\underline{10}, \mbox{ etc.}
\end{eqnarray}

In the case of $SO(10)$ the level two
representations   can be obtained by the
suitable direct products:
\begin{eqnarray}
 \underline{45}, \underline{54}, \underline{120},
\underline{126}, \underline{210}, \underline{144} \subset
\underline{10}\times\underline{10},
\underline{\bar{16}}\times\underline{10},
\underline{\bar{10}}\times\underline{16}, \underline{16}\times
\underline{16}, \underline{\bar{16}}\times\underline{16}.
\end{eqnarray}

The level-two
representations of $E(6)$ are the corresponding factors of the
decomposition of the direct
products:
\begin{eqnarray}
 \underline{78}, \underline{351}, \underline {351}^{'},
\subset
\underline{\bar{27}}\times\underline{27}\mbox{ or }\underline{27}
\times\underline{27}.
\end{eqnarray}

The only exception from this rule is the $E(8)$
group, two level-two representations ($\underline {248}$ and
$\underline{3875}$) of which cannot be constructed as a product of level-one
representations \cite{24i}.

Secondly, the diagonal (symmetric) subgroup $G^{\rm symm}$ of $G \times G $
effectively corresponds to the level-two WSCA
$g(x=1)\oplus g(x=1)$ \cite{25i,26i} because taking the $G\times G$
representations in the form $(R_G,R^{'}_G)$ of the $G\times G$,
where $R_G$ and $R_G^{'}$ belong to the level-one of G,
one obtains representations of the form $R_G\times R^{'}_G$
when one considers only the diagonal subgroup of $G\times G$.
This observation is crucial, because such a construction
allows one to obtain level-two representations. (This construction
has implicitly been used in \cite{26i} (see also \cite{25i}
where we have  constructed some examples
of GUST with gauge symmetry realized
as a diagonal subgroup of direct product
of two rank eight groups $U(8) \times U(8) \subset SO(16) \times SO(16)$.)

In strings, however, not all level-two representations can be obtained in that
way because, as we will demonstrate, some of them become massive (with masses
of order of the Planck scale). The condition ensuring that states in the
string spectrum transforming as a representation $R$ are massless reads:
\begin{equation}\label{eq4}
h(R) = \frac{Q_R}{2 k + Q_{ADJ}} =
\frac{Q_R}{2 Q_M} \leq 1 ,
\end{equation}
where $Q_i$ is the quadratic Casimir invariant of the corresponding
representations, and M has been already defined before (see eq. \ref{eq2}).
Here the conformal weight is defined by $L_0 |R{\rangle } = h(R)|R{\rangle }$,
\begin{eqnarray}\label{eq 100}
L_0 =\frac{1}
{2k + Q_{\psi}}\times  \biggl (\sum_{a=1}^{\dim g}(J_0^aJ_0^a +
2 \sum_{n=1}^{\infty}{J_{-n}^aJ_n^a})\biggr),
\end{eqnarray}
where $J_n^a|R{\rangle }=0$ for $n>0$, $J_0^a|R{\rangle }= t^a|R{\rangle }$ .
The condition (\ref {eq4}), when combined with (\ref{eq1}), gives a
restriction on the rank of GUT's group ($r \leq 8$), whose representations
can accommodate chiral matter
fields. For example, for antisymmetric representations of $SU(n=l+1)$ we have
the following values correspondingly : $h = p (n-p) /(2 n )$.
More exactly , for SU(8) group: $h(\underline 8)=7/16$,
$h(\underline{28})=3/4$, $h(\underline{56})=15/16$, $h(\underline {70})=1 $;
for SU(5), correspondingly $h(\underline 5)=2/5$ and $h(\underline{10})=3/5$;
for SU(3) group
$h(\underline 3)=1/3$ although  for adjoint representation of SU(3) -
$h(\underline 8)=3/4$;
for SU(2) doublet representation we have $h(\underline 2)=1/4$.
For vector representation of orthogonal series $D_l$, $h= 1/2$,
and, respectively, for spinor, $h(spinor)= l/8$.

There are some another important cases. The values of conformal
weights for $ G =SO(16)$ or $E(6) \times SU(3)$, representations
$\underline{128}$, $(\underline{27}, \underline {3})$
($h(\underline{128})= 1$, $h(\underline{27}, \underline{3}) = 1$)
respectively, satisfy both conditions. Obviously, these (important for
incorporation of chiral matter) representations will exist in the
level-two WSCA of the symmetric subgroup of the group $G\times G$.

In general, condition (\ref{eq4}) severely constrains  massless string states
transforming as $(R_G(x = 1), {R_G}^{'}(x = 1))$ of the direct product
$G\times G$. For example, for $SU(8)\times SU(8)$ and for $SU(5)\times SU(5)$
constructed from $SU(8)\times SU(8)$ only representations of the form
\begin{eqnarray}\label{eq5}
R_{N,N} =  \bigl((\underline {N},\underline {N}) +
h.c.\bigr ),\,\,\,\,\,
\bigl((\underline {N},\underline {\bar N}) + h.c.\bigr);\,\,\,
\end{eqnarray}
with $h(R_{N,N}) = (N-1)/N$, where $N\,=\,8$ or $5$ respectively can be
massless. For $SO(2N) \times SO(2N)$ massless states are contained
only in representations
\begin{eqnarray}\label{eq6}
R_{v,v} = (\underline{2N} ,\underline{2N})
\end{eqnarray}
with $h(R_{v,v}) = 1$. Thus, for the GUSTs based on a diagonal subgroup
$G^{\rm symm}\subset G\times G$, $G^{\rm symm}$ -- high dimensional
representations,
which are embedded in $R_G(x = 1) \times {R^{'}}_G(x = 1)$ are also severely
constrained by  condition (\ref{eq4}).

For spontaneous breaking of $G\times G$ gauge symmetry down to $G^{\rm symm}$
(rank $G^{\rm symm}$ = rank $G$) one can use the direct product of
representations $R_G(x=1)\times R_G(x=1)$, where $R_G(x = 1)$ is the
fundamental representation of $G = SU(N)$ or vector representation of
$G = SO(2N)$. Furthermore, $G^{\rm symm}\subset G\times G$ can subsequently
be broken down to a smaller dimension gauge group (of the same rank as
$G^{\rm symm}$) through the VEVs of the adjoint representations which can
appear as a result of $G\times G$ breaking. Alternatively, the real Higgs
superfields (\ref{eq5}) or (\ref{eq6}) can directly break the $G\times G$
gauge symmetry down to a $G_1^{\rm symm}\subset G^{\rm symm}$ (rank
$G_1^{\rm symm}\leq$ rank $G^{\rm symm}$). For example when
$G = SU(5)\times U(1)$ or $SO(10)\times U(1)$,  $G\times G$ can
directly be broken in this way down to
$SU(3^c)\times G^I_{EW}\times G^{II}_{EW}\times\ldots$.

The above examples show clearly, that within the framework of GUSTs with the
WSCA one can get interesting gauge symmetry breaking chains including the
realistic ones when $G\times G $ gauge symmetry group is considered.
However the lack of the higher dimension representations (which are forbidden
by \ref{eq4}) on the level-two WSCA prevents the construction of the realistic
fermion mass matrices.
That is why we consider an extended grand unified string model of rank eight
$SO(16)$ or $E(6) \times SU(3)$ of $E(8)$.

The full chiral $SO(10)\times SU(3)\times U(1)$ matter multiplets can be
constructed from $SU(8)\times U(1)$--multiplets
\begin{eqnarray}\label{eq7}
(\underline {8} + \underline {56} + \underline {\bar{8}}
+ \underline {\bar{56}})~ = ~\underline {128}
\end{eqnarray}
of $SO(16)$. In the 4-dimensional heterotic superstring with free complex
world sheet fermions, in the spectrum of the Ramond sector there can appear
also
representations which are factors in the decomposition of
$\underline{128^{'}}$,
in particular, $SU(5)$-decouplets $(\underline {10} + \underline {\bar {10}})$
from $(\underline{28}+\underline{\bar{28}})$ of $SU(8)$. However their
$U(1)_5$ hypercharge does not allow to use
them for $SU(5)\times U(1)_5$--symmetry
breaking. Thus, in this approach we have only singlet and
$(\underline{5} + \underline{\bar5})$ Higgs fields which can break the grand
unified $SU(5)\times U(1)$ gauge symmetry. Therefore it is necessary (as we
already explained) to construct rank eight GUST based on a diagonal subgroup
$G^{\rm symm}\subset G\times G$ primordial symmetry group, where in each
rank eight group $G$ the Higgs fields will appear only in singlets and in the
fundamental representations as in (see \ref{eq5}).

A comment concerning $U(1)$ factors can be made here. Since the available
$SU(5)\times U(1)$ decouplets have non-zero hypercharges with respect
to $U(1)_5$ and $U(1)_H$, these $U(1)$ factors may remain unbroken down to
the low energies in the model considered which seems to be very interesting.

\section{Modular Invariance in GUST Construction with
Non-Abelian Gauge Family Symmetry}\label{sec5}
\subsection{Spin-Basis in the Free World-Sheet Fermion Sector.}
\label{sbsec50}

 The GUST model is completely defined by a set $\Xi$ of spin
boundary conditions for all these world-sheet fermions.
In a diagonal basis the vectors of $\Xi$ are determined by the values
of phases $\alpha(f)$ $\in$(-1,1]
fermions $f$ acquire ~($f\longrightarrow -\exp({i\pi\alpha(f)}) f $) ~when
parallel transported around the string.
To construct the GUST according to the
scheme outlined at the end of the previous section we consider
three different bases each of them  with six elements
$B= {b_1, b_2, b_3, b_4 \equiv S, b_5, b_6 }$. (See Tables \ref{tabl1},
\ref{Basis 1i}, \ref{bas1ii}, \ref{basis2}, \ref{basis 3}, \ldots)

Following \cite{19i} (see Appendix A) we  construct the canonical
basis in such a way that the
vector $\bar1$, which belongs to $\Xi$, is the first element $b_1$ of the
basis.
The basis vector $b_4=S$ is the generator of supersymmetry \cite{20i}
responsible for the conservation of the space-time $SUSY$.

In this chapter we have chosen a basis in which all left movers
$({\psi}_{\mu}; {\chi}_i, y_i, {\omega}_i; i=1,\ldots,6)$ (on which the
world sheet supersymmetry is realized nonlinear\-ly) as well as 12 right
movers $({\bar\varphi}_k; k=1,\ldots,12)$ are
real whereas (8 + 8) right movers $\bar{\Psi}_A$, $\bar{\Phi}_M$ are complex.
Such a construction corresponds to $SU(2)^6$ group of automorphisms of the
left supersymmetric sector of a string. Right- and left-moving real
fermions can be used for breaking $G^{\rm comp}$ symmetry \cite{20i}. In
order to have a possibility to reduce the rank of the compactified group
$G^{\rm comp}$, we have to select the spin boundary conditions for the maximal
possible number, $N_{LR}$ = 12, of left-moving, ${\chi}_{3,4,5,6}$,
$y_{1,2,5,6}$, ${\omega}_{1,2,3,4}$, and right-moving,
{}~${\bar{\phi}}^{1,\ldots,12}$~
(${\bar{\phi}}^p~=~{\bar{\varphi}}_p$,~$p=1,\ldots,12$) real fermions.
The KMA based on $16$ complex right moving fermions gives rise to the
"observable" gauge group $G^{\rm obs}$ with:
\begin{equation}\label{eq8}
\rank (G^{\rm obs})  \leq 16.
\end{equation}

The study of the Hilbert spaces of the string theories
is connected to the problem of finding all possible choices
of the GSO coefficients ${\cal C}
\left[
\begin{array}{c}
{\alpha} \\
{\beta}
\end{array}\right]$
(see Appendix A),
such that the one--loop partition function
\begin{equation}
Z=\sum_{ \alpha , \beta} {\cal C}
\left[
\begin{array}{c}
{\alpha} \\
{\beta}
\end{array}\right] \prod_f Z
\left[
\begin{array}{c}
{\alpha}_f \\
{\beta}_f
\end{array}\right]
\label{Z}
\end{equation}
and its multiloop counterparts are all modular invariant.
In this formula ${\cal C}
\left[
\begin{array}{c}
{\alpha} \\
{\beta}
\end{array}\right]$ are GSO coefficients,
$\alpha$ and $\beta$ are $(k+l)$--component spin--vectors
$\alpha=[\alpha(f_1^r), \ldots , \alpha(f_k^r);
\alpha(f_1^c), \ldots , \alpha(f_l^c)]$,
the components $\alpha_f$, $\beta_f$ specify the spin
structure of the $f$th fermion and $Z[\ldots]$ -- corresponding one-fermion
partition functions on torus:
$Z[\ldots]=\mbox{Tr exp}[2\pi iH_{({\rm sect.})}]$.

The physical states in the Hilbert space of a given sector $\alpha$ are
obtained
acting on the vacuum ${|0{\rangle }}_{\alpha}$ with the bosonic and fermionic
operators
with frequencies
\begin{eqnarray}\label{eq9}
n(f) = 1/2 + 1/2 \alpha(f),\:\:\:   n(f^*) = 1/2  -1/2 \alpha(f^*)
\end{eqnarray}
and subsequently applying the generalized GSO projections. The physical states
satisfy the Virasoro condition:
\begin{eqnarray}\label{eq10}
M_L^2 = - 1/2 + 1/8 \:({\alpha}_L \cdot {\alpha}_L) + N_L =
-1 +1/8\: ({\alpha}_R \cdot {\alpha}_R) +N_R = M_R^2,
\end{eqnarray}
where $\alpha=({\alpha}_L,{\alpha}_R)$ is  a sector in the set $\Xi$,
$N_L = {\sum }_{L} ({\rm frequencies})$ and  $N_R = {\sum}_{R} ({\rm freq.})$.

We keep the same sign convention for the fermion number operator  $F$
as in \cite{20i}. For complex fermions we have $F_{\alpha}(f)~=~1$,
{}~$F_{\alpha}(f^*)~=~-1$ with the exception of the periodic fermions
for which we get $F_{{\alpha}=1}(f) ~=~-1/2(1-{\gamma}_{5f})$, where
{}~${\gamma}_{5f}|\Omega{\rangle }~=~|\Omega{\rangle }$,
{}~${\gamma}_{5f}b^+_o|\Omega{\rangle }~=~-b^+_o|\Omega{\rangle }$.

The full Hilbert space of the string theory is constructed as a direct sum of
different sectors ${\sum}_{i} {m_ib_i}$, ($m_i=0,1,\ldots,N_i$), where the
integers $N_i$ define additive groups $Z(b_i)$ of the basis vectors $b_i$.
The generalized GSO projection leaves in sectors $\alpha$ those states, whose
$b_i$-fermion number satisfies:
\begin{equation}\label{eq11}
\exp(i \pi b_i F_{\alpha})
= {\delta}_{\alpha}
{\cal C}^*
\left[
\begin{array}{c}
{\alpha} \\
b_i
\end{array}\right],
\end{equation}
where the space-time phase
${\delta}_{\alpha}=\exp(i\pi{\alpha}({\psi}_{\mu}))$
is equal $-1$ for the Ramond sector and +1 for the Neveu-Schwarz sector.

\subsection{$SU(5)\times U(1)\times SU(3)\times U(1)$ -- Model~1.}
\label{sbsec51}
Model~1 is defined by 6 basis vectors given in Table \ref{tabl1} which
generates the $Z_2\times Z_4\times Z_2\times Z_2\times Z_8\times Z_2$
group under addition.

\begin{table}[b]
\caption{\bf Basis of the boundary conditions for all world-sheet fermions.
Model~1.}
\label{tabl1}
\footnotesize
\begin{center}
\begin{tabular}{|c||c|ccc||c|cc|}
\hline
Vectors &${\psi}_{1,2} $ & ${\chi}_{1,\ldots,6}$ & ${y}_{1,\ldots,6}$ &
${\omega}_{1,\ldots,6}$
& ${\bar \varphi}_{1,\ldots,12}$ &
${\Psi}_{1,\ldots,8} $ &
${\Phi}_{1,\ldots,8}$ \\ \hline
\hline
$b_1$ & $1 1 $ & $1 1 1 1 1 1$ &
$1 1 1 1 1 1 $ & $1 1 1 1 1 1$ &  $1^{12} $ & $1^8 $ & $1^8$ \\
$b_2$ & $1 1$ & $1 1 1 1 1 1$ &
$0 0 0 0 0 0$ &  $0 0 0 0 0 0 $ &  $ 0^{12} $ &
${1/2}^8 $ & $0^8$  \\
$b_3$ & $1 1$ & $ 1 1 1 1 0 0 $ & $0 0 0 0 1 1 $ & $ 0 0 0 0 0 0 $ &
$0^4  1^8 $ & $ 0^8 $ &  $ 1^8$ \\
$b_4 = S $ & $1 1$ &
$1 1 0 0 0 0 $ & $ 0 0 1 1 0 0 $ & $ 0 0 0 0 1 1 $ &
$ 0^{12} $ & $ 0^8 $ & $ 0^8 $ \\
$  b_5 $ & $ 1 1 $ & $ 0 0  1 1 0 0 $ &
$0 0 0 0 0 0 $ &  $1 1  0 0 1 1$ &  $ 1^{12} $ &
${ 1/4}^5  {-3/4}^3 $ & $ {-1/4}^5\ {3/4}^3  $  \\
$  b_6 $ & $ 1 1 $ & $ 1 1  0 0  0 0 $ &
$ 0 0  0 0 1 1 $ &  $0 0  1 1 0 0$ &  $ 1^2 0^4 1^6 $ &
$ 1^8  $ & $ 0^8  $  \\
\hline \hline
\end {tabular}
\end{center}
\normalsize
\end{table}

In our approach the basis vector $b_2$ is constructed as a complex vector
with the $1/2$ spin-boundary conditions for the right-moving fermions
${\Psi}_A$, $A = 1,\ldots,8$. Initially  it generates chiral  matter fields
in the $\underline{8}+\underline{56}+\underline{\bar{56}}+\underline{\bar{8}}$
representations of $SU(8)\times U(1)$, which subsequently are decomposed under
$SU(5)\times U(1)\times SU(3)\times U(1)$ to which $SU(8)\times U(1)$ gets
broken by applying the $b_5$ $GSO$ projection.

Generalized GSO projection coefficients are originally  defined up to fifteen
signs  but some of them are fixed by the supersymmetry conditions.
Below, in Table \ref{tabl2}, we present a set of numbers
$$
\gamma\left[\begin{array}{c}b_i\\b_j\end{array}\right]=\frac{1}{i \pi}
\log{\cal C}\left[\begin{array}{c}b_i\\b_j\end{array}\right].
$$
which we use as a basis for our GSO projections.

\begin{table}[b]
\caption{\bf The choice of the GSO basis $\gamma [b_i, b_j]$. Model~1.
($i$ numbers rows and $j$ -- columns)}
\label{tabl2}
\footnotesize
\begin{center}
\begin{tabular}{|c||c|c|c|c|c|c|}
\hline
& $b_1$ & $b_2$ & $b_3$ & $b_4$ & $b_5$ & $b_6$\\ \hline
\hline
$b_1$ & $0$ &    $1$ & $1$ & $1$ &    $1$ & $0$\\
$b_2$ & $1$ &  $1/2$ & $0$ & $0$ &  $1/4$ & $1$\\
$b_3$ & $1$ & $-1/2$ & $0$ & $0$ &  $1/2$ & $0$\\
$b_4$ & $1$ &    $1$ & $1$ & $1$ &    $1$ & $1$\\
$b_5$ & $0$ &    $1$ & $0$ & $0$ & $-1/2$ & $0$\\
$b_6$ & $0$ &    $0$ & $0$ & $0$ &    $1$ & $1$\\
\hline \hline
\end {tabular}
\end{center}
\normalsize
\end{table}

In our case of the ~${Z_2}^4\times {Z_4}\times {Z_8}$ ~model, we initially
have $256\times2$ sectors. After applying the GSO-projections we get only
$49\times2$ sectors containing massless states, which depending on the
vacuum energy values, $E^{vac}_L$ and $E^{vac}_R$, can be naturally divided
into some classes  and which determine the GUST representations.

Generally RNS (Ramond -- Neveu-Schwarz) sector (built on vectors $b_1$
and $S=b_4$) has high symmetry including $N=4$ supergravity and gauge
$SO(44)$ symmetry. Corresponding gauge bosons are constructed as follows:
\begin{eqnarray}\label{eq12}
&{\psi}_{1/2}^{\mu}{|0{\rangle }}_L \otimes  {\Psi}_{1/2}^I
 {\Psi}_{1/2}^{J} |0{\rangle }_R ,\nonumber\\
 &{\psi}_{1/2}^{\mu}{|0{\rangle }}_L \otimes  {\Psi}_{1/2}^I
 {\Psi}_{1/2}^{*J} |0{\rangle }_R ,\:\:I,~J~=1,~\dots,22 &
\end{eqnarray}
 While $U(1)_J$ charges for Cartan subgroups is given by formula
 $Y=\frac{\alpha}{2}+F$ (where $F$ --- fermion number, see (\ref{eq11})),
 it is obvious that states (\ref{eq12}) generate root lattice for
 $SO(44)$:
 \begin{eqnarray}
\pm \varepsilon_I \pm \varepsilon_J \ \ (I\neq J);\qquad
  \pm \varepsilon_I \mp \varepsilon_J
\end{eqnarray}
 The other vectors break $N=4$ SUSY to $N=1$ and gauge group $SO(44)$
 to $SO(2)^3_{1,2,3}\times SO(6)_4\times {\left[ SU(5)\times U(1)
 \times SU(3)_H\times U(1)_H\right]}^2$, see Figure \ref{fig1}.

 Generally, additional basis vectors can generate extra vector bosons and
 extend gauge group that remains after applying GSO-projection to
 RNS-sector. In our case dangerous sectors are: $2b_2+nb_5,\ n=0,2,4,6;
 \  2b_5;6b_5$. But our choice of GSO coefficients cancels all the vector
 states in these sectors. Thus gauge bosons in this model  appear
 only from RNS-sector.

\begin{table}[t]
\caption{\bf The list of quantum numbers of the states. Model~1.}
\label{tabl3}
\footnotesize
\noindent \begin{tabular}{|c|c||c|cccc|cccc|} \hline
N$^o$ &$ b_1 , b_2 , b_3 , b_4 , b_5 , b_6 $&
$ SO_{hid}$&$ U(5)^I $&$ U(3)^I $&$ U(5)^{II} $&$
U(3)^{II} $&$ {\tilde Y}_5^I $&$ {\tilde Y}_3^I $&$ {\tilde Y}_5^{II} $&$
{\tilde Y}_3^{II}$ \\ \hline \hline
1 & RNS &&5&$\bar 3$&1&1&--1&--1&0&0 \\
  &     &&1&1&5&$\bar 3$&0&0&--1&--1 \\
  &0\ 2\ 0\ 1\ 2(6)\ 0&&5&1&5&1&--1&0&--1&0 \\
${\hat \Phi }$
  &&&1&3&1&3&0&1&0&1 \\
  &&&5&1&1&3&--1&0&0&1 \\
  &&&1&3&5&1&0&1&--1&0 \\ \hline \hline
2 &0\ 1\ 0\ 0\ 0\ 0&&1&3&1&1&5/2&--1/2&0&0 \\
  &&&$\bar 5$&3&1&1&--3/2&--1/2&0&0 \\
  &&&10&1&1&1&1/2&3/2&0&0 \\
${\hat \Psi }$
  &0\ 3\ 0\ 0\ 0\ 0&&1&1&1&1&5/2&3/2&0&0 \\
  &&&$\bar 5$&1&1&1&--3/2&3/2&0&0 \\
  &&&10&3&1&1&1/2&--1/2&0&0 \\ \hline
3 &0\ 0\ 1\ 1\ 3\ 0&$-_1\ \pm_2$&1&1&1&3&0&--3/2&0&--1/2 \\
  &0\ 0\ 1\ 1\ 7\ 0&$-_1\ \pm_2$&1&$\bar 3$&1&1&0&1/2&0&3/2 \\
${\hat \Psi }^H$
  &0\ 2\ 1\ 1\ 3\ 0&$+_1\ \pm_2$&1&$\bar 3$&1&3&0&1/2&0&--1/2 \\
  &0\ 2\ 1\ 1\ 7\ 0&$+_1\ \pm_2$&1&1&1&1&0&--3/2&0& 3/2 \\ \hline
4 &1\ 1\ 1\ 0\ 1\ 1&$\mp_1\ \pm_3$&1&1&1&$\bar 3$&0&--3/2&0&1/2 \\
  &1\ 1\ 1\ 0\ 5\ 1&$\mp_1\ \pm_3$&1&$\bar 3$&1&1&0&1/2&0&--3/2 \\
${\hat \Phi }^H$
  &1\ 3\ 1\ 0\ 1\ 1&$\pm_1\ \pm_3$&1&$\bar 3$&1&$\bar 3$&0&1/2&0&1/2 \\
  &1\ 3\ 1\ 0\ 5\ 1&$\pm_1\ \pm_3$&1&1&1&1&0&--3/2&0&--3/2 \\ \hline
5 &0\ 1(3)\ 1\ 0\ 2(6)\ 1&$-_1\ \pm_3$&1&3($\bar 3$)&1&1&$\pm$5/4&$\pm$1/4
&$\pm$5/4&$\mp$3/4 \\
  &&$+_1\ \pm_3$&5($\bar 5$)&1&1&1&$\pm$1/4&$\mp$3/4&$\pm$5/4&$\mp$3/4 \\
${\hat \phi }$
  &0\ 1(3)\ 1\ 0\ 4\ 1&$-_1\ \pm_3$&1&1&1&3($\bar 3$)&$\pm$5/4&$\mp$3/4
&$\pm$5/4&$\pm$1/4 \\
  &&$+_1\ \pm_3$&1&1&5($\bar 5$)&1&$\pm$5/4&$\mp$3/4&$\pm$1/4&$\mp$3/4 \\
\hline
6 &1\ 2\ 0\ 0\ 3(5)\ 1&$\pm_1\ -_4$&1&1&1&1&$\pm$5/4&$\pm$3/4
&$\mp$5/4&$\mp$3/4 \\
  &1\ 1(3)\ 0\ 1\ 5(3)\ 1&$+_1\ \mp_4$&1&1&1&1&$\pm$5/4&$\pm$3/4
&$\pm$5/4&$\pm$3/4 \\
${\hat \sigma }$
  &0\ 0\ 1\ 0\ 2(6)\ 0&$\mp_3\ +_4$&1&1&1&1&$\pm$5/4&$\mp$3/4
&$\pm$5/4&$\mp$3/4 \\ \hline
\end{tabular}
\normalsize
\end{table}

In NS sector the $b_3$ GSO projection leaves $(5,\bar{3})+(\bar{5},3)$ Higgs
superfields:
\begin{equation}\label{eq14}
\chi^{1,2}_{1/2}|\Omega{\rangle }_L\otimes {\Psi}_{1/2}^a
{\Psi}_{1/2}^{i*};\,\, {\Psi}_{1/2}^{a*}
{\Psi}_{1/2}^{i} |\Omega{\rangle }_R\ \ \mbox{and exchange}\  \Psi
 \longrightarrow\Phi, \end{equation}
where $a,\:b=1,\dots,\:5,\ \ i,\:j=1,2,3$.

Four $(3_H + 1_H)$ generations of chiral matter fields from
$~({SU(5)\times SU(3)})_I$ group forming $SO(10)$--multiplets
$(\underline 1, \underline 3) + (\underline {\bar 5},\underline 3) +
(\underline {10}, \underline 3)$ ; $( \underline 1,\underline 1) +
(\underline {\bar 5},\underline 1) + (\underline { {10}},\underline 1)$
are contained in $b_2$ and $3b_2$ sectors. Applying $b_3$ $GSO$
projection to the $3b_2$ sector yields the following massless states:

\begin{eqnarray}\label{eq15}
b_{\psi_{12}}^+ b_{{\chi}_{34}}^{+} b_{{\chi}_{56}}^+
|\Omega {\rangle }_L& \otimes &  \Biggl \{   {\Psi}_{3/4}^{i*}~,~~
  {\Psi}_{1/4}^{a}   {\Psi}_{1/4}^{b}   {\Psi}_{1/4}^{c},~
  {\Psi}_{1/4}^{a}   {\Psi}_{1/4}^{i}   {\Psi}_{1/4}^{j}~
\Biggr \} ~|\Omega{\rangle }_R, \nonumber \\
b_{{\chi}_{12}}^{+} b_{{\chi}_{34}}^{+} b_{{\chi}_{56}}^{+}
|\Omega {\rangle }_L& \otimes &  \Biggl \{   {\Psi}_{3/4}^{a*}~,~~
  {\Psi}_{1/4}^{a}   {\Psi}_{1/4}^{b}   {\Psi}_{1/4}^{i},~~
  {\Psi}_{1/4}^{i}   {\Psi}_{1/4}^{j}   {\Psi}_{1/4}^{k}~
\Biggr \}~|\Omega{\rangle }_R
\end{eqnarray}
with the space-time chirality  ${\gamma}_{5 {\psi}_{12}}~=-~1$
and  ${\gamma}_{5 {\psi}_{12}}~=~1$, respectively.
In these formulae the Ramond creation operators
$b_{\psi_{1,2}}^+$ and $b_{{\chi}_{\alpha, \beta}}^+$ of the zero modes
are built of a pair of real fermions (as indicated by double indices):
${\chi}_{\alpha, \beta}$,~ $(\alpha,\beta)$~= ~$(1,2)$, ~$(3,4)$, ~$(5,6)$.
Here, as in (\ref{eq14}) indices take values $a,b$~=~1,\ldots,5 ~and
$i,j$~=~1,2,3 respectively.

We stress that without using the ~$b_3$~ projection we would get matter
supermultiplets belonging to real representations only i.e. "mirror"
particles would remain in the spectrum. The ~$b_6$~ projection instead,
eliminates all chiral matter superfields from $U(8)^{II}$ group.
It is interesting, that without $b_6$-vector the Model~1 is
fictitious U(1)-anomaly \cite{anomaly} fully free.

Since the matter fields form the chiral multiplets of $SO(10)$, it is possible
to write down  $U(1)_{Y_5}$--hypercharges of massless states. In order to
construct the right electromagnetic charges for matter fields we must define
the hypercharges operators for the observable $U(8)^{I}$ group as

\begin{equation}\label{eq16}
Y_5=\int^\pi_0 d\sigma\sum_a \Psi^{*a}\Psi^a ,\,\,\,\,\,
Y_3=\int^\pi_0 d\sigma\sum_i \Psi^{*i}\Psi^i
\end{equation}
and analogously for the $U(8)^{II}$ group.

Then the orthogonal combinations
\begin{equation}\label{eq17}
 \tilde Y_5 = {1\over 4}(Y_5 + 5Y_3), \,\,\,\,\,
 \tilde Y_3 = {1\over 4}(Y_3 - 3Y_5),
\end{equation}
play the role of the hypercharge operators of $U(1)_{Y_5}$ and
$U(1)_{Y_H}$ groups,
respectively. In  Table \ref{tabl3} we give the hypercharges
 $\tilde Y_5^{I},\tilde Y_3^{I}, \tilde Y_5^{II},\tilde Y_3^{II}$.

 The full list of states in this model is given in  Table \ref{tabl3}.
 For fermion states only sectors with positive (left) chirality are
written. Superpartners arise from sectors with $S=b_4$-component
changed by 1. Chirality under hidden $SO(2)^3_{1,2,3}\times SO(6)_4$ is
defined as $\pm_1,\ \pm_2,\ \pm_3,\ \pm_4$ respectively. Lower signs in item
5 and 6 correspond to sectors with components given in brackets.

In Model~1 there exists a
possibility to break the GUST group $(U(5)\times U(3))^I\times (U(5)\times
U(3))^{II}$
down to the symmetric group by the ordinary Higgs mechanism:
\begin{equation}\label{eq18}
G^I\times G^{II} \rightarrow {G}^{\rm symm}
\rightarrow \ldots
\end{equation}
To achieve such breaking one can use nonzero vacuum expectation values of the
tensor Higgs fields (see Table \ref{tabl3}, row No 1),
contained in the $2b_2 + 2(6)b_5 (+S)$ sectors which transform
under the $(SU(5)\times U(1)\times SU(3)\times U(1))^{\rm symm}$ group in the
following way
$$(\underline 5,\underline 1;\underline 5,\underline 1)_{(-1,0;-1,0)}$$.
(see the section 8).

For the electromagnetic charge we get
\begin{eqnarray}
\label{eq23}
Q_{em} &=& Q^{II} - Q^I = \nonumber\\
&=& (T^{II}_5 - T^{I}_5) + \frac{2}{5}(\tilde Y^{II}_5 - \tilde Y^{I}_5) =
\bar T_5 + \frac{2}{5}\bar Y_5,
\end{eqnarray}
where $ T_5 = diag (\frac{1}{15},\frac{1}{15},\frac{1}{15}, \frac{2}{5},
-\frac{3}{5})$. Note, that this charge quantization does not lead to exotic
states with fractional electromagnetic charges \\(e.g. $Q_{em} =\pm 1/2,
\pm 1/6$).

In turn, the Higgs fields
from the NS sector
\begin{eqnarray}\label{eq26}
(\underline 5, \underline {\bar 3})_{(-1,-1)} +
(\underline {\bar 5}, \underline 3)_{(1,1)}
\end{eqnarray}
are obtained from N=2 SUSY vector representation ~$\underline{63}$
of $SU(8)^{I}$ (or $SU(8)^{II}$) by applying the $b_5$ GSO projection.
These Higgs fields 
can be used for constructing
chiral fermion (see Table \ref{tabl3}, row No 2) mass matrices.

The $b$ spin boundary conditions (Tabl.\ref{tabl1}) generate chiral matter and
Higgs fields with the $GUST$ gauge symmetry
$G_{\rm comp}\times (G^I\times G^{II})_{\rm obs}$
(where $G_{\rm comp} = {U(1)}^3\times SO(6) $ and $G^{I,II}$ have been already
defined). The chiral matter spectrum, which we denote as
${\hat \Psi}_{(\Gamma, N)}$ ~~with
($\Gamma = \underline 1,\underline {\bar 5},
\underline{10}; ~~N=\underline3, \underline1$),~~consists of ~~
$N_g = 3_H + 1_H $~~ families. See Table \ref{tabl3}, row No 2 for the
$((SU(5)\times U(1))\times (SU(3)\times U(1))_H)^{\rm symm}$ quantum numbers.

The $SU(3)_H$ anomalies of the matter fields (row No 2) are naturally
canceled by the chiral "horizontal" superfields  forming two sets:
${\hat \Psi}^H_{(1,N;1,N)}$ and ${\hat \Phi}^H_{(1, N;1, N)}$,
{}~$\Gamma = \underline 1$,~~$N = \underline 1, \,  \underline 3$,~
(with both ${SO(2)}$ chiralities, see Table \ref{tabl3}, row No 3, 4
respectively).

The horizontal fields (No 3, 4) cancel all $SU(3)^{I}$ anomalies introduced
by the chiral matter spectrum (No 2) of the  $(U(5)\times U(3))^{I}$
group (due to $b_6$ GSO  projection the chiral fields of
the $(U(5)\times U(3))^{II}$ group disappear from the final string spectrum).
Performing the decomposition of fields (No 3) under $(SU(5)\times
SU(3))^{\rm symm}$
we get (among other) three "horizontal" fields ${\hat \Psi }^H$:
\begin{eqnarray}\label{eq27}
2\times (\underline 1,\underline {\bar 3})_{(0,-1)}\quad ,\qquad
(\underline 1,\underline 1)_{(0,-3)}\quad ,\qquad
(\underline 1,\underline {\bar 6})_{(0,1)}\quad ,
\end{eqnarray}
coming from
${\hat \Psi}^H_{(\underline1,\bar{\underline 3};\underline1,\underline1)}$,
(and
${\hat \Psi}^H_{(\underline1,\underline1;\underline1,\underline {3})}$),
${\hat \Psi}^H_{(\underline1,\underline1;\underline1,\underline1)}$ and
${\hat \Psi}^H_{(\underline1,\underline {\bar3};\underline1,\underline 3)}$
respectively  which
make the low energy spectrum of the resulting model
${SU(3)_H}^{\rm symm}$-
anomaly free. The other fields ${\hat \Phi }^H$ arising from
rows No 4, Table \ref{tabl3}  form
anomaly-free representations of $(SU(3)_H \times U(1)_H)^{\rm symm}$:
\begin{eqnarray}\label{eq28}
2\times (\underline 1,\underline 1)_{(0,0)}\quad ,\qquad
(\underline 1,\underline {\bar3})_{(0,2)}~+~
(\underline 1,\underline 3)_{(0,-2)}\quad ,\qquad
(\underline 1,\underline {8})_{(0,0)}\quad.
\end{eqnarray}

The  superfields ~${\hat \phi}_{(\Gamma, N)} + h.c.$, where
($\Gamma = \underline 1, \underline 5$; $N = \underline 1,\underline 3$), from
the Table \ref{tabl3}, row No.\ 5 forming
representations of $(U(5) \times U(3))^{I.II}$ have either $Q^I$ or $Q^{II}$
exotic fractional charges.
Because of the strong $G^{\rm comp}$ gauge forces these fields may
develop the double scalar condensate $ {{\langle }\hat \phi \hat \phi{\rangle
}}$, which can also serve for $U(5)\times U(5)$ gauge symmetry breaking.
For example, the composite
condensate ${{\langle }{\hat \phi}_{(5,1;1,1)} {\hat \phi}_{(1,1;\bar
5,1)}{\rangle }}$ can
break the $U(5)\times U(5)$ gauge symmetry down to the symmetric diagonal
subgroup with generators of the form
\begin{eqnarray}\label{eq29}
{\triangle}_{\rm symm} (t) = t \times 1 + 1 \times t,
\end{eqnarray}
so for the electromagnetic charges  we would have the form
\begin{eqnarray}\label{eq30}
Q_{em} = Q^{II} + Q^I.
\end{eqnarray}
leading again to no exotic, fractionally charged states
in the low-energy string spectrum.

The superfields which transform nontrivially under the compactified group
$G^{\rm comp} = SO(6)\times {SO(2)}^{\times 3}$,~~(denoted as $\hat{\sigma}$),
and which are singlets of $(SU(5)\times SU(3))\times (SU(5)\times SU(3))$,
arise
in three sectors, see Table \ref{tabl3}, row No 6.
The superfields $\hat\sigma$ form the spinor representations $\underline4+
\underline {\bar4}$ of $SO(6)$ and they are also spinors of one of the $SO(2)$
groups.
With respect to the diagonal $G^{\rm symm}$ group with generators given by
(\ref{eq29}), some $\hat {\sigma}$-fields
are of zero hypercharges and can, therefore, be used for breaking the
$SO(6)\times {SO(2)}^{\times 3}$ group.

Note, that for the fields $\hat {\phi}$ and for the fields $\hat {\sigma}$
any other electromagnetic charge quantization different from (\ref{eq23}) or
(\ref{eq30}) would lead to "quarks" and "leptons" with the exotic fractional
charges, for example,
for the $\underline 5$- and $\underline 1$-multiplets according to the values
of hypercharges (see Table \ref{tabl3}, row No 6) the generator $Q^{II}$  (or
$Q^{I}$) has the
eigenvalues \\
$(\pm 1/6,\pm 1/6,\pm 1/6,\pm 1/2,\mp1/2)$ or $\pm 1/2$, respectively.

\noindent

Let us remark that the Model 1 can be built in another basis of the boundary 
conditions
for world-sheet fermions, for example, see the next tables down.

  \begin{table}[t]
\caption{\bf Basis of the boundary conditions for all world-sheet fermions.
Model $\protect\makebox{\bf\rm 1}'$, $Z_2\times Z_{6}\times Z_{12}\times Z_2
\times Z_2\times Z_2$.}
\label{Basis 1i}
\footnotesize
\begin{center}
\begin{tabular}{|c||c|ccc||c|cc|}
\hline
Vectors &${\psi}_{1,2} $ & ${\chi}_{1,\ldots,6}$ & ${y}_{1,\ldots,6}$ &
${\omega}_{1,\ldots,6}$
& ${\bar \varphi}_{1,\ldots,12}$ &
${\Psi}_{1,\ldots,8} $ &
${\Phi}_{1,\ldots,8}$ \\ \hline
\hline
$b_1$ & $1 1 $ & $1 1 1 1 1 1$ &
$1 1 1 1 1 1 $ & $1 1 1 1 1 1$ &  $1^{12} $ & $1^8 $ & $1^8$ \\
$b_2$ & $1 1$ & $1 1 1 1 1 1$ &
$0 0 0 0 0 0$ &  $0 0 0 0 0 0 $ &  $ 0^{12} $ &
$1 1 1 1 1 {1/3}^3 $ & $0^8$  \\
$b_3$ & $1 1$ & $ 1 1 0 0 1 1 $ & $0 0 0 0 0 0  $ & $ 0 0 1 1 0 0 $ &
$0^8  1^4 $ & $ {1/2}^5 {1/6}^3 $ &  $ {-1/2}^5 {1/6}^3$ \\
$b_4 = S $ & $1 1$ &
$1 1 0 0 0 0 $ & $ 0 0 1 1 0 0 $ & $ 0 0 0 0 1 1 $ &
$ 0^{12} $ & $ 0^8 $ & $ 0^8 $ \\
$  b_5 $ & $ 1 1 $ & $ 0 0 0 0 0 0 $ &
$1 1 0 0 0 0 $ &  $ 0 0 1 1 1 1 $ &  $ 1^{6} 0^2 1^2 0^2$ &
${ 1}^5  {0}^3 $ & $ {0}^5 {1}^3  $  \\
$  b_6 $ & $ 1 1 $ & $ 1 1  0 0  0 0 $ &
$ 0 0  0 0 1 1 $ &  $0 0  1 1 0 0$ &  $ 1^6 0^4 1^2 $ &
$ 0^8  $ & $ 1^8  $  \\
\hline \hline
\end {tabular}
\end{center}
\normalsize
\end{table}

\begin{table}[b]
\caption{\bf The choice of the GSO basis $\gamma [b_i, b_j]$. Model
$\protect\makebox{\bf\rm 1}'$. ($i$ numbers rows and $j$ -- columns)}
\label{tabl GSO 1i}
\begin{center}
\begin{tabular}{|c||c|c|c|c|c|c|}
\hline
& $b_1$ & $b_2$ & $b_3$ & $b_4$ & $b_5$ & $b_6$\\ \hline
\hline
$b_1$ & $0$ &    $1$         & $1/2$        & $0$ &  $0$ & $1$\\
$b_2$ & $0$ &  $2/3$ & $-1/6$       & $1$ &  $1$ & $1$\\
$b_3$ & $0$ & $1/3$          &$5/6$ & $1$ &  $0$ & $0$\\
$b_4$ & $0$ &    $0$         & $0$          & $0$ &  $0$ & $0$\\
$b_5$ & $0$ &    $1$         & $1$       & $1$ &  $1$ & $1$\\
$b_6$ & $1$ &    $0$         & $0$        & $1$ &  $1$ & $0$\\
\hline \hline
\end {tabular}
\end{center}
\end{table}

\begin{table}[b]
\caption{\bf Basis of the boundary conditions for all world-sheet fermions.
Model $\protect\makebox{\bf\rm 1}''$.}
\label{bas1ii}
\footnotesize
\begin{center}
\begin{tabular}{|c||c|ccc||c|cc|}
\hline
Vectors &${\psi}_{1,2} $ & ${\chi}_{1,\ldots,6}$ & ${y}_{1,\ldots,6}$ &
${\omega}_{1,\ldots,6}$
& ${\bar \varphi}_{1,\ldots,12}$ &
${\Psi}_{1,\ldots,8} $ &
${\Phi}_{1,\ldots,8}$ \\ \hline
\hline
$b_1$ & $1 1 $ & $1 1 1 1 1 1$ &
$1 1 1 1 1 1 $ & $1 1 1 1 1 1$ &  $1^{12} $ & ${\hat 1}^8 $ & ${\hat 1}^8$ \\
$b_2 = S $ & $1 1$ &
$1 1 0 0 0 0 $ & $ 0 0 1 1 0 0 $ & $ 0 0 0 0 1 1 $ &
$ 0^{12} $ & $ {\hat 0}^8 $ & $ {\hat 0}^8 $ \\
$b_3$ & $0 0$ & $ 0 0 1 1 1 1 $ & $0 0 0 0 1 1 $ & $ 0 0 1 1 0 0 $ &
$1^8  0^4 $ & $ {\hat 1}^8 $ &  $ {\hat 0}^8$ \\
$b_4$ & $1 1$ & $1 1 1 1 1 1$ &
$0 0 0 0 0 0$ &  $0 0 0 0 0 0 $ &  $ 0^{12} $ &
${\hat{1/2}}^8 $ & ${\hat 0}^8$  \\
$  b_5 $ & $ 1 1 $ & $ 0 0  1 1 0 0 $ &
$0 0 0 0 0 0 $ &  $1 1  0 0 1 1$ &  $ 1^{6} 0^2 1^2 0^2 $ &
${\hat{ 1/4}}^5  {\hat{-3/4}}^3 $ & $ {\hat{-1/4}}^5\ {\hat{3/4}}^3  $  \\
$  b_6 $ & $ 0 0 $ & $ 0 0 1 1 0 0 $ &
$ 1 0  0 0 0 0 $ &  $1 0  1 1 0 0$ &  $ 1^4 0^4 1 0 1 0 $ &
$ {\hat 1}^8  $ & $ {\hat 1}^8  $  \\
\hline \hline
\end {tabular}
\end{center}
\normalsize
\end{table}

\begin{table}[t]
\caption{\bf The choice of the GSO basis $\gamma [b_i, b_j]$.
Model $\protect\makebox{\bf\rm 1}''$. ($i$ numbers rows and $j$ -- columns)}
\label{gso1ii}
\footnotesize
\begin{center}
\begin{tabular}{|c||c|c|c|c|c|c|}
\hline
& $b_1$ & $b_2$ & $b_3$ & $b_4$ & $b_5$ & $b_6$\\ \hline
\hline
$b_1$ & $0$ &    $1$ & $0$ & $1$ &    $0$ & $1$\\
$b_2$ & $1$ &    $1$ & $1$ & $1$ &    $1$ & $1$\\
$b_3$ & $0$ &    $1$ & $1$ & $1$ &  $-1/2$ & $1$\\
$b_4$ & $1$ &    $0$ & $0$ &$1/2$&  $1/4$ & $0$\\
$b_5$ & $0$ &    $0$ & $0$ & $1$ &    $0$ & $1$\\
$b_6$ & $1$ &    $1$ & $1$ &$1/2$&  $1/2$ & $0$\\
\hline \hline
\end {tabular}
\end{center}
\normalsize
\end{table}

\begin{table}[t]
\caption{\bf The list of quantum numbers of the states.
Model~$\mbox{\rm 1}''$.}
\label{spectr1}
\footnotesize
\noindent \begin{tabular}{|c|c||c|cccc|cccc|} \hline
N$^o$ &$ b_1 , b_2 , b_3 , b_4 , b_5 , b_6 $&
$ SO(4)\times U(1)^2_{1,2}$&$ U(5)^I $&$ U(3)^I $&$ U(5)^{II} $&$
U(3)^{II} $&$ {\tilde Y}_5^I $&$ {\tilde Y}_3^I $&$ {\tilde Y}_5^{II} $&$
{\tilde Y}_3^{II}$ \\ \hline \hline
1 & RNS &&5&$\bar 3$&1&1&--1&--1&0&0 \\
  &     &&1&1&5&$\bar 3$&0&0&--1&--1 \\
  &     &$+1_1\ +1_2$&1&1&1&1&0&0&0&0 \\
  &     &$+1_1\ -1_2$&1&1&1&1&0&0&0&0 \\
  &     &&1&1&1&1&0&0&0&0 \\
  &     &&1&1&1&1&0&0&0&0 \\
  &0\ 0\ 0\ 2\ 2(6)\ 0&&5&1&5&1&--1&0&--1&0 \\
${\hat \Phi }$
  &&&1&3&1&3&0&1&0&1 \\
  &&&5&1&1&3&--1&0&0&1 \\
  &&&1&3&5&1&0&1&--1&0 \\ \hline \hline
2 &0\ 0\ 0\ 1\ 0\ 0&&1&3&1&1&5/2&--1/2&0&0 \\
  &&&$\bar 5$&3&1&1&--3/2&--1/2&0&0 \\
  &&&10&1&1&1&1/2&3/2&0&0 \\
${\hat \Psi }^I$
  &0\ 0\ 0\ 3\ 0\ 0&&1&1&1&1&5/2&3/2&0&0 \\
  &&&$\bar 5$&1&1&1&--3/2&3/2&0&0 \\
  &&&10&3&1&1&1/2&--1/2&0&0 \\ \hline
3 &0\ 0\ 0\ 3\ 2\ 0&&1&1&1&$\bar 3$&0&0&--5/2&1/2 \\
  &&&1&1&5&$\bar 3$&0&0&3/2&1/2 \\
  &&&1&1&$\bar {10}$&1&0&0&--1/2&--3/2 \\
${\hat \Psi }^{II}$
  &0\ 0\ 0\ 1\ 6\ 0&&1&1&1&1&0&0&--5/2&--3/2 \\
  &&&1&1&5&1&0&0&3/2&--3/2 \\
  &&&1&1&$\bar{10}$&$\bar{3}$&0&0&--1/2&1/2 \\ \hline
4 &000231\ (111071)&$1,\ \pm 1/2_1$&1&1&1&3&0&--3/2&0&--1/2 \\
  &000271\ (111031)&$1,\ \pm 1/2_1$&1&$\bar 3$&1&1&0&1/2&0&3/2 \\
${\hat \Psi }^H$
  &000031\ (111271)&$1,\ \pm 1/2_1$&1&$\bar 3$&1&3&0&1/2&0&--1/2 \\
  &000071\ (111231)&$1,\ \pm 1/2_1$&1&1&1&1&0&--3/2&0& 3/2 \\ \hline

5 &$\frac{\mbox{000070\ (111230)}}{\mbox{000010\ (111250)}}$
&$\pm_{SO(4)},\ \mp 1/2_1$&1&1&1&1&$\pm$5/4&$\pm$3/4
&$\mp$5/4&$\mp$3/4 \\
${\hat \sigma }$
&$\frac{\mbox{010310\ (101150)}}{\mbox{010170\ (101330)}}$
&$\pm_{SO(4)},\ \pm 1/2_1$&1&1&1&1&$\pm$5/4&$\pm$3/4
&$\pm$5/4&$\pm$3/4 \\
\hline
\end{tabular}
\normalsize
\end{table}

\subsection{$SU(5)\times U(1)\times SU(3)\times U(1)$ -- Model~2.}
Consider then another ${\left[U(5)\times U(3)\right]}^2$ model which after
breaking gauge symmetry by Higgs mechanism leads to the spectrum similar
to Model~1.

This model is defined by basis vectors given in  Table \ref{basis2}
with the $Z^4_2\times Z_6\times Z_{12}$ group under addition.

\begin{table}[tp]
\caption{Basis of the boundary conditions for Model~2.}
\label{basis2}
\begin{center}
\begin{tabular}{|c||c|ccc||c|cc|}
\hline
Vectors &${\psi}_{1,2} $ & ${\chi}_{1,\ldots,6}$ & ${y}_{1,\ldots,6}$ &
${\omega}_{1,\ldots,6}$
& ${\bar \varphi}_{1,\ldots,12}$ &
${\Psi}_{1,\ldots,8} $ &
${\Phi}_{1,\ldots,8}$ \\ \hline
\hline
$b_1$      & $1 1$ & $1^6$   & $1^6$   & $1^6$   &   $1^{12}$       & $1^8$
     & $1^8$ \\
$b_2$      & $1 1$ &  $1^6$  & $0^6$   & $0^6$   &   $0^{12}$       & $1^5$
$1/3^3$  & $0^8$  \\
$b_3$      & $1 1$ & $1^2 0^2 1^2$ & $0^6$   & $0^2 1^2 0^2$ &  $0^8\  1^4 $ &
$1/2^5\ 1/6^3$ & $-1/2^5\ 1/6^3 $ \\
$b_4 = S $ & $1 1$ & $1^2\ 0^4$ & $0^2 1^2 0^2$ & $0^4\ 1^2$ &  $0^{12}$
& $ 0^8 $        & $ 0^8 $ \\
$b_5 $     & $1 1$ & $1^4\ 0^2$ & $0^4\ 1^2$ & $0^6$   &  $1^8\ 0^4$   & $1^5\
0^3$    & $ 0^5\ 1^3  $  \\
$b_6 $     & $1 1$ & $0^2 1^2 0^2$ & $1^2\ 0^4$ & $0^4\ 1^2$ &  $1^2 0^2 1^6
0^2$     & $ 1^8  $       & $ 0^8  $  \\
\hline \hline
\end {tabular}
\end{center}
\end{table}
GSO coefficients are given in Table \ref{GSO2}.

\begin{table}
\caption{\bf The choice of the GSO basis $\gamma [b_i, b_j]$. Model~2.
($i$ numbers rows and $j$ -- columns)}
\label{GSO2}
\footnotesize
\begin{center}
\begin{tabular}{|c||c|c|c|c|c|c|}
\hline
& $b_1$ & $b_2$ & $b_3$ & $b_4$ & $b_5$ & $b_6$\\ \hline
\hline
$b_1$ & $0$ &    $1$ & $1/2$ & $0$ &    $0$ & $0$\\
$b_2$ & $0$ &    $2/3$ & $-1/6$ & $1$ &    $0$ & $1$\\
$b_3$ & $0$ &    $1/3$ & $5/6$ & $1$ &  $0$ & $0$\\
$b_4$ & $0$ &    $0$ & $0$ &$0$&  $0$ & $0$\\
$b_5$ & $0$ &    $1$ & $-1/2$ & $1$ &    $1$ & $1$\\
$b_6$ & $0$ &    $1$ & $1/2$ &$1$&  $0$ & $1$\\
\hline \hline
\end {tabular}
\end{center}
\normalsize
\end{table}

The given model corresponds to the following chain of the gauge
symmetry breaking:
$$E^2_8\longrightarrow SO(16)^2\longrightarrow U(8)^2
\longrightarrow [U(5)\times U(3)]^2\ . $$
Here the breaking of $U(8)^2-$group to $[U(5)\times U(3)]^2$
is determined by basis vector $b_5$, and the breaking
of N=2 SUSY$\longrightarrow$N=1 SUSY
is determined by basis vector $b_6$.

It is interesting to note that in the absence of vector $b_5$
$U(8)^2$ gauge group is restored by sectors $4b_3,\ 8b_3,\ 2b_2+c.c.$
and $4b_2+c.c.$
\begin{table}[tp]
\caption{The list of quantum numbers of the states. Model~2.}
\label{sost2}
\noindent \begin{tabular}{|c|c||c|cccc|cccc|} \hline
N$^o$ &$ b_1 , b_2 , b_3 , b_4 , b_5 , b_6 $&$SO_{hid}$&$ U(5)^I $&$ U(3)^I $&$
U(5)^{II} $&$
U(3)^{II} $&$  Y_5^I $&$ Y_3^I $&$ Y_5^{II} $&$
Y_3^{II}$ \\ \hline \hline
1 & RNS & $6_1\ 2_2$ & 1&1&1&1&0&0&0&0 \\
  && $2_3\ 2_4$ & 1&1&1&1&0&0&0&0 \\
  &&& 5&1&$\bar 5$&1&1&0&--1&0 \\
  &0\ 0\ 4\ 1\ 0\ 0&&1&3&1&3&0&--1&0&--1 \\
  &0\ 0\ 8\ 1\ 0\ 0&&1&$\bar 3$&1&$\bar 3$&0&1&0&1 \\ \hline\hline
2 &0\ 1\ 0\ 0\ 0\ 0&&5&$\bar 3$&1&1&--3/2&--1/2&0&0 \\
  &&&1&$\bar 3$&1&1&5/2&--1/2&0&0 \\
  &0\ 3\ 0\ 0\ 0\ 0&&$\bar {10}$&1&1&1&1/2&3/2&0&0 \\ \hline
3 &0\ 1\ 10\ 0\ 0\ 0&&1&1&$\bar {10}$&3&0&0&1/2&1/2 \\
  &0\ 3\ 6\ 0\ 0\ 0&&1&1&5&1&0&0&--3/2&--3/2 \\
  &&&1&1&1&1&0&0&5/2&--3/2 \\ \hline
4 &0\ 2\ 3\ 0\ 0\ 0&$-_3\ \pm_4$&1&3&1&1&--5/4&--1/4&5/4&3/4 \\ \hline
5 &0\ 0\ 3\ 0\ 0\ 0&$+_3\ \pm_4$&1&1&$\bar 5$&1&--5/4&3/4&1/4&3/4 \\ \hline
6 &0\ 0\ 9\ 0\ 0\ 0&$+_3\ \pm_4$&1&1&5&1&5/4&--3/4&--1/4&--3/4 \\ \hline
7 &0\ 4\ 9\ 0\ 0\ 0&$-_3\ \pm_4$&1&$\bar 3$&1&1&5/4&1/4&--5/4&--3/4 \\ \hline
8,9 &0\ 5\ 0\ 1\ 0\ 1&$-_1\ \pm_3$&1&3&1&1&0&--1&0&0 \\
  &0\ 3\ 0\ 1\ 0\ 1&$+_1\ +_3$&5&1&1&1&1&0&0&0 \\
  &&$+_1\ -_3$&$\bar 5$&1&1&1&--1&0&0&0 \\
  &&$-_1\ +_3$&1&1&5&1&0&0&1&0 \\
  &&$-_1\ -_3$&1&1&$\bar 5$&1&0&0&--1&0 \\
  &0\ 5\ 8\ 1\ 0\ 1&$+_1\ +_3$&1&1&1&$\bar 3$&0&0&0&1 \\ \hline
10 &0\ 3\ 3\ 0\ 0\ 1&$+_1\ \pm_4$&1&1&1&1&--5/4&3/4&5/4&3/4 \\ \hline
11 &1\ 0\ 3\ 0\ 0\ 1&$\pm_2\ -_3$&1&1&5&1&--1/4&3/4&--5/4&--3/4 \\
   &1\ 2\ 11\ 0\ 0\ 1&$\pm_2\ -_3$&1&1&1&$\bar 3$&--5/4&3/4&--5/4&1/4 \\ \hline
12 &1\ 0\ 9\ 0\ 0\ 1&$\pm_2\ +_3$&$\bar 5$&1&1&1&1/4&--3/4&5/4&3/4 \\
   &1\ 4\ 9\ 0\ 0\ 1&$\pm_2\ +_3$&1&$\bar 3$&1&1&5/4&1/4&5/4&3/4 \\ \hline
13 &0\ 0\ 0\ 1\ 1\ 1&$\pm_2\ +_3$&1&1&1&1&0&--3/2&0&3/2 \\
   &0\ 2\ 0\ 1\ 1\ 1&$\pm_2\ -_3$&1&3&1&1&0&1/2&0&3/2 \\
   &0\ 2\ 8\ 1\ 1\ 1&$\pm_2\ -_3$&1&1&1&$\bar 3$&0&--3/2&0&--1/2 \\
   &0\ 4\ 8\ 1\ 1\ 1&$\pm_2\ +_3$&1&3&1&$\bar 3$&0&1/2&0&--1/2 \\
   &1\ 0\ 3\ 1\ 1\ 1&$+_1\ +_3$&1&1&1&1&5/4&3/4&--5/4&3/4 \\
   &1\ 0\ 9\ 1\ 1\ 1&$+_1\ +_3$&1&1&1&1&--5/4&--3/4&5/4&--3/4 \\
   &1\ 3\ 3\ 0\ 1\ 1&$-_1\ -_3$&1&1&1&1&--5/4&--3/4&--5/4&3/4 \\
   &1\ 3\ 9\ 0\ 1\ 1&$-_1\ +_3$&1&1&1&1&5/4&3/4&5/4&--3/4 \\ \hline
\end{tabular}
\end{table}

The full massless spectrum  for the given model is given in Table \ref{sost2}.
By analogy with Table \ref{tabl3}
only fermion states with positive chirality
are written and obviously vector supermultiplets are absent.
Hypercharges are determined by formula:
$$ Y_n=\sum_{k=1}^{n}(\alpha_k/2 + F_k)\ . $$

The given model possesses  the hidden gauge symmetry
$SO(6)_1\times SO(2)^3_{2, 3, 4}$.
The corresponding chirality is given in column $SO_{hid.}$.
The sectors  are divided by horizontal lines and
without including the $b_5-$vector form $SU(8)-$multiplets.

For example, let us consider row No 2.
In sectors $b_2$, $5b_2$ in addition to states $(1, \bar{3})$ and
$(5, \bar{3})$ the (10, 3)--state appears, and in the sector $3b_2$
besides the $(\bar{10}, 1)-$ the states (1, 1) and $(\bar{5}, 1)$
survive too. All these states form $\bar{8}+56$ representation
of the $SU(8)^I$ group.

Analogically we can get the full structure of the theory according
to the $U(8)^I\times U(8)^{II}-$group.
(For correct restoration of the $SU(8)^{II}-$group we must invert
3 and $\bar{3}$ representations.)

In Model~2 matter fields appear both in $U(8)^I$ and $U(8)^{II}$ groups.
This is the main difference between this model and Model~1.
However, note that in the
Model~2 similar to the Model~1 all gauge fields appear in RNS--sector only
and $10 +\bar{10}$ representation (which can be the Higgs field
for gauge symmetry breaking) is absent.

Scheme of the breaking of the gauge group to the symmetric subgroup, which is
similar to the scheme of model 1, works for Model 2 too. In this case
vector-like multiplets
$(\underline5,\underline 1;\underline {\bar 5}, \underline1)$
from R-NS sector and
$(\underline1,\ \underline3;\ \underline1,\ \underline3)$
from $4b_3$ $(8b_3)$ play the role of Higgs fields.
Then generators of the symmetric subgroup and electromagnetic
charges of particles are determined by formulas:
\begin{eqnarray}
\Delta^{(5)}_{\rm sym}&=&t^{(5)}\times 1\ \oplus\ 1\times t^{(5)} \nonumber \\
\Delta^{(3)}_{\rm sym}&=&(-t^{(3)})\times 1\ \oplus\ 1\times t^{(3)}
\nonumber \\
Q_{em}=t^{(5)}_5-2/5\,Y^5&,&\mbox{where}
\ t^{(5)}_5=(1/15,\ 1/15,\ 1/15,\ 2/5,\ -3/5) \label{eq}
\end{eqnarray}

After this symmetry breaking matter fields (see Table \ref{sost2}
rows No 2, 3) as usual for flip models take place in representations
of the $U(5)-$group and form four generations
$(\underline1 +\underline5 +\underline{\bar{10}};\ \underline{\bar3}
+\underline1)_{\rm sym}$.
And Higgs fields form adjoint representation of the symmetric group,
similar to Model~1, which is necessary for breaking of the gauge
group to the Standard group. Besides, due to quantization of the
electromagnetic charge according to the formula (\ref{eq})
 sates with exotic charges in low energy
spectrum also do not appear in this model.
In this model U(1)-group in hidden sector has anomalies which is broken by
Dine-Seiberg-Witten mechanism \cite{Dine'}. The corresponding D-term
could break supersymmetry at very large scale however  it is possible to
show that there exist D-flat directions with
respect to some non-anomalous gauge symmetries, canceling the anomalous
D-term and restoring supersymmetry. As result the corresponding VEVs imply
that the final symmetry will be less than the original one, which includes the
$$(SU(5)\times U(1))^{\rm sym}\times SU(3_H)\times U(1_H)^{\rm sym}$$
 observable gauge symmetry. Let us note that
the models 5 and 3 do not contain the anomalous U(1)-groups.

\subsection{$SO(10) \times SU(4)$ -- Model~3.}
As an illustration we can consider the GUST construction involving $SO(10)$
as GUT gauge group.  We consider the set  consisting of seven vectors $B=
{b_1, b_2, b_3, b_4 \equiv S, b_5, b_6,b_7 }$ given in Table \ref{basis 3}.

\begin{table}[b]
\caption{Basis of the boundary conditions for the Model~3.}
\label{basis 3}
\begin{center}
\begin{tabular}{|c||c|ccc||c|cc|}
\hline
Vectors &${\psi}_{1,2} $ & ${\chi}_{1,\ldots,6}$ & ${y}_{1,\ldots,6}$ &
${\omega}_{1,\ldots,6}$
& ${\bar \varphi}_{1,\ldots,12}$ &
${\Psi}_{1,\ldots,8} $ &
${\Phi}_{1,\ldots,8}$ \\ \hline
\hline
$b_1$ & $1 1 $ & $1 1 1 1 1 1$ &
$1 1 1 1 1 1 $ & $1 1 1 1 1 1$ &  $1^{12} $ & $1^8 $ & $1^8$ \\
$b_2$ & $1 1$ & $1 1 1 1 1 1$ &
$0 0 0 0 0 0$ &  $0 0 0 0 0 0 $ &  $ 0^{12} $ &
$1^5 {1/3}^3 $ & $0^8$  \\
$b_3$ & $1 1$ & $ 0 0 0 0 0 0 $ & $1 1 1 1 1 1 $ & $ 0 0 0 0 0 0 $ &
$0^8 1^4$ & $ 0^5 1^3 $ &  $ 0^5 1^3$ \\
$b_4 = S $ & $1 1$ &
$1 1 0 0 0 0 $ & $ 0 0 1 1 0 0 $ & $ 0 0 0 0 1 1 $ &
$ 0^{12} $ & $ 0^8 $ & $ 0^8 $ \\
$  b_5 $ & $ 1 1 $ & $ 1 1 1 1 1 1 $ &
$0 0 0 0 0 0 $ &  $0 0 0 0 0 0$ &  $ 0^{12} $ &
$ 0^8 $ & $ 1^5 {1/3}^3  $  \\
$  b_6 $ & $ 1 1 $ & $ 0 0  1 1  0 0 $ &
$ 1 1  0 0 0 0 $ &  $0 0  0 0 1 1$ &  $ 1^2 0^2 1^6 0^2 $ &
$ 1^8  $ & $ 0^8  $  \\
$  b_7 $ & $ 1 1 $ & $ 0 0  1 1  0 0 $ &
$ 1 0  0 0 0 0 $ &  $1 0  0 0 1 1$ &  $ 1^2 1 0 1^2 1^2 1 0 0^2 $ &
$ 0^8  $ & $ 1^8  $  \\
\hline \hline
\end {tabular}
\end{center}
\end{table}
GSO projections are given in Table \ref{GSO 3}.

It is interesting to note that in this model the horizontal gauge symmetry
$U(3)$ extends to $SU(4)$. Vector bosons which needed for this appear
in sectors $2b_2\ (4b_2)$ and $2b_5\ (4b_5)$.

\begin{table}[t]
\caption{The choice of the GSO basis $\gamma [b_i, b_j]$. Model~3.
($i$ numbers rows and $j$ -- columns)}
\label{GSO 3}
\begin{center}
\begin{tabular}{|c||c|c|c|c|c|c|c|}
\hline
& $b_1$ & $b_2$ & $b_3$ & $b_4$ & $b_5$ & $b_6$ & $b_7$\\ \hline
\hline
$b_1$ & $0$ &    $1$ & $0$ & $0$ &    $1$ & $0$ & $0$\\
$b_2$ & $0$ &  $2/3$ & $1$ & $1$ &    $1$ & $1$ & $0$\\
$b_3$ & $0$ &    $1$ & $0$ & $1$ &    $1$ & $1$ & $0$\\
$b_4$ & $0$ &    $0$ & $0$ & $0$ &    $0$ & $0$ & $0$\\
$b_5$ & $0$ &    $1$ & $1$ & $1$ &  $2/3$ & $0$ & $0$\\
$b_6$ & $0$ &    $1$ & $0$ & $1$ &    $1$ & $1$ & $0$ \\
$b_7$ & $0$ &    $1$ & $1$ & $1$ &    $0$ & $1$ & $1$ \\
\hline \hline
\end {tabular}
\end{center}
\end{table}

The spectrum of the Model 3 is as follows:

1. Model possesses the 
$[U(1)\times SO(6)]_{Hid.}\times [SO(10)\times SU(4)]^2$  gauge
group, the $U(1)$ group is anomaly free;

2. The matter fields,  $(16,\ 4;\ 1,\ 1)$,
 are from $3b_2$ and $5b_2$ sectors;

3. There are  Higgs fields from RNS-sectors -
$(\pm 1)_1 (6)_6(1,\ 1;\ 1,\ 1)$ , $(10,\ 1;\ 10,\ 1)$
and 2 total singlets;

4. There are some  Higgs fields from $mb_2+nb_5$ sectors,
where $m, n =2- 4$ : $(1,\ 6;\ 1,\ 6)$;

5. Another additional fields are $(-)_6(10,\ 1;\ 1,\ 1)$ ,
$(+)_6(1,\ 1;\ 10,\ 1)$ ,
$(-)_6(1,\ 1;\ 1,\ 6)$ , $(+)_6(1,\ 6;\ 1,\ 1)$  \\
$(\pm1/2)_1(1,\ \bar 4;\ 1,\ \bar 4)$ , $(\pm1/2)_1(1,\ \bar 4;\ 1,\ 4)$ ,
$2\times (+)_6(1,\ 1;\ 1,\ 1)$ , $(\pm 1)_1(-)_6(1,\ 1;\ 1,\ 1)$.

The condition of generation chirality in this model results in the choice
of Higgs fields as vector representations of SO(10)
 ($\underline{16}+\underline{\bar{16}}$ are absent). According to
conclusion (\ref{eq6}) the only Higgs fields $(\underline{10}, \underline{1};
\underline{10}, \underline{1})$ of $(SO(10)\times SU(4))^{\times 2}$
appear in the model (from RNS--sector) which can be used for
GUT gauge symmetry.

\subsection{Model~4 with Vector-Like Horizontal Interactions.}

\begin{table}[t]
\caption{\bf Basis of the boundary conditions for all world-sheet fermions.
Model~4 , $Z_2\times Z_{30} \times Z_{30} \times Z_2 \times Z_2 \times Z_2$.}
\label{basis v }
\begin{center}
\begin{tabular}{|c||c|ccc||c|cc|}
\hline
Vectors &${\psi}_{1,2} $ & ${\chi}_{1,\ldots,6}$ & ${y}_{1,\ldots,6}$ &
${\omega}_{1,\ldots,6}$
& ${\bar \varphi}_{1,\ldots,12}$ &
${\Psi}_{1,\ldots,8} $ &
${\Phi}_{1,\ldots,8}$ \\ \hline
\hline
$b_1$ & $1 1 $ & $1 1 1 1 1 1$ &
$1 1 1 1 1 1 $ & $1 1 1 1 1 1$ &  $1^{12} $ & $1^8 $ & $1^8$ \\
$b_2$ & $1 1$ & $1 1 1 1 1 1$ &
$0 0 0 0 0 0$ &  $0 0 0 0 0 0 $ &  $ 0^{12} $ &
${1/5}^5 {1/3}^3 $ & ${0}^8$  \\
$b_3$ & $1 1$ & $  0 0 1 1 0 0$ & $0 0 0 0 0 0  $ & $ 1 1 0 0 1 1 $ &
$0^{12}  $ & $ {0}^8 $ &  $ {1/5}^5 {1/3}^3$ \\
$b_4 = S $ & $1 1$ &
$1 1 0 0 0 0 $ & $ 0 0 1 1 0 0 $ & $ 0 0 0 0 1 1 $ &
$ 0^{12} $ & $ 0^8 $ & $ 0^8 $ \\
$  b_5 $ & $ 0 0 $ & $ 0 0 0 0 1 1 $ &
$1 1 1 1 0 0$ &  $ 1 1 1 1 1 1 $ &  $ 1^{6} 0^6$ &
${ 0}^5  {1}^3 $ & ${1}^8  $  \\
$  b_6 $ & $ 1 1 $ & $  0 0  0 0 0 0$ &
$ 1 1 1 1 1 1$ &  $0 0  0 0 0 0$ &  $ 1^2 0^4 1^2 0^4$ &
$ 1^5 0^3  $ & $ 1^5 0^3  $  \\
\hline \hline
\end {tabular}
\end{center}
\end{table}

\begin{table}[t]
\caption{The choice of the GSO basis $\gamma [b_i, b_j]$. Model~4
($i$ numbers rows and $j$ -- columns)}
\label{GSO v}
\begin{center}
\begin{tabular}{|c||c|c|c|c|c|c|}
\hline
& $b_1$ & $b_2$ & $b_3$ & $b_4$ & $b_5$ & $b_6$\\ \hline
\hline
$b_1$ & $0$ &    $0$         & $0$        & $0$ &  $0$ & $1$\\
$b_2$ & $1$ &  $13/15$ & $1$       & $1$ &  $1$ & $1$\\
$b_3$ & $1$ & $0$          &$13/15$ & $1$ &  $1$ & $1$\\
$b_4$ & $0$ &    $0$         & $0$          & $0$ &  $0$ & $0$\\
$b_5$ & $0$ &    $1$         & $1$       & $0$ &  $1$ & $1$\\
$b_6$ & $1$ &    $1$       & $1$        & $1$ &  $1$ & $0$\\
\hline \hline
\end {tabular}
\end{center}
\end{table}

The spectrum of the Model 4 is as follows:

1. The gauge group of the model is 
 $[U(1)\times SO(4)]^2_{Hid.}\times [SO(10)\times SU(4)]^2$,
 the $U(1)$ group is anomaly free;

2. The matter fields,  $(16, 4; 1, 1)$ +$(16,\bar 4;1,1)$;

3. There are  Higgs fields -
  $(10, 1; 10, 1)$,
   $\pm 1_1 (1,1,1,6)$, $\pm 1_4 (1,6,1,1) $
 $\pm 1 \pm _{SO(4)}(16,1,1,1,)$;
 $\pm 1 - _{SO(4)}(16,1,1,1,)$;
 $\pm 1 - _{SO(4)}(1,1,16,1,)$.

\section{Self Dual Charge Lattice.
More Explicit Methods of Model Building.}

\subsection{Self-Duality of the Charge Lattice and Possible Gauge Groups.}

Here we present some results based on the important feature of the  charge
lattice that is self-duality.

As was shown in \cite{18i} the  charge lattice ${\bf Q}$
 shifted by a constant vector ${\bf S}$ is an odd
self-dual Lorentzian lattice
constituted by 32-components vectors with components
$$
Q_i=\frac{\alpha_i}{2}+F_i,
$$
where $\alpha_i$ is a boundary condition for
$i$th fermion and $F_i$ is the  corresponding fermion number in a
particular string state. Vector $S$ takes care of the space-time
spin-statistic and in  the case of heterotic string is $(1,0,0,\dots,0)$.

Let us first recall some basic features of the Lie algebras representations,
weights and roots. In the Cartan basis the adjoint representation of a Lie
algebra consist of the Abelian Cartan subalgebra of the Cartan generators
$H^i$, a set of step operators $E^\alpha$ that is chosen so that
$$ [H^i,E^\alpha]=\alpha^i E^\alpha,$$
where vector $\alpha$ is called the root. For each root $\alpha$ there is
one and only one step operator $E^\alpha$. So the root system uniquely
define the algebra. For the simply laced algebras the important feature
is that if $\alpha,\beta$ are roots and $(\alpha,\beta)=-1$ then
$\alpha+\beta$ is also a root. Again for the simply laced algebras
$\alpha^2=2$ ($\alpha^2=0$ for the roots that corresponds to Cartan
generators).

For any representation of a Lie algebra we define  weight vectors $\mu$ for
the basis $| \mu \rangle$ of a representation:
$$H^i | \mu \rangle = \mu^i | \mu \rangle$$
In the case of $\alpha^2=2$ for any weight $\mu$ and any root $\alpha$
we have
$$(\alpha \cdot \mu) \in Z$$
Hence the weight lattice $\Lambda_W$ is dual to the root lattice $\Lambda$.
Conversely we can define weight lattice $\Lambda_W$ as a dual to
the root lattice.

If we take the set of vectors as a basis of any representation
then the weights will be the scalar products with vectors representing
simple roots. So we describe a Lie algebra by a set of vectors (roots) with
scalar products defined by Dynkin diagram, and any representation
of the algebra is given by a set of vectors, whose scalar products
with simple roots forms weights of a representation.

Suppose that the weights of a two representations
are $\lambda^1_i$ and $\lambda^2_i$.
Hence the scalar product of that weights is determined by inverse
Cartan matrix
$$ (\lambda^1 \cdot \lambda^2) = \lambda^1_i K^{-1}_{ij} \lambda^2_j$$

Returning to the weight lattice note that if a vector of a highest
weight belongs to the weight lattice (i.e. has integer scalar products
with all the simple roots) then the whole representation belongs
to the lattice since it is derived from the highest weight by subtracting
corresponding root vectors. Indeed if $\lambda$ has integer scalar product
with roots and $\alpha$ is a root, $\alpha^2=2, \ (\alpha\cdot\lambda)=1$
then $\lambda-\alpha$ also has integer scalar product and has the same square
as a $\lambda$. Hence we have to pay attention to the highest weights.

The charge lattice of a string model is a more complicated object
since we have Lorentzian scalar product. However right parts of charge
vectors are weights and roots of a Lie algebra.
The most fundamental feature of the charge lattice is its self duality.
This feature allow us to carry out thorough analysis of the possible
spectrum of the string models.

The scheme of our analysis is as follows. Suppose first that we
can build a model with a given gauge group. Then all states in the
spectrum form vectors of a charge lattice. These vectors consist of
the weights of some representations of the gauge group and root vectors
of the adjoint representation. At the starting point we have at least
root vectors in the lattice. Then since the lattice is self-dual we
have to include some weights. Obviously the weights of any representation
has integer scalar products with root vectors (even in the case of
Lorentzian scalar product since root vectors have zero left part,
see below). So we should consider all the highest weights for a given
gauge group. Eventually we should find a set of  highest weights which
have integer Lorentzian scalar product with each other as it is demanded
by self-duality. Fortunately the phenomenology demands that we have to
have some representations in the spectrum to incorporate matter.
So there not too much possibilities for the remaining weights.

As we will see below the self duality of the charge lattice
 apply serious restriction on the
possible gauge group and matter spectrum of the GUST. In this section we
will consider only models that permit bosonization which means that
we write all fermions in terms of complex fermions and consequently can
construct the charge lattice. Also we will restrict ourselves considering
only models that have only periodic or antiperiodic boundary conditions
for left moving fermions (supersymmetric sector). Other possible forms of
the left sector can be treated by the similar way but our case is more
convenient in sense of building a $N=1$ SUSY model.

Before analyzing particular GUST with appropriate gauge group we will
consider some common features of a class of the lattices that we define
above.

Let us give some definitions. Below we refer to $Q$ as a shifted
charge lattice. Indices $R$ and $L$ refer to left and right
(10- and 22-component correspondingly) parts of
a vector in the charge lattice, $\Lambda$ denotes any 32-component
vector in the charge lattice, $(\Lambda_L,\Lambda_R)$ denotes 32-component
vector, $(\Lambda\cdot\Lambda^\prime)$ means appropriate
(Lorentzian or Euclidean in the case of only left or right parts of
vectors) scalar product.
Denote $Q_R=\left\{ \Lambda_R : \exists \Lambda_L
 :  (\Lambda_L,\Lambda_R) \in Q \right\}$, i.e. $Q_R$  is a set of only
 right parts of vectors from $Q$.
 Denote
 $$(\Lambda_R \cdot Q_R) \sim \frac{1}{n} \Leftrightarrow$$
$$\exists {\Lambda^\prime}_R\in
Q_R:(\Lambda_R\cdot{\Lambda^\prime}_R)=\frac{k}{n}, k\in Z, \gcd(k,n)=1 $$
This means that we consider a vector $\Lambda_R$ that has non-integer
scalar product of the form $k/n$ with at least one vector in the $Q_R$.

Actually now we can determine whether a vector $\Lambda$ represent gauge
boson, chiral fermion or vector-like state in the following way.
Suppose that for a given $\Lambda_R$ we have
$(\Lambda_R \cdot Q_R) \sim 0$ (i.e. scalar product is integer). Hence
the vector $(0,\Lambda_R)$ should belong to the lattice (because
self-dual lattice contains all the vectors that have integer scalar
product with the lattice). Obviously, the vector $(0,\Lambda_R)$ represent
the gauge boson (taking into account the shift of the charge lattice).
Consider then $\Lambda_R$, $(\Lambda_R \cdot Q_R) \sim 1/2$.
Hence if a vector $(\Lambda_L,\Lambda_R)\in Q$ then obviously
$(\Lambda_L,-\Lambda_R)\in Q$. This means that we have vector-like state.
Finally, if $(\Lambda_R \cdot Q_R) \sim 1/n, \, n>2$ then we conclude
that $(\Lambda_L,-\Lambda_R)\notin Q$, hence $(\Lambda_L,\Lambda_R)$
represent chiral fermion.

Notice that the above arguments are valid in the case of string models
that does not permit bosonization, because we rely only upon the
self-duality of the lattice.

Below we will restrict ourselves to models with only periodic or
antiperiodic complex fermions in the left part.
Since we take all
$\alpha_i$ in the left sector to be 0 or 1 then all of the scalar products
of the left parts of the lattice vectors will be in the form
of $n/4,\ \ n\in Z$.  So the scalar product of the right parts
must has the same form in order to obtain the integer scalar product.
Thus  we have 3 classes of vectors:
\begin{itemize}
\item Gauge bosons that are space-time vectors, $(\Lambda_R \cdot Q_R)\sim 0$
\item Vector-like states (Higgs fields), $(\Lambda_R \cdot Q_R) \sim 1/2$
\item Chiral fermions (matter fields), $(\Lambda_R \cdot Q_R) \sim 1/4$
\end{itemize}

Finally notice that the scalar products of the right part is the  scalar
products of the weight vectors of particular gauge group representation.
The structure of a representation is well known and for particular gauge
group we can determine the representation that give appropriate model
spectrum and necessary scalar products.

As a first simple example that illustrates all above discussion of this
section we consider a model with $E_6\times SU(3)$  gauge group. Notice
that both $E_6$ and $SU(3)$ groups have only representations with scalar
products of the weight vectors of the form $\frac{n}{3}$. If we wish to
obtain a model which  has representation for matter then we have to
include representation $({\bf 27,3})$ in the spectrum so that this states
are space-time chiral fermions. But to make this representation to be
chiral fermions one has to include another weight vector  that give scalar
product of the right part equal to $\frac{k}{4},\ k\neq 0 \bmod 2$. Since
there is no such weight vector among weight vector of $E_6$ and $SU(3)$
representations then it is impossible to build a model with $E_6\times SU(3)$
gauge group and chiral fermions in the representations appropriate for
the GUST.
However if one take spin structure vector with $\frac23$ in the
left part then it will be possible to build a model. However the model will
have $N=2$ space-time supersymmetry \cite{Lop}
 which means that there are no chiral fermions.

Let us go further with $E_6$ group. Consider the horizontal gauge group
$SU(2)\times U(1)$ in addition to $E_6$, i.e. we have GUST with
$E_6\times SU(2) \times U(1)$ gauge group. We demand that the
representation $({\bf 27,2})_{1/\sqrt{6}}$ should be in the spectrum
(${1/\sqrt{6}}$ means $U(1)$ charge calculated from the condition
that $\Lambda^2_R=2$). We need this representation to be chiral
to incorporate matter fields.

Consider then the representation $({\bf 27,1})_{-\sqrt{2/3}}$ (the value
of the $U(1)$ charge is calculated from the square condition and
using the condition of integer scalar product with other vectors).
Since we already have $({\bf 27,2})_{1/\sqrt{6}}$ in the lattice, we
can calculate the form of other allowed representations.
Consider representation $({\bf X,Y})_\alpha$. The condition of the
scalar product of the right parts reads
$$ \frac{n}{3}+\frac{s}{2}+\alpha \sqrt{\frac{1}{6}}=\frac{m}{4},$$
where $n,m,s$ are integer, and $n/3$ represent scalar product of
weight vectors of ${\bf 27}$ and ${\bf X}$, and $s/2$ represent scalar
product of weight vectors of ${\bf 2}$ and ${\bf Y}$.
So we can determine possible $\alpha$:
$$ \alpha= \sqrt{6} \left( \frac{m-2s}{4}-\frac{n}{3} \right) $$
In the case of scalar products of $({\bf 27,1})_{-\sqrt{2/3}}$ and
$({\bf X,Y})_\alpha$ we have
$$ \frac{n}{3}-\alpha \sqrt{\frac{2}{3}}=n+s-\frac{m}{2}$$
Thus we see that $({\bf 27,1})_{-\sqrt{2/3}}$ could not be chiral,
and we have only 2 generations in this case.

Now we will present a more complicated example, namely the
$SO(10)\times SU(3)\times U(1)$ gauge group.
We demand that the model spectrum includes
space-time chiral fermions in the representation $({\bf
16,3})_{\sqrt{1/12}}$ of  the $SO(10)\times SU(3)\times U(1)$ gauge group.

Thus we have a weight vector that we have to include in the lattice  in
addition to the root lattice of $SO(10)\times SU(3)\times U(1)$.
The we can find all
the weights (actually the components of the weights that have
nonvanishing scalar products with root vector and weight of
matter representation and that
correspond to the $SO(10)\times SU(3)\times U(1)$
representation) that give appropriate scalar product with weight of matter
representation.

Like as in the $E_6 \times SU(2) \times U(1)$ we can determine the
form of representation allowed by the scalar product with
$({\bf 1,3})_{\sqrt{4/3}}$. For any representation
$({\bf X,Y})_\alpha$ we have
$$ \frac{s}{4}+\frac{n}{3}+\alpha \sqrt{\frac{1}{12}}=\frac{m}{4},$$
where $n,m,s$ are integer, and $s/4$ represent scalar product of
weight vectors of ${\bf 16}$ and ${\bf X}$, and $n/3$ represent scalar
product of weight vectors of ${\bf 3}$ and ${\bf Y}$.
Hence
$$ \alpha= \sqrt{\frac34} \left( m-s-\frac{4}{3}n \right) $$
Consider then the representation $({\bf 1,3})_{\sqrt{4/3}}$.
It appears that it has integer scalar
product with all other vectors that are allowed by weight of
$({\bf 16,3})_{\sqrt{1/12}}$.
Indeed, for the scalar products of $({\bf 16,3})_{\sqrt{1/12}}$ and
$({\bf X,Y})_\alpha$ we have
$$ \frac{n}{3}+\alpha \sqrt{\frac{4}{3}}=m-n-s$$
It means that the  representation  $({\bf 1,3})_{\sqrt{4/3}}$ will be a
space-time gauge bosons that extend the initial
$SO(10)\times SU(3)\times U(1)$ gauge
group to the $SO(10)\times SU(4)$. Now we can see why all attempts to
build a model with $SO(10)\times SU(3)\times U(1)$
gauge group and chiral matter in
$({\bf 16,3})_{\sqrt{1/12}}$ representation failed.
All models obviously must have
$SO(10)\times SU(4)$ gauge group with corresponding matter
representations.

Concluding we summarize the predictions
 obtained by exploiting the self-duality
of the lattice in the case of using complex fermion:
\begin{itemize}
\item It is impossible to build realistic model with $E_6\times SU(3)$
gauge group and $N=1$ SUSY.
\item Model with $E_6\times SU(2) \times U(1)$ gauge group will
have chiral fermions in $({\bf 27,2})$
and vector-like representation $({\bf 27,1})$. Hence in this case
we have only 2 generations of matter.
\item It is impossible to build realistic model with
$SO(10)\times SU(3)\times U(1)$ gauge group.
Only $SO(10)\times SU(4)$ gauge group is allowed hence again giving
4 generations.
\end{itemize}

\subsection{Building GSO-Projectors for a Given Algebra}
As we follow certain breaking chain of $E_8$ then it is very naturally
to take $E_8$ construction as a starting point. Note that root lattice
of $E_8$ arises from two sectors: NS sector gives {\underline{120}} of
$SO(16)$ while sector with $1^8$ gives {\underline{128}}   of $SO(16)$.
This corresponds to the following choice of simple roots
$$ \pi_i=-e_i+e_{i+1}\ , \mbox{ where } i=1,\ldots ,7\ ; $$
$$ \pi_8=\frac{1}{2}(e_1+e_2+e_3+e_4+e_5-e_6-e_7-e_8)$$
Basing on this choice of roots it is very clear how to build basis
of simple roots
for any subalgebra of $E_8$. One can just find out appropriate
vectors $\pi_i$ of the form as in $E_8$ with needed scalar products
or build weight diagram and break it in a desirable fashion to find
roots corresponding to certain representation in terms of $E_8$
roots.
After the basis of simple roots is written down one can build
GSO-projectors in a following way.
GSO-projection is defined by operator $(b_i\cdot F)$ acting on given state.
The goal is to find those $b_i$ that allow only states from algebra
lattice to survive. Note that $F_i=\gamma_i-\alpha_i/2$ ($\gamma_i$ ---
components of a root in basis of $e_i$),
so value of GSO-projector for sector $\alpha$ depends on $\gamma_i$ only.
So, if scalar products of all simple roots that arise from a given sector
 with vector $b_i$ is equal mod 2 then they surely will survive
 GSO-projection. Taking several such vectors $b_i$ one can eliminate
 all extra states that do not belong to a given algebra.

Suppose that simple roots of the algebra are in the form
$$ \pi_i=\frac{1}{2}(\pm e_1\pm e_2\pm e_3\pm e_4\pm e_5\pm e_6\pm e_7\pm
e_8)$$
$$ \pi_j=(\pm e_k\pm e_m)$$
In this choice we have to find vectors $b$ which gives 0 or 1 in a scalar
 product with all simple roots. Note that
$(b\cdot\pi_i)=(b\cdot\pi_j)\ \mbox{mod}2$ for
 all $i,j$ so $c_i=(b\cdot\pi_i)$ either all equal 0 mod 2 or equal 1 mod 2.
 ( for $\pi_j=(\pm e_k\pm e_m)$ it should be 0 mod 2 because they are arise
   from NS sector )
 Value 0 or 1 is taken because if root $\pi \in $ algebra lattice then
 $-\pi$ is a root also. With such choice of simple roots and scalar products
 with $b$ all states from sector like $1^8$ will have the same
 projector value. Roots like $\pm e_i\pm e_j$ rise from NS sector
 and are sum of roots like
$ \pi_i=\frac{1}{2}(\pm e_1\pm e_2\pm e_3\pm e_4\pm e_5\pm e_6\pm e_7\pm e_8)$
and therefore have scalar products equal to 0 mod 2 as is needed for NS sector.
Now vectors $b$ are obtained very simple. Consider
$$ c_i=(b\cdot \pi_i)=b_jA_{ji} $$
where $A_{ji}=(\pi_i)_j$ --- matrix of roots component in $e_j$ basis.
Hence $b=A^{-1}\cdot c$ where either all $c_j=0 \bmod 2$   or $c_j=1 \bmod 2$.
One has to try some combination of $c_j$ to obtain appropriate set of $b$.
The next task is to combine those $b_i$ that satisfy modular invariance
rules and do not give extra states to the spectrum.

\subsection{Breaking a Given Algebra Using GSO-Projectors}
 It appears that this method of constructing GSO-projectors allows
to break a given algebra down to its subalgebra.
Consider root system of a simple Lie algebra. It is well known that if
$\pi_1,\,\pi_2\in\Delta$, where $\Delta$ is a set of positive roots then
$(\pi_1-\pi_2(\pi_1\cdot\pi_2))\in\Delta$ also. For simply laced
algebras it means that if $\pi_i,\,\pi\in\Delta$ and $(\pi_i\cdot\pi)=-1$
where $\pi_i$ is a simple root then $\pi+\pi_i$ is a root also.
This rule is hold automatically in string construction: if a sector
gives some simple roots then  all roots of algebra and only them also exist
(but part of them may be found in another sector). Because square of every
root represented by a state is 2 then if $(\pi_i\cdot\pi)\neq-1$ then
$(\pi+\pi_i)^2\neq2$. So one must construct GSO-projectors checking
 only simple roots.
On the other hand if one cut out some of simple roots then algebra
will be broken. For example if a vector $b$ has non-integer scalar product
with simple root $\pi_1$ of $E_6$ then we will obtain algebra
$SO(10)\times U(1)$ ( $(b\cdot\pi_1)$ even could be 1 if others products
are equals 0 mod 2).
More complicated examples are $E_6\times SU(3)$ and
 $SU(3)\times SU(3)\times SU(3)\times SU(3)$. For the former we must
forbid the $\pi_2$ root but permit it to form $SU(3)$ algebra.
Note that in $E_8$ root system there are two roots with $3\pi_2$.
We will use them for $SU(3)$. So the product $(b\cdot\pi_2)$ must be 2/3
while  others must be  0 mod 2.
We can also get GSO-projectors for all interesting subgroups of
$E_8$ in such a way but so far choosing of constant for scalar products
( $c_i$ in a previous subsection ) is rather experimental so
it is more convenient to follow certain breaking chain.
Below we will give some results for $E_6\times SU(3)$,
 $SU(3)\times SU(3)\times SU(3)\times SU(3)$ and
 $SO(10)\times U(1)\times SU(3)$.
We will give algebra basis and vectors that give GSO-projection needed
for  obtaining this algebra.
$E_6\times SU(3)$. This case follow from $E_8$ using root basis from a
previous subsection and choosing
$$c_i=(-2,-\frac{2}{3}, 0,2,-2,-2,2,0)$$
This gives GSO-projector of the form
$$b_1=(1,1,\frac{1}{3},\frac{1}{3},\frac{1}{3},\frac{1}{3},
\frac{1}{3},\frac{1}{3})$$
Basis of simple roots arises from sector with $1^8$ in right part and reads
\begin{eqnarray}
 \pi_1&=&\frac{1}{2}(+ e_1+ e_2+ e_3 +e_4 +e_5 +e_6 +e_7 +e_8)\nonumber\\
 \pi_2&=&\frac{1}{2}( +e_1 +e_2 -e_3 -e_4 -e_5 -e_6 -e_7 -e_8)\nonumber\\
 \pi_3&=&\frac{1}{2}( +e_1 -e_2 -e_3 -e_4 +e_5 +e_6 -e_7 +e_8)\nonumber\\
 \pi_4&=&\frac{1}{2}(- e_1 +e_2 +e_3 -e_4 -e_5 +e_6 +e_7 -e_8)\nonumber\\
 \pi_5&=&\frac{1}{2}( +e_1 -e_2 +e_3 +e_4 +e_5 -e_6 -e_7 -e_8)\nonumber\\
 \pi_6&=&\frac{1}{2}( -e_1 +e_2 -e_3 +e_4 -e_5 +e_6 -e_7 +e_8)\nonumber\\
 \pi_7&=&\frac{1}{2}( +e_1 -e_2 +e_3 -e_4 -e_5 -e_6 +e_7 +e_8)\nonumber\\
 \pi_8&=&\frac{1}{2}( -e_1 +e_2 -e_3 -e_4 +e_5 -e_6 +e_7 +e_8)\nonumber\\
\end{eqnarray}
$SO(10)\times U(1)\times SU(3)$. This case follow from $E_6\times SU(3)$.
In addition to $b_1$ we must find a vector that cut out $\pi_3$. Using
$$c_i=(0,0,1,0,0,0,0,0)$$
and inverse matrix of $E_6\times SU(3)$ basis we get GSO-projector of the form
$$b_2=(0,0,\frac{1}{3},-\frac{2}{3},\frac{1}{3},\frac{1}{3},
-\frac{2}{3},\frac{1}{3})$$
Basis of simple roots is the same as for $E_6\times SU(3)$ excluding $\pi_3$.

 $SU(3)\times SU(3)\times SU(3)\times SU(3)$.
Using  $E_6\times SU(3)$ basis inverse matrix with
$$c_i=(1,-1,-1,\frac{1}{3},1,\frac{1}{3},-1,-1)$$
We get GSO-projector of the form
$$b_2=(-\frac{1}{3},\frac{1}{3},1,1,\frac{1}{3},\frac{1}{3},
-\frac{1}{3},-\frac{1}{3})$$
Easy to see that such a $c_i$ cut out $\pi_4$ and $\pi_6$ roots but due to
appropriate combination in $E_6$ root system two $SU(3)$ groups will remain.
Basis of simple roots is
\begin{eqnarray}
 \pi_1&=&\frac{1}{2}(+ e_1+ e_2+ e_3 +e_4 +e_5 +e_6 +e_7 +e_8)\nonumber\\
 \pi_2&=&\frac{1}{2}( +e_1 +e_2 -e_3 -e_4 -e_5 -e_6 -e_7 -e_8)\nonumber\\
 \pi_3&=&\frac{1}{2}( +e_1 -e_2 +e_3 -e_4 -e_5 -e_6 +e_7 +e_8)\nonumber\\
 \pi_4&=&\frac{1}{2}(- e_1 +e_2 +e_3 -e_4 +e_5 +e_6 -e_7 -e_8)\nonumber\\
 \pi_5&=&\frac{1}{2}( +e_1 -e_2 +e_3 +e_4 +e_5 -e_6 -e_7 -e_8)\nonumber\\
 \pi_6&=&\frac{1}{2}( -e_1 +e_2 -e_3 -e_4 +e_5 -e_6 +e_7 +e_8)\nonumber\\
 \pi_7&=&\frac{1}{2}( -e_1 +e_2 +e_3 +e_4 -e_5 -e_6 +e_7 -e_8)\nonumber\\
 \pi_8&=&\frac{1}{2}( +e_1 -e_2 -e_3 -e_4 +e_5 +e_6 +e_7 -e_8)\nonumber\\
\end{eqnarray}
Using all this methods we could construct a model described in the next
section.

\newcommand{\thr}{\frac{1}{3}}
\newcommand{\tthr}{\frac{2}{3}}

\subsection{$E_6 \times SU(3)$ Three Generations Model -- Model~5}
 This model illustrates a branch of $E_8$ breaking
$E_8\rightarrow E_6\times SU(3)$ and is an interesting result on a way to
obtain three generations with gauge horizontal symmetry. Basis of the boundary
conditions (see Table \ref{basis4my}) is rather simple
but there are some subtle points.
In \cite{Lop} the possible left parts of basis vectors were worked out,
see it for details. We just use the notation given in \cite{Lop}
(~hat on left part means complex fermion, other fermions on the left
sector are real, all of the right movers are complex)
 and an example of commuting set of vectors.
\begin{table}[htb]
\caption{Basis of the boundary conditions for the Model~5.}
\label{basis4my}
\footnotesize
\begin{center}
\begin{tabular}{|c||c|cc||c|cc|}
\hline
Vectors &${\psi}_{1,2} $ & ${\chi}_{1,\ldots,9}$ &
${\omega}_{1,\ldots,9}$
& ${\bar \varphi}_{1,\ldots,6}$ &
${\Psi}_{1,\ldots,8} $ &
${\Phi}_{1,\ldots,8}$ \\ \hline
\hline
$b_1$ & $1 1 $ & $1^9$ &
$1^9$ & $1^{6} $ & $1^8 $ & $1^8$ \\
$b_2$ & $1 1$ & $\widehat\thr,1;-\widehat\tthr,0,0,\widehat\tthr$ &
$\widehat\thr,1;-\widehat\tthr,0,0,\widehat\tthr$ &  $ \tthr^3\:-\tthr^3$ &
$0^2\:-\tthr^6 $ & $1^2\:\thr^6$  \\
$b_3$ & $0 0$ & $0^9$ &
$0^9$ & $ 0^6$ &
$1^8$&$0^8$  \\
$b_4$ & $1 1$ & $\widehat{1},1;\widehat0,0,0,\widehat0$ &
$\widehat1,1;\widehat0,0,0,\widehat0$ &$0^6$ & $ 0^8 $ & $0^8$ \\
\hline \hline
\end {tabular}
\end{center}
\normalsize
\end{table}
A construction of an $E_6\times SU(3)$ group caused us to use rational
for left boundary conditions. It seems that it is the only way to
obtain such a gauge group with appropriate matter contents.
The model has $N=2$ SUSY. We can also construct model with
$N=0$ but according to \cite{Lop} using vectors that can give rise
to $E_6\times SU(3)$ (with realistic matter fields)
one cannot obtain $N=1$ SUSY.
\begin{table}[htb]
\caption{The choice of the GSO basis $\gamma [b_i, b_j]$. Model~5.
($i$ numbers rows and $j$ -- columns).}
\label{GSO4}
\footnotesize
\begin{center}
\begin{tabular}{|c||c|c|c|c|}
\hline
& $b_1$ & $b_2$ & $b_3$ & $b_4$   \\ \hline
\hline
$b_1$ & $0$ &  $1/3$ & $1  $ & $1$ \\
$b_2$ & $1$ &  $  1$ & $1  $ & $1$ \\
$b_3$ & $1$ &    $1$ & $0  $ & $1$ \\
$b_4$ & $1$ &   $1/3$ & $1  $ & $1$ \\
\hline \hline
\end {tabular}
\end{center}
\normalsize
\end{table}
Let us give a brief review of the model contents. First notice
that all superpartners of states in sector $\alpha$ are found in
sector $\alpha+b_4$ as in all previous models. Although the same sector
may contain, say, matter fields and gauginos simultaneously.
The observable gauge group
 $(SU(3)^I_H\times E^I_6)\times (SU(3)^{II}_H\times E^{II}_6)$ and hidden
group $SU(6)\times U(1)$ are rising up from sectors NS, $b_3$ and $3b_2+b_4$.
Matter fields in representations $({\bf 3},{\bf 27})+
(\overline{\bf 3},\overline{\bf 27})$
for each $SU(3)_H\times E_6$ group are found in sectors $3b_2$, $b_3+b_4$
and $b_4$. Also there are some interesting states in sectors $b_2,\:b_2+b_3,\:
2b_2+b_3+b_4,\:2b_2+b_4$ and $5b_2,\:5b_2+b_3,\:4b_2+b_3+b_4,\:4b_2+b_4$
that form representations $(\overline{\bf 3},{\bf 3})$
and $({\bf 3},\overline{\bf 3})$
of the $SU(3)^I_H\times SU(3)^{II}_H$ group. This states are singlets
under both $E_6$ groups but have nonzero $U(1)_{hidden}$ charge.
We suppose that the model permits further breaking of $E_6$ down to other
grand unification groups, but problem with breaking supersymmetry ${N=2}
\rightarrow N=1$ is a great obstacle on this way.

\section{N=2 Superconformal World-Sheet Supersymmetry in String
Scattering Amplitudes.  Superpotential.}
\subsection{The Global $U(1)_J$-Symmetry in  String Amplitude.}
The ability of giving a correct description of the fermion masses
and mixings will, of
course, constitute the decisive criterion for selection of a model of this
kind. Therefore, within our approach one has to
\begin{enumerate}
\item study the possible nature of the $G_H$ horizontal gauge symmetry
($N_g=3_H$ or $3_H+1_H$),
\item investigate the possible cases for
$G_H$-quantum numbers for quarks (anti-quarks) and leptons (anti-leptons),
 whether one can obtain vector-like or axial-like structure (or even
chiral $G_{HL}\times G_{HR}$ structure) for the horizontal interactions.
\item find the structure of the sector of the matter fields which
are needed for
the $SU(3)_H$ anomaly cancelation (chiral neutral "horizontal" or "mirror"
fermions),
\item write out all possible renormalizable and relevant non-renormalizable
contributions to the superpotential $W$ and their consequences for fermion
mass matrices depending on the ways of the horizontal symmetry breakings.
\end{enumerate}

In the superstring models the admissible superpotential is defined
by both gauge invariance and by two dimensional superconformal model
on worldsheet. This gives strong additional restriction for superpotential.
Each term of superpotential must correspond to non-zero correlator
of vertex operators on worldsheet.

The conformal invariance allows us to construct vertex
operators in different pictures, carrying different ghost numbers.
A canonical vertex for a space-time boson (fermion) is in the $-1(-1/2)$
picture. The conformal field representing the ghost charge q is written
by $e^{qc}$ with conformal dimension $(-1/2q^2-q,0)$.

A modular invariant,
N=1 space-time supersymmetric theory also contains a hidden global N=2
world sheet superconformal symmetry  \cite{bank'}, which distinguishes
now three components of the N=1 supercurrent
 $T_F^+$, $T_F^-$ and $T_F^0$ with $\pm 1, 0$ charge under
the $U_J(1)$ group with  conserved current $J$ of the N=2 worldsheet algebra.
This conserved current may play a key role in
constructing of realistic phenomenology. So, all vertex operators
have a definite $U(1)$ charge.
For $J(z)$ we have
\medskip
\begin{eqnarray}
J(z) \,=\,i \partial _z \,(S_{12}\,+\,S_{34}\,+\,S_{56}\,),
\end{eqnarray}
\medskip
where $S_{ij}$ are the bosonized components of supersymmetry generator
(which corresponds to basis vector $S: \chi^I=1$) \cite{20'}.

The general form of vertex operator for bosonic and fermionic components of
the chiral superfield is given by, respectively:

\begin{equation}
\label{bos}
V_{(-1)}^b(z)=e^{-c}e^{i\alpha S_{12}} e^{i\beta S_{34}} e^{i\gamma S_{56}}
G e^{i\frac{1}{2} K X} e^{i\frac{1}{2} K \bar X}
\end{equation}
and
\begin{equation}
\label{ferm}
V_{\alpha (-1/2)}^f(z)=e^{-c/2}S_{\alpha}e^{i(\alpha-1/2)
 S_{12}}e^{i(\beta-1/2) S_{34}}e^{i(\gamma-1/2) S_{56}}
G e^{i\frac{1}{2} K X} e^{i\frac{1}{2} K \bar X}
\end{equation}
where $\alpha, \beta, \gamma = 0,\pm1/2,\pm1)$.
The conformal fields for left $\chi$-fermions (excitations of SUSY-generated
basis vector $S=b_4$) are written explicitly. $G$ contains
remaining left and right conformal fields.

The $U_{J}(1)$ charge for bosonic (fermionic) vertex operator in the canonical
$-1(-1/2)$ picture is
$\alpha+ \beta + \gamma =1$ , ($\alpha+ \beta + \gamma -3/2= -1/2 )$.
This condition and requirement of correct conformal dimension $(1;1)$
($\alpha^2 +\beta^2 +\gamma^2 \leq 1$) for vertex operators gives us
that the only solution are permutations of
\begin{equation}
\label{permut}
(1,0,0)\ \ (NS)\ \mbox{ and }\ (1/2,1/2,0)\ \  (R).
\end{equation}
These are $\alpha, \beta, \gamma $--charges for canonical bosonic vertex.

Since we are dealing with the supersymmetric theories, we can consider
canonical bosonic vertices ($-1$ picture) what define all remaining
vertices in different pictures (see for example (\ref{ferm})).
Let us define a vertex to be a R[NS]-vertex if it [not] 
contains $S_{ij}-$fields
(for $\chi$-fermions) with charges $\alpha, \beta, \gamma = \pm1/2$
in canonical bosonic picture.
It means that for NS sectors all left part equals to zero, and all
the remaining sectors we label as R (for massless states).

Due to space-time supersymmetry for reconstruction of the superpotential
we need the N-point functions with two fermionic vertex only.
The tree-level string amplitude is given by
\begin{equation}
\label{amplituda}
A_N = \frac{g^{(N-2)}}{(2\pi)^{(N-3)}}\sqrt{2}\int\prod^{N-3}_{i=1} d^2 z_i
\langle V_{1(-1/2)}^f V_{2(-1/2)}^f V_{3(-1)}^b V_{4(0)}^b\ldots
V_{N(0)}^b\rangle
\end{equation}
The choice of pictures is determined by the requirement that the total ghost
charge must be $-2$.

Of course,
for nonzero amplitude, the total charge must be equal to zero
(after all picture-changings have been done) not only
for the $U_J(1)$ global symmetry, but also for every global left $U_R(1)_i$
symmetry.
In particular, requirement of conservation for three left-moving
$U(1)_{\alpha ,\beta ,\gamma}$ R-symmetries gives very useful
selections rules for nonzero amplitudes (see below).

\subsection{Renormalizable Three Point Amplitudes}
\label{ren.vert.}

Before calculations let us describe some selection rules which follow
from $U(1)_{\alpha ,\beta ,\gamma}$ conservation.
According to the formula (\ref{amplituda}) for renormalizable case
(three-point amplitude) we have two canonical fermionic vertices and
one canonical bosonic vertex.
From (\ref{ferm}) it follows that after transition from bosonic vertex
to fermionic vertex each $\alpha, \beta, \gamma$--charge decreases
by $1/2$. On the other hand we know the admissible charges of
canonical bosonic vertex (\ref{permut}).
So, we obtain the following admissible combinations of charges for
canonical bosonic vertices:

\medskip

\begin{tabular}{|c||ccc|ccc|ccc|}\hline
&\multicolumn{3}{c|}{$(NS)^3$}&\multicolumn{3}{c|}{$(R)^2(NS)$}&
\multicolumn{3}{c|}{$(R)^3$} \\ \hline\hline
$\alpha$ & 1 & 0 & 0 & 1/2 & 1/2 & 0 & 1/2 & 1/2 & 0 \\ \hline
$\beta $ & 0 & 1 & 0 & 1/2 & 1/2 & 0 & 1/2 &  0  & 1/2 \\ \hline
$\gamma$ & 0 & 0 & 1 &  0  &  0  & 1 &  0  & 1/2 & 1/2 \\ \hline
\end{tabular} 

\medskip

The case $R(NS)^2$ is forbidden. For understanding of this fact let us
transform two vertices (R and NS) from bosonic to fermionic (\ref{ferm})
and sum up their charges. Then two of three charges will be half integer
and remaining one canonical bosonic $NS$-vertex
(and also the bosonic $NS$-vertices in zero picture) can not compensate them.
In general, this argument takes place for all operators of the form
$R\times (NS)^k$. These operators vanish.

Now let us calculate the renormalizable superpotential of Model 1.
For example take the contribution  $R^2(NS)$ to
the three point fermionic-fermionic-bosonic function
 $$\Psi^1_{(1,3)} \Psi^2_{(\bar 5 ,1)} \Phi^3_{(5,\bar 3 )}, $$
where the fields 1 and 2 are from Ramond-sector and 3 is from NS-sector.
The charges  $\alpha , \beta , \gamma$ correspond to  positions
2, 6, 10 of Model 1. (see $b_4$).
In  boson sector in $\Psi^1_{(1,3)}$,
$ \Psi^2_{(\bar 5 ,1)}$  the left Ramond fermions' position numbers are
3, 4, 6, 10. The field $ \Phi^3_{(5,\bar 3 )} $ from  NS-sector
has in boson sector the excitation by world sheet left fermion No 2.
The nonvanishing $U(1)_R$--charges of these three vertex operators
are: $\beta_1=\gamma_1=\beta_2=\gamma_2=1/2$
and $\alpha_3=1$.

So for the corresponding vertex we have:
\begin{equation}\label{vertex1}
V^f_{1(-1/2)}=e^{-c/2}S_{\alpha}e^{-i/2\ H_2}\Sigma^{+}_3
\Sigma^{+}_4 e^{i/2\ K_1X} \bar G_1 e^{i/2\ K_1\bar X}
\end{equation}
\begin{equation}\label{vertex2}
V^f_{2(-1/2)}=e^{-c/2}S_{\alpha}e^{-i/2\ H_2}\Sigma^{-}_3
\Sigma^{-}_4 e^{i/2\ K_2X} \bar G_2 e^{i/2\ K_2\bar X}
\end{equation}
\begin{equation}\label{vertex3}
V^b_{3(-1)}=e^{-c}e^{i H_2} e^{i/2\ K_3X} \bar G_3 e^{i/2\ K_3\bar X}
\end{equation}
where $\Sigma^{\pm}_{k}\equiv e^{\pm\frac{i}{2}H_{k}}$
($H_k$ is the boson which corresponds to left complex fermion No $k$).
The right-moving operator ${\bar G}_j =\prod_i  e^{iW_i^1\cdot J_i}(j) $
represents the gauge degrees of freedom.

We see that the correlator of these vertex operators is not equal to zero
 (for details see \cite{20'}).
 So one can easily  get the superpotential $W_1$ (Higgs-matter)
(with  the corresponding coefficient $g\sqrt{2}$):

\begin{eqnarray}\label{eq32}
W_1&=& g\sqrt{2} \biggl[ {\hat \Psi}_{(1,3)}
{\hat \Psi}_{({\bar 5},1)}
{\hat\Phi}_{(5,{\bar 3})} +
 {\hat \Psi}_{(1,1)} {\hat \Psi}_{({\bar 5},3)}
{\hat\Phi}_{(5,{\bar 3})} + \nonumber\\
&+& {\hat \Psi}_{(10,3)} {\hat \Psi}_{({\bar 5},3)}
{\hat\Phi}_{({\bar5},3)} +
{\hat \Psi}_{(10,3)} {\hat \Psi}_{(10,1)}
{\hat\Phi}_{(5,{\bar 3})} \biggr]
\end{eqnarray}

From the above form of the Yukawa couplings it
follows that two (chiral) generations
have to be very light (comparing to $M_W$ scale).

The $SU(3)_H$ anomalies of the matter fields (row No 2) are
naturally canceled by
the chiral "horizontal" superfields  forming two sets:
${\hat \Psi}^H_{(1,N;1,N)}$ and ${\hat \Phi}^H_{(1, N;1, N)}$,
where $N = \underline 1, \,  \underline 3$,~
(with both ${SO(2)}$ chiralities, see Table \ref{tabl3}, row No 3, 4
respectively).
The superpotential, $W_2$, consists of the following $(R)^2\times (NS)$-terms
and two $R^3$-terms:
\begin{eqnarray}\label{eq34}
W_2&=& g\sqrt{2} \biggl\{\biggl[
{\hat \Phi}^H_{(1,1;1,\bar3)}{\hat \Phi}^H_{(1,\bar 3;1,1)}
{\hat \Phi}_{(1,3;1,3)} +
{\hat \Phi}^H_{(1,1;1,1)}{\hat \Phi}^H_{(1,\bar 3;1,\bar 3)}
{\hat \Phi}_{(1,3;1,3)} + \nonumber\\
&+& {\hat \Phi}^H_{(1,\bar 3;1,\bar 3)}{\hat \Phi}^H_{(1,\bar 3;1,\bar 3)}
{\hat \Phi}_{(1,\bar 3;1,\bar 3)} +
 {\hat \Psi}^H_{(1,{\bar 3};1,1)}{\hat \Psi}^H_{(1,{\bar 3};1, 3)}
{\hat \Phi}_{(1,{\bar 3};1,{\bar 3})} +\nonumber\\
&+&{\hat \Psi}^H_{(1,1;1,3)}
{\hat \Psi}^H_{(1,{\bar 3};1,3)}
{\hat \Phi}_{(1, 3;1, 3)}\biggr]
+\biggl[ {\hat \Phi}_{(5, \bar 3 ;1, 1)} {\hat \phi}_{(-_1 +_3)(1, 3;1, 1)}
{\hat \phi}_{(+_1 -_3)(\bar 5, 1;1, 1)} +\nonumber\\
&+& {\hat \Phi}_{(1, 1 ;5, \bar 3 )} {\hat \phi}_{(-_1 +_3)(1, 1;1, 3)}
{\hat \phi}_{(+_1 -_3)(1, 1;\bar 5 , 1)} + \mbox{conj. repr-s}\biggr] +
\nonumber\\
&+& {\hat \sigma }^1_{(-_1 -_4 )} {\hat \sigma }^2_{(+_1 +_4 )}
{\hat \Psi }_{(1,1;1,1)}
+ {\hat \sigma }^2_{(+_1 -_4 )} {\hat \sigma }^3_{(+_3 +_4 )}
{\hat \Phi }^H_{(-_1 -_3)} \biggr\}
\end{eqnarray}

From (\ref{eq34}) it follows that some of the horizontal fields in sectors
(No 3, 4) remain massless at the tree-level.
This is a remarkable prediction: "horizontal" fields are "sterile",
e.g. they interact with the ordinary
chiral
matter fields only through the $U(1)_H$ and $SU(3)_H$
gauge boson
and therefore this "sterile" matter
is of an  interest in the context of the
experimental searches on accelerators or in astrophysics.
The Higgs fields  could give the following
$(NS)^3$ contributions to the renormalizable superpotential:
\begin{eqnarray}\label{w3}
W_3=\sqrt{2}g\biggl\{
{\hat\Phi }_{ (5,1;1,3)}  {\hat\Phi }_{ (\bar 5,1;\bar 5,1)}
{\hat\Phi }_{ (1,1;5,\bar 3)} +
{\hat\Phi }_{ (5,1;1,3)}  {\hat\Phi }_{ (1,\bar 3;1,\bar 3)}
{\hat\Phi }_{ (\bar 5,3;1,1)} && \nonumber\\
+ {\hat\Phi }_{ (1,3;5,1)}  {\hat\Phi }_{ (\bar 5,1;\bar 5,1)}
{\hat\Phi }_{ (5,\bar 3;1,1)} +
{\hat\Phi }_{ (1,3;5,1)}  {\hat\Phi }_{ (1,\bar 3;1,\bar 3)}
{\hat\Phi }_{ (1,1;\bar 5,3)} +\mbox{conj. repr-s}
\biggr\}&&
\end{eqnarray}

So, $W_1 +W_2 +W_3$ is the most general renormalizable superpotential
which includes all nonzero three-point (F-type) expectation values
of the vertex operators for corresponding 2-dimensional conformal
model.

\subsection{Non-Renormalizable Four, Five, and Six Point Contributions
to the Superpotential.}

Following the formula (\ref{amplituda}), beginning with
the 4-th, vertices have to be written in noncanonical form (in picture 0).
The formula of changing pictures from  $q$ to $q+1$ is\cite{20'}:
\begin{equation}
\label{pic.ch.3}
V_{q+1}(z)=\lim_{w\to z} e^c(w) T_F(w) V_q(z) \label{lim}
\end{equation}
However, the sum of $U(1)_J$-charges for first three vertices
(\ref{amplituda}) equals to zero and $U(1)_J$-charge of canonical
($-1$ picture) bosonic vertex is $+1$.
So, the effective contribution in formula (\ref{lim}) is going only
from  $T_F^{-1}$ and for complex case one can obtain:
\begin{equation}
T_F^{-1}=\frac{i}{2\sqrt{2}} \sum_k e^{-iH_k}
\biggl[ (1-i)e^{iH_{k'}} e^{iH_{\bar k'}}
+(1+i)e^{iH_{k'}} e^{-iH_{\bar k'}} +(1+i)e^{-iH_{k'}} e^{iH_{\bar k'}}
+(1-i)e^{-iH_{k'}} e^{-iH_{\bar k'}} \biggr]
\end{equation}
We take that for complex worldsheet fermions
${\hat\chi}_k^{(*)}=\frac{1}{\sqrt{2}}(\chi_n \pm i\chi_{n+1})=e^{\pm iH_k}$ ,
where $k= 1 +(n+1)/2$ ;
$H_{k'}$ and $H_{\bar k'}$ -- similarly   for $y_n$ and $\omega_n$.
The multipliers  for $\chi$ have been written
explicitly and for $y (\omega)$ we have also:
$\Sigma^{\pm}_{k'}\equiv e^{\pm\frac{i}{2}H_{k'}}$.
For Model~1 complex triplets $({\hat\chi}_j , {\hat y}_j , {\hat\omega}_j)$
correspond to complex fermions with numbers:
$(2,5,8)$; $(6,3,9)$; $(10,4,7)$.

Gauge invariance leads to more possible contributions to
4-,5-,6-, \ldots-point  terms, see Tables \ref{table1f}, \ref{table2f},
and \ref{table3f}. It is very important to take into account that
some of them are equal zero because of global symmetries coming from the
left superstring sector.

Let us study the $\underline{(R)^4}$ amplitude,
which is the only nonvanishing type of four-point operators \cite{20'}.
Firstly, let us study $U(1)_R^3$ constraints. We know, that for the 
first and the second vertices of (\ref{amplituda}) each "canonical"
charge is decreased by $1/2$
(in sum $-1$), third canonical vertex remains the same, and for
other canonical bosonic vertex(es) one of the three charges
is decreased by $1$ (in virtue of $T_F^{-1}$ which consists three
different parts, each with one nonzero $-1$ charge).
Moreover, it is important, that if in canonical vertex $V^b_{-1}$
one of charges $\alpha $, $\beta $, $\gamma $ is equal to zero,
then for vertex $V^b_0$ in zero-picture this charge is equal to 0 too,
since the only nontrivial OPE in formula (\ref{pic.ch.3}) is
$e^c (w) e^{-c} (z)\sim (w -z)\rightarrow 0 $.
For example, further we will not consider the charges combinations
with one total charge (of $\alpha $, $\beta $, $\gamma $) equals to zero,
since such combinations are forbidden by charge conservation.

So, the only admissible charges combination for $(R)^4$
(for canonical bosonic vertices) is:

\medskip

\begin{tabular}{|c||cccc|c|}\hline
&\multicolumn{4}{c|}{$(R)^4$}& Tot. \\ \hline\hline
$\alpha$ &  0  & 1/2 & 1/2 &  0  & 1 \\ \hline
$\beta $ & 1/2 &  0  &  0  & 1/2 & 1 \\ \hline
$\gamma$ & 1/2 & 1/2 & 1/2 & 1/2 & 2 \\ \hline
\end{tabular} 

\medskip

(Of course all permutations of four vertices and permutations
of $\alpha , \beta , \gamma$--charges are admissible also.)
Other combinations with total charges (0, 2, 2), (1, 3/2, 3/2) and
(1/2, 3/2, 2)
(and permutations) are forbidden by $U(1)_R^3$ conservation.

Now we can investigate superpotential of fourth order.
We need to calculate the correlator:
$\langle V^f_{1(-1/2)} V^f_{2(-1/2)} V^b_{3(-1)} V^b_{4(0)}\rangle $.
So for R-case we need to change  vertex $V^b_R$ from picture --1 to picture 0.
Making use of (\ref{lim}) to:
$$V^b_{R(-1)}=e^{-c}e^{i/2\ H_k}e^{i/2\ H_l}
\Sigma^{\pm}_{k'}\Sigma^{\pm}_{l'}
e^{i/2\ KX} \bar G_R e^{i/2\ K\bar X}$$
one can get:
\begin{eqnarray}
&&V^b_{R(0)}=\frac{i}{2\sqrt{2}}\biggl\{ e^{-i/2\ H_k} e^{i/2\ H_l}
\biggl( \Sigma_{k'}^{\mp} [(1\pm i)e^{iH_{\bar k'}}
+ (1\mp i)e^{-iH_{\bar k'}}]\biggr) \Sigma^{\pm}_{l'} \nonumber\\
&&+ e^{i/2\ H_k} e^{-i/2\ H_l} \Sigma^{\pm}_{k'}
\biggl( \Sigma_{l'}^{\mp} [(1\pm i)e^{iH_{\bar l'}}
+ (1\mp i)e^{-iH_{\bar l'}}]\biggr)  \biggr\}
\times e^{i/2\ KX} \bar G_R e^{i/2\ K\bar X}
\end{eqnarray}

For Model~1 many $(R)^4$ terms are permitted by gauge invariance
(see table \ref{table1f}).
However, the marked by "plus" terms only correspond to the total 
charges combination
(2, 1, 1). Other terms correspond to (0, 2, 2) combination and vanish.

Let us for example consider one of these marked by "plus" terms.
The vertex operators consist of
\begin{eqnarray}
(1,3,1,1)_{Mat.} & \sim e^{-c/2} e^{-i/2\ H_2}\Sigma^+_3\Sigma^+_4
 &  (fer.-1/2) \nonumber\\
{}_{(+_1,-_3)}(1,\bar 3,1,1)_{No 4} & \sim e^{-c/2} e^{-i/2\ H_6}
\Sigma^+_5\Sigma^+_7
 &  (fer.-1/2) \nonumber\\
(-_1,-_4)_{No 6} & \sim e^{-c} e^{i/2\ H_2}e^{i/2\ H_{10}}\Sigma^+_4\Sigma^-_5
 &  (bos.-1) \nonumber\\
(+_3, +_4)_{No 6} & \sim\frac{i\sqrt{2}}{4}\biggl\{
e^{-i/2\ H_6}e^{i/2\ H_{10}}\Sigma^+_3\Sigma^+_7
[(1-i)e^{iH_9} + (1+i)e^{-iH_9}] & \nonumber\\
& + \underline{e^{i/2\ H_6}e^{-i/2\ H_{10}}\Sigma^-_3\Sigma^-_7}
[(1+i)e^{iH_4} + \underline{(1-i)e^{-iH_4}}] \biggr\} & (bos.\ 0) \nonumber
\end{eqnarray}
Where have omitted the factor $e^{i/2\ K_iX} \bar G_R e^{i/2\ K_i\bar X}$.
The correlator of the product of the underlined terms and the first three
vertices is nonvanishing.
Let us give more detailed calculations.
\begin{equation}
\label{A4}
A_4 = \frac{g^2\sqrt{2}}{2\pi}\int d^2 z\ \langle \mbox{left part}\rangle
\ \langle \bar G_1 \bar G_2 \bar G_3 \bar G_4 \rangle
\ \langle\prod_i e^{i/2\ K_iX} e^{i/2\ K_i\bar X} \rangle
\end{equation}
where
$$
\langle \mbox{left part}\rangle ={\small \frac{1+i}{2\sqrt{2}}}\ z_{12}^{-3/4}
z_{13}^{-1/2} z_{14}^{-3/4} z_{23}^{-3/4} z_{24}^{-1/2} z_{34}^{-3/4}
$$
and for the right part we have
\begin{equation}
\label{Rpart}
\langle \bar G_1 \bar G_2 \bar G_3 \bar G_4 \rangle =
\prod_i\langle e^{iW_i^1\cdot J_i}(1) e^{iW_i^2\cdot J_i}(2)
e^{iW_i^3\cdot J_i}(3) e^{iW_i^4\cdot J_i}(4) \rangle
=\prod_{k < l} ({\bar z}_{kl})^{\sum_i W_i^k\cdot W_i^l} \prod_i C_i^{1234}
\end{equation}
where the $ C_i^{1234}$ guarantees the gauge invariance.
From the conformal invariance we have constraint:
$\frac{1}{2}\sum_i (W_i^l)^2 =1\ \forall\ l $, and from the gauge invariance
we have constraint:
$\sum_l W_i^l =0\ \forall\ i $, where $i=1,\ldots ,22$ and $l=1,\ldots ,N$.
So, we can obtain:
$\sum_{k < l}\sum_i W_i^k W_i^l = -N$.

Let us write down (charges) weight $W^l$-vectors of our vertices for some
representatives of gauge multiplets. \\
\begin{tabular}{|c||ccc|ccc|c|ccc|c|c|}\hline
 &\multicolumn{3}{c|}{$U(1)^3$} &\multicolumn{3}{c|}{SO(6)} &
U(5) &\multicolumn{3}{c|}{U(3)} & U(5) & U(3) \\ \hline\hline
$W^1$ & &$0^3$& & &$0^3$& &$1/4^5$&1/4&1/4&5/4&$0^5$&$0^3$ \\ \hline
$W^2$ &1/2&0&--1/2& &$0^3$& &$-1/8^5$&3/8&3/8&--5/8&$3/8^5$&$-1/8^3$ \\ \hline
$W^3$ &--1/2&0&0&--1/2&1/2&1/2&$1/8^5$& &$-3/8^3$& &$-1/8^5$&$3/8^3$ \\ \hline
$W^4$ &0&0&1/2&1/2&--1/2&--1/2&$-1/4^5$& &$-1/4^3$& &$-1/4^5$&$-1/4^3$\\ \hline
\end{tabular} \\
Now from formulae (\ref{A4}-\ref{Rpart}) we can obtain, 
that  there is the following factor in the amplitude $A_4$:
$$
|z_{12}|^{-3/2} |z_{13}|^{-1} |z_{14}|^{-3/2}
|z_{23}|^{-3/2} |z_{24}|^{-1} |z_{34}|^{-3/2}\ .
$$
If we fix the gauge: $z_1=\infty$, $z_2=z$, $z_3=1$, $z_4=0$
and take into account the Faddeev-Popov minideterminant (\ref{det})
then we will end up with integral
$$
I_4=\int d^2 z |z|^{-1} |z-1|^{-3/2}=8\pi\sum_{k=0}^{\infty}
\frac{1}{1+4k}\biggl[\frac{(2k-1)!!}{(2k)!!}\biggr]^2
=\frac{128}{\pi}\Gamma^4(5/4) = 27.50
$$

Now, we should take zero-momentum amplitude and take into account the factor
$\biggl( \frac{2\sqrt{8\pi}}{gM_{Pl}}\biggr)$ which reduces dimension.
For other underlined terms calculations are analogous.

So, in Model~1 the only non-vanishing nonrenormalizable four-point
superpotential $W_4=(R)^4$ is as follows
\small
\begin{eqnarray}
W_4 &=& \frac{2g I_4}{\sqrt{\pi}M_{Pl}}\biggl\{ {\hat\Psi}_{(1,3;1,1)}
{\hat\Phi}^H_{ (+_1,-_3)(1,\bar 3;1,1)}
{\hat\sigma}_{1\ (-_1,-_4)} {\hat\sigma}_{3\ (+_3,+_4)} \nonumber\\
&+& [{\hat\phi}_{1\ (-_1 ,+_3)(1,3;1,1)}
{\hat\phi}_{1\ (-_1 ,-_3)(1,\bar 3 ;1,1)}
+{\hat\phi}_{3\ (-_1 ,+_3)(1,1;3,1)}
{\hat\phi}_{3\ (-_1 ,-_3)(1,1 ;\bar 3,1)}]\times {\hat\sigma}_{2\ (+_1 ,-_4)}
{\hat\sigma}_{2\ (+_1 ,+_4)}
 \nonumber\\
&+& {\hat\phi}_{1 (-_1 ,-_3)(1,\bar 3 ;1,1)}
{\hat\phi}_{3 (-_1 ,+_3)(1,1;1,3)} [
{\hat\Phi}^H_{2 (+_1 ,-_3)(1,\bar 3;1,1)}
{\hat\Phi}^H_{3 (+_1 ,+_3)(1,\bar 3;1,\bar 3)} +
{\hat\Psi}^H_{3 (+_1 ,+_2)(1,\bar 3;1,3)}
{\hat\Psi}^H_{3 (+_1 ,-_2)(1,\bar 3;1, 3)} ]
 \nonumber\\
&+& {\hat\sigma}_{1\ (+_1 ,-_4)}
{\hat\sigma}_{1\ (-_1 ,+_4)} {\hat\sigma}_{3\ (-_3 ,+_4)}
{\hat\sigma}_{3\ (+_3 ,+_4)}
+ {\hat\phi}_{1\ (-_1 ,+_3)(1,3 ;1,1)}
{\hat\phi}_{3\ (-_1 ,-_3)(1,1;1,\bar 3)}\times \\
&\times & [{\hat\Psi}^H_{3\ (+_1 ,-_2)(1,\bar 3;1,3)}
{\hat\Psi}^H_{4\ (+_1 ,+_2)}
+
{\hat\Psi}^H_{3\ (+_1 ,+_2)(1,\bar 3;1,3)}
{\hat\Psi}^H_{4\ (+_1 ,-_2)}
+
{\hat\Phi}^H_{1\ (+_1 ,-_3)(1,1;1,\bar 3)}
{\hat\Phi}^H_{3\ (+_1 ,+_3)(1,\bar 3;1,\bar 3)}]
\biggr\} \nonumber
\end{eqnarray}
\normalsize
\bigskip

Let us consider five point contributions to the superpotential.
We know that operators of the form  $R\times (NS)^4$ vanish (see section
\ref{ren.vert.}). Also operators $(NS)^5$ and $R^3\times (NS)^2$ are
forbidden\cite{20'}.

So, the admissible terms in superpotential of five order are:
$R^5$, $R^4\times (NS)$ and $R^2\times (NS)^3$.

Firstly, let us consider $\underline{R^4\times (NS)}$ operators.
Only the following combinations of $U(1)_R^3$ charges for the canonical boson 
vertices are admissible: \\
\small
\begin{tabular}{|c||ccccc|c||ccccc|c||ccccc|c|}\hline
&\multicolumn{5}{c|}{$R^4\times (NS)$ \underline{Case 1.}}& T. &
\multicolumn{5}{c|}{$R^4\times (NS)$ Case 2.}& T. & 
\multicolumn{5}{c|}{$R^4\times (NS)$ \underline{Case 3.}}& T. 
\\ \hline\hline
$\alpha$ & 1/2 & 1/2 & 1/2 & 1/2 & 1 & 3 
& 1/2 & 1/2 & 1/2 & 1/2 & 0 & 2
& 1/2 & 1/2 & 1/2 & 1/2 & 0 & 2
\\ \hline
$\beta $ & 1/2 & 1/2 & 0 & 0  & 0 & 1 & 1/2 & 1/2 & 0 & 0  & 0 & 1 
& 1/2 & 1/2 & 1/2 & 1/2 & 0 & 2
\\ \hline
$\gamma$ & 0 & 0 & 1/2 & 1/2 & 0 & 1 & 0 & 0 & 1/2 & 1/2 & 1 & 2 
& 0 & 0 & 0 & 0 & 1 & 1
\\ \hline
\end{tabular} 
\normalsize

\medskip

(otherwise we will have half integer total charges or the case
which have been described before $R^4$).

However, the amplitude with charges combination like in the Case 2 vanishes.
Really, if we write down the last three vertices in pictures
$-1/2$, $-1/2$, $-1$ correspondingly, then the $\gamma$-charge of
these three operators will be $+1$. This charge can not be compensated
by the charge of $T_F^{-1}$ operator since the first two operators
have "canonical" $\gamma $-charges equal to zero.

So, only the contributions to the superpotential from terms 
with $U(1)_R^3$ charges like Cases 1 and 3 are admissible.

In Table \ref{table2f} there are given all gauge invariant terms
$R^4\times (NS)$. There we denote non-zero and zero states after global
invariance considerations with signs `+' and `--'.
Note, that in Model~1 all admissible terms correspond to Case 1.

Let us write down the non-vanishing nonrenormalizable
five-point superpotential $R^4\times (NS)$ for Model~1:

\begin{eqnarray}
W_5 &&=\frac{2\sqrt{2}g I_5}{\pi M_{Pl}^2} \biggl\{ {\hat\Phi }_{5\ (5,1;1,3)}
\times [ {\hat\Psi }_{(10,3;1,1)}  {\hat\Psi }_{(10,1;1,1)}
\nonumber\\
&& +{\hat\Psi }_{(1,3;1,1)}  {\hat\Psi }_{(\bar 5 ,1;1,1)}
+{\hat\Psi }_{(\bar 5 ,3;1,1)}  {\hat\Psi }_{(1,1;1,1)}
+{\hat\phi }_{1\ (1,3;1,1)}  {\hat\phi }_{2\ (1,\bar 5 ;1,1)} ]\times
\nonumber\\
&&\times [{\hat\Psi }^H_{1 (1,1;1,3)} {\hat\Psi }^H_{3 (1,\bar 3;1,3)}
+ {\hat\Phi }^H_{1 (1,1;1,\bar 3)} {\hat\Phi }^H_{2 (1,\bar 3;1,1)}
+ {\hat\Phi }^H_{3 (1,\bar 3;1,\bar 3)} {\hat\Phi }^H_{4 (1,1;1,1)}]
+{\hat\Phi }_{5 ({\bar 5},1;1,{\bar 3})} \times
\nonumber\\
&& \times [{\hat\Psi }_{(10,3;1,1)} {\hat\Psi }_{(\bar 5,3;1,1)}
+{\hat\phi }_{1 (1,\bar 3 ;1,1)}  {\hat\phi }_{2 (1, 5 ;1,1)}]
\times
[{\hat\Psi }^H_{2 (1,\bar 3;1,1)} {\hat\Psi }^H_{3 (1,\bar 3;1,3)} +
{\hat\Phi }^H_{3 (1,\bar 3 ;1,\bar 3 )}
{\hat\Phi }^H_{3 (1,\bar 3 ;1,\bar 3 )}]
\nonumber\\
&&+{\hat\Phi }_{6 (1,3;5,1)} {\hat\phi }_{3(1,1;1,3)}
{\hat\phi }_{4(1,1;\bar 5 ,1)}\times
[{\hat\Psi }^H_{1 (1,1;1,3)} {\hat\Psi }^H_{3 (1,\bar 3 ;1,3)} +
{\hat\Phi }^H_{1 (1,1;1,\bar 3)} {\hat\Phi }^H_{2 (1,\bar 3 ;1,1)} +
{\hat\Phi }^H_{3 (1,\bar 3 ;1,\bar 3)} {\hat\Phi }^H_{4 (1,1 ;1,1)}]
\nonumber\\
&&+{\hat\Phi }_{6 (1,\bar 3 ;\bar 5 ,1)} {\hat\phi }_{3(1,1;1,\bar 3 )}
{\hat\phi }_{4(1,1; 5 ,1)}\times
[{\hat\Psi }^H_{2 (1,\bar 3 ;1,1)} {\hat\Psi }^H_{3 (1,\bar 3 ;1,3)} +
{\hat\Phi }^H_{3 (1,\bar 3 ;1,\bar 3)} {\hat\Phi }^H_{3 (1,\bar 3 ;1,\bar 3)}]
\nonumber\\
&&+{\hat\Phi }_{5 (5,1;1, 3)} {\hat\Psi }_{({\bar 5},1;1,1)}
{\hat\Phi }^H_{1(1,1;1,\bar 3 )}{\hat\sigma }_1 {\hat\sigma }_3
\nonumber\\
&&+{\hat\Phi }_{5 ({\bar 5},1;1,{\bar 3})} {\hat\Psi }_{({\bar 5},1;1,1)}
{\hat\Psi }_{(10,1;1,1)}\times
[{\hat\Psi }^H_{1 (1,1;1,3)} {\hat\Psi }^H_{4 (1,1;1,1)} +
{\hat\Phi }^H_{1 (1,1;1,\bar 3)} {\hat\Phi }^H_{1 (1,1;1,\bar 3)}] \biggr\}
\end{eqnarray}
where the value of the parameter $I_5$ is different for different terms.

For example, let us present explicit calculation for the last term.
So, we have for $(\bar 5,1,1,1)_{Mat.}$
\begin{equation}
V^f_{1\ R(-1/2)}\sim e^{-c/2}S_{\alpha }e^{-i/2\ H_2}
\Sigma^{-}_{3}\Sigma^{-}_{4}\ ,
\end{equation}
for $(10,1,1,1)_{Mat.}$
\begin{equation}
V^f_{2\ R(-1/2)}\sim e^{-c/2}S_{\beta }e^{-i/2\ H_2}
\Sigma^{+}_{3}\Sigma^{+}_{4}\ ,
\end{equation}
for  $(\bar 5,1,1,\bar 3)_{Hig.}$
\begin{equation}
V^b_{5\ NS(-1)}\sim e^{-c}e^{i H_{10}}
\end{equation}

The rest two bosonic fields we must write in
non-canonical picture 0. At first we write it in canonical form and then we
will go to 0-picture.
For the fields $(+_1,-_3)(1,1,1,\bar3)_{No4}$ in canonical picture we
have:
$$V^b_{R(-1)}\sim e^{-c}e^{i/2\ H_2}e^{i/2\ H_{10}}
\Sigma^{+}_{5}\Sigma^{+}_{7}$$
Let us go to picture 0:
\begin{eqnarray}
&&V^b_{3\ R(0)}\sim\biggl\{ e^{-i/2\ H_2} e^{i/2\ H_{10}}
\biggl( \Sigma_{5}^{-} [(1+i)e^{iH_{8}}
+ (1- i)e^{-iH_{8}}]\biggr) \Sigma^{+}_{7} \nonumber\\
&&+ e^{i/2\ H_2} e^{-i/2\ H_{10}} \Sigma^{+}_{5}
\biggl( \Sigma_{7}^{-} [ \underline{ (1+ i)e^{iH_{4}} }
+ \underline{\underline{ (1-i)e^{-iH_{4}} }} ]\biggr)  \biggr\}
\end{eqnarray}
At least for the field  $(-_1,+_3)(1,1,1,\bar 3)_{No4}$
canonical form similar to the previous case, only we must change the sign
of $\Sigma$:
\begin{eqnarray}
&&V^b_{4\ R(0)}\sim\biggl\{ e^{-i/2\ H_2} e^{i/2\ H_{10}}
\biggl( \Sigma_{5}^{+} [(1-i)e^{iH_{8}}
+ (1+ i)e^{-iH_{8}}]\biggr) \Sigma^{-}_{7} \nonumber\\
&&+ e^{i/2\ H_2} e^{-i/2\ H_{10}} \Sigma^{-}_{5}
\biggl( \Sigma_{7}^{+} [ \underline{\underline{ (1- i)e^{iH_{4}} }}
+ \underline{ (1+i)e^{-iH_{4}} } ]\biggr)  \biggr\}
\end{eqnarray}
Now we should calculate the correlator of the product of all 
five vertices written.
The more nontrivial factor in calculations is:
\begin{eqnarray}
&&(1+i)^2 \langle e^{-i/2\ H_4}(1) e^{i/2\ H_4}(2)
e^{i H_4}(3) e^{-i H_4}(4) \rangle
+(1-i)^2 \langle e^{-i/2\ H_4}(1) e^{i/2\ H_4}(2)
e^{-i H_4}(3) e^{i H_4}(4) \rangle \nonumber \\
&&=2i\ z_{12}^{-1/4} z_{34}^{-1}\
\frac{z_{14}z_{23} -z_{13}z_{24}}{(z_{13}z_{24}z_{14}z_{23})^{1/2}}
\end{eqnarray}
The total left-moving part of vertices gives
\begin{equation}
2i\ z_{12}^{-1} z_{13}^{-3/4} z_{14}^{-3/4} z_{15}^{-1/2} z_{23}^{-3/4}
z_{24}^{-3/4} z_{25}^{-1/2} z_{34}^{-1} z_{35}^{-1/2} z_{45}^{-1/2}
\cdot (z_{14}z_{23} -z_{13}z_{24})
\end{equation}
And after taking into account the right-moving part we obtain
\begin{equation}
2i\ |z_{13}|^{-3/2} |z_{14}|^{-3/2} |z_{15}|^{-1} |z_{23}|^{-3/2}
|z_{24}|^{-3/2} |z_{25}|^{-1} |z_{35}|^{-1} |z_{45}|^{-1}
\cdot\biggl(\frac{z_{14}z_{23} - z_{13}z_{24}}{z_{12}z_{34}} \biggr)
\end{equation}
For simplicity of calculations we want to fix $z_5=\infty$.
So, if we invert the order of vertices ($1\leftrightarrow 5$,
$2\leftrightarrow 4$) and fix the T-ordering gauge:
$z_1=\infty$, $z_4=1$, $z_5=0$, then the term
in brackets equal to $-1$,
and after taking into account the Jacobian (\ref{det})
we end up with integral
$$
I_5=\int\int d^2 z_2 d^2 z_3 |z_2|^{-3/2} |z_3|^{-3/2}
|z_2 -1|^{-3/2} |z_3 -1|^{-3/2} =I_4^2 =(27.50)^2
$$

\bigskip

Now let us consider the $\underline{R^5}$ terms.
Obviously, there is only one admissible $U(1)^3_R$ charges combination
of canonical bosonic vertex operators:

\medskip

\begin{tabular}{|c||ccccc|c|}\hline
&\multicolumn{5}{c|}{$R^5$} & Tot. \\ \hline\hline
$\alpha$ & 1/2 & 1/2 & 1/2 & 1/2 & 0 & 2 \\ \hline
$\beta $ & 1/2 & 1/2 & 1/2 & 0  & 1/2 & 2 \\ \hline
$\gamma$ & 0 & 0 & 0 & 1/2 & 1/2 & 1  \\ \hline
\end{tabular} 

\medskip

(otherwise we will have half integer total charges).
In particular this means that the terms like $R^5$ must include
simultaneously $R$-vertices of all three possible types
(three permutations (1/2, 1/2, 0)).
For Model~1 this means, that 
the superfield ${\hat\sigma}_2$ must participate
in $R^5$ interaction (from sector No 6), since this is the only field
with Ramond world-sheet fermions No 2 and 6 ($\chi$-fermions).
We did not find such gauge invariance terms.

\bigskip

Finally, let us consider terms of the form $\underline{ R^2\times (NS)^3 }$.
Firstly, it is obviously that there are four possible charge combinations
in correspondence with rough charges calculation
(we do not take into account the cases with zero total charges):

\medskip

\begin{tabular}{|c||ccccc|c||ccccc|c|}\hline
&\multicolumn{5}{c|}{$R^2\times (NS)^3$ Case 1.}& Tot. &
\multicolumn{5}{c|}{$R^2\times (NS)^3$ Case 2.}& Tot. \\ \hline\hline
$\alpha$ & 1/2 & 1/2 & 1 & 0 & 0 & 2 & 1/2 & 1/2 & 1 & 0 & 0 & 2
\\ \hline
$\beta $ & 1/2 & 1/2 & 0 & 1  & 0 & 2 & 1/2 & 1/2 & 0 & 0  & 0 & 1 \\ \hline
$\gamma$ & 0 & 0 & 0 & 0 & 1 & 1 & 0 & 0 & 0 & 1 & 1 & 2 \\ \hline
\end{tabular}

\begin{tabular}{|c||ccccc|c||ccccc|c|}\hline
&\multicolumn{5}{c|}{$R^2\times (NS)^3$ Case 3.}& Tot. &
\multicolumn{5}{c|}{$R^2\times (NS)^3$ Case 4.}& Tot. \\ \hline\hline
$\alpha$ & 1/2 & 1/2 & 1 & 1 & 0 & 3 & 1/2 & 1/2 & 0 & 0 & 0 & 1
\\ \hline
$\beta $ & 1/2 & 1/2 & 0 & 0  & 0 & 1 & 1/2 & 1/2 & 0 & 0  & 0 & 1 \\ \hline
$\gamma$ & 0 & 0 & 0 & 0 & 1 & 1 & 0 & 0 & 1 & 1 & 1 & 3 \\ \hline
\end{tabular}

\medskip

We will change pictures for fourth and fifth (NS) vertex operators.
Then for the Cases 1,2 there will be unconserved 
$U(1)_{\alpha}$ charges and for Case 3 -- $U(1)_{\gamma}$. 
So, these combinations are forbidden.

For example, let us consider the following term from Model~1:
$$\Psi^1_{(1,3;1,1)} \Psi^2_{(\bar 5 ,1;1,1)} \Phi^3_{(1,\bar 3;1,\bar 3 )}
\Phi^4_{(5,1; 5,3 )}  \Phi^5_{(1,1;\bar 5,3 )}$$
which corresponds to the Case 3.
We write the vertices of fields
$\Psi^1_{(1,3;1,1)} , \Psi^2_{(\bar 5 ,1;1,1)} , \Phi^3_{(1,\bar 3;1,\bar 3 )}$
analogous with the example (\ref{vertex1}-\ref{vertex3}).
Then we should write the 
boson vertices for "NS"-fields $\Phi^4_{(5,1; 5,3 )} ,
\Phi^5_{(1,1;\bar 5,3 )} $
in the $0-$picture according to the \cite{20'}.
These fields have the excitation by world-sheet left fermion No 6 and
No 2 correspondingly.
So, we have:
\begin{equation}
V^f_{1(-1/2)}=e^{-c/2}S_{\alpha}e^{-i/2\ H_2}\Sigma^{+}_3
\Sigma^{+}_4 e^{i/2\ K_1X} \bar G_1 e^{i/2\ K_1\bar X}
\end{equation}
\begin{equation}
V^f_{2(-1/2)}=e^{-c/2}S_{\alpha}e^{-i/2\ H_2}\Sigma^{-}_3
\Sigma^{-}_4 e^{i/2\ K_2X} \bar G_2 e^{i/2\ K_2\bar X}
\end{equation}
\begin{equation}
V^b_{3(-1)}=e^{-c}e^{i H_6} e^{i/2\ K_3X} \bar G_3 e^{i/2\ K_3\bar X}
\end{equation}
\begin{equation}
\label{V4}
V^b_{4(0)}\sim \frac{i}{2\sqrt{2}}\biggl[ (1-i)e^{iH_3}e^{iH_9}
+(1+i)e^{iH_3}e^{-iH_9} +(1+i)e^{-iH_3}e^{iH_9}
+(1-i)e^{-iH_3}e^{-iH_9}\biggr]
\end{equation}
\begin{equation}
\label{V5}
V^b_{5(0)}\sim \frac{i}{2\sqrt{2}}\biggl[ (1-i)e^{iH_5}e^{iH_8}
+(1+i)e^{iH_5}e^{-iH_8} +(1+i)e^{-iH_5}e^{iH_8}
+(1-i)e^{-iH_5}e^{-iH_8}\biggr]
\end{equation}
The full product of left-moving part will be proportional to
$$
\sim \biggl\langle e^{-i/2\ H_2}(1) e^{-i/2\ H_2}(2) e^{i H_6}(3)
\biggl\rangle =0
$$
The situation is analogous for all $R^2 (NS)^3$ of this type.

Finally, the situation in the Case~4 is as follows.
The full product of the left-moving part will be proportional to
\begin{equation}
\label{corr}
\sim[(1-i)^2+(1+i)^2]\ \biggl\langle e^{i/2\ H}(x) e^{-i/2\ H}(y)
\biggl\rangle
\end{equation}
However,
$(1-i)^2+(1+i)^2=0$.
Really, the full result will be proportional to the correlator of two 
identical square
brackets from fourth and fifth vertices like (\ref{V4} or \ref{V5})
and this is equal to zero.

So, we have shown that the terms like $R^2\times (NS)^3$ are forbidden
in the case of complex left-moving fermions.

In conclusion, let us write down some of the non-renormalizable six point
$R^6$ terms for superpotential of our Model~1 (the full possible $R^6$
terms following only from gauge invariance are given in Table \ref{table3f}):
\begin{eqnarray}
&&W_6 \sim {\hat\Psi}^{Mat.}_{(10,3 ;1,1)}
{\hat\Psi}^{Mat.}_{(10,3 ;1,1)}
{\hat\phi}_{2\ (+_1 ,+_3)(5,1 ;1,1)}
{\hat\phi}_{3\ (-_1 ,-_3)(1,1 ;1,\bar 3 )}\times \nonumber\\
&&\times [
{\hat\Phi}^H_{3\ (+_1 ,+_3)(1,\bar 3 ;1,\bar 3 )}
{\hat\Phi}^H_{3\ (-_1 ,-_3)(1,\bar 3 ;1,\bar 3 )}
+{\hat\Psi}^H_{2\ (-_1 ,+_2)(1,\bar 3 ;1,1)}
{\hat\Psi}^H_{3\ (+_1 ,-_2)(1,\bar 3 ;1, 3 )} + \nonumber\\
&&\ \ + {\hat\Psi}^H_{2\ (-_1 ,-_2)(1,\bar 3 ;1,1)}
{\hat\Psi}^H_{3\ (+_1 ,+_2)(1,\bar 3 ;1, 3 )}] \ .
\end{eqnarray}

\subsection {The Fermion Mass Matrices for Model 1.}
The studying of the mass matrices of the matter fields restricted only
to the renormalizable superpotential proves that 2 families
remain massless on that scale while the others 2 families
 have degenerated masses. However taking into account the last term from
$W_5$ with large VEV of the 3-fields condensate
leads to disappearing of the degeneration.
In particular
the masses of $b$ and $b'$ quarks are different.

We give the fermion mass matrices 
with taking into account both contributions from five-point and from
six-point terms ($W_5$ and $W_6$):

\begin{equation}
\langle\Phi_{(5;\bar 3);4}^{1,2,3}\rangle =H^{1,2,3}\quad ,\qquad
\langle\Phi_{(\bar 5 ; 3);4}^{1,2,3}\rangle ={\bar H}^{1,2,3}
\end{equation}
Here and further the up-indices $1,2,3$ are horizontal generations
indices.

Mass matrices for ($e$, $d$) and $u$ (in the flipped case) are correspondingly
\begin{displaymath}
\left( \begin{array}{cccc}
h^{11}_1 & h_1^{12} & h_1^{13} & H^1+\bar H^1_5 \\
h^{21}_1 & h_1^{22} & h_1^{23} & H^2+\bar H^2_5 \\
h^{31}_1 & h_1^{32} & h_1^{33} & H^3+\bar H^3_5 \\
H^1+\bar H^1_5 & H^2+\bar H^2_5 & H^3+\bar H^3_5 & h_3 
\end{array}\right) \quad ,
\end{displaymath}
\begin{displaymath}
\left( \begin{array}{cccc}
H_5^{11} & {\bar H}^3 +H_5^{12} &-{\bar H}^2 +H_5^{13} & h_2^1 \\
 -{\bar H}^3 +H_5^{21} & H_5^{22} & {\bar H}^1 +H_5^{23} & h_2^2 \\
 {\bar H}^2+H_5^{31} & -{\bar H}^1 +H_5^{32} & H_5^{33} & h_2^3 \\
h_2^1 & h_2^2 & h_2^3 & H_5
\end{array}\right)
\end{displaymath}
Where $H$, $H_5$ and $h_{1,2,3}$ with up generations indices correspond
to contributions from renormalizable and non-renormalizable five-, six-point
terms of superpotential to the fermion mass matrices:
$$
{\bar H}_5^i=A^i \langle {\hat\Phi }_{5\ (5,1;1,3)}\rangle_0 \ ,\ 
H_5^{ij}=A^{ij}\langle {\hat\Phi }_{5\ ({\bar 5},1;1,{\bar 3})} \rangle_0 \ ,\
H_5 =A\langle {\hat\Phi }_{5\ ({\bar 5},1;1,{\bar 3})} \rangle_0 \ ,
$$
$$
h_1^{ij}=A^{ij} \langle \phi_{2\ (5,1,1,1)}\rangle_0 
\langle\phi_{3\ (1,1,1,\bar 3)} \rangle_0 \ ,\ 
h_2^i =A^i \langle \phi_{3(1,1,1,3)}\rangle_0 
\langle\phi_{2(\bar 5,1,1,1)}\rangle_0 \ ,
$$
$$
h_3 =A \langle \phi_{2\ (5,1,1,1)}\rangle_0 
\langle\phi_{3\ (1,1,1,\bar 3)}\rangle_0 \ .
$$
where
$$
A^i =\langle {\hat\Psi }^H_{1\ (1,1;1,3)}\rangle_0 
\langle {{\hat\Psi }^{H\ \ i}_{3\ (1,\bar 3;1,3)}}\rangle_0
+\langle {\hat\Phi }^H_{1\ (1,1;1,\bar 3)}\rangle_0 
\langle {{\hat\Phi }^{H\ \ i}_{2\ (1,\bar 3;1,1)}}\rangle_0
+\langle {{\hat\Phi }^{H\ \ i}_{3\ (1,\bar 3;1,\bar 3)}}\rangle_0
\langle {\hat\Phi }^H_{4\ (1,1;1,1)}\rangle_0 \ ,
$$
$$
A^{ij}=\langle {{\hat\Psi }^{H\ \ i}_{2\ (1,\bar 3;1,1)}} \rangle_0
\langle {{\hat\Psi }^{H\ \ j}_{3\ (1,\bar 3;1,3)}}\rangle_0 +
\langle {{\hat\Phi }^{H\ \ i}_{3\ (1,\bar 3 ;1,\bar 3 )}}\rangle_0
\langle {{\hat\Phi }^{H\ \ j}_{3\ (1,\bar 3 ;1,\bar 3 )}}\rangle_0 \ ,
$$
$$
A=\langle {\hat\Psi }^H_{1\ (1,1;1,3)}\rangle_0 
\langle {\hat\Psi }^H_{4\ (1,1;1,1)}\rangle_0 +
\langle {\hat\Phi }^H_{1\ (1,1;1,\bar 3)} \rangle_0
\langle {\hat\Phi }^H_{1\ (1,1;1,\bar 3)}\rangle_0 \ .
$$
Here we do not indicate the quantum numbers of hidden gauge groups.
These quantum number can be easy reconstructed from the table of states
of the Model~1. We use a short description for products of superfields.
In particular the products like ${\Psi}^H_k {\Psi}^H_l $ or
${\Phi}^H_k {\Phi}^H_l $ with $k\not= l$ means two terms with opposite
"hidden" quantum numbers.
For non-flipped case we only have to replace 
$u\leftrightarrow d,e$.

We can try to use some of the Horizontal fields $\Psi^H_i$, $\Phi^H_i$
and ${\phi}_i$
(see spectrum of model 1) for partial breaking the horizontal symmetry 
$U(3)^I_{H} \times U(3)^{II}_{H}$ on
the $M_{\rm GUT}$ scale and conserving the space-time N=1 supersymmetry.
In this case the contribution of the non-renormalizable 5- and 6-point
terms could be important for the construction of realistic fermion mass
matrices. It is very important to note that in this case the form of mass
matrix ansatz will depend on the way of breaking primordial horizontal
symmetry $U(3)^I_H \times U(3)^{II}_H$ on the $M_{\rm GUT}$-scale.
Below, we give one example of the breaking of the Horizontal
symmetry $SU(3)_{H}^I \times U(1)_{H}^I \times SU(3)_{H}^{II} \times
U(1)^I \to SU(2)_{H}^I \times SU(2)_{H}^{II} \times U(1)_{H}'$
on the high energy scale, $M_{U}\sim M_{\rm GUT}$, conserving the
$N=1$ space-time supersymmetry. Also the hidden group $SU(4)\times U(1)^3$
is broken on this mass scale. In our example we considered the possibility
to give big VEVs $\sim M_{\rm GUT}$ only to the following scalar fields 
in model 1 : $\Psi^H_{1,2,4}$, $\Phi^H_{1,2,4}$, $\phi_1$, $\phi_3$ and 
all sigma. Really we can consider
also some another  examples, with conserving the part of the Horizontal
symmetry to the low energy ($M_{U} \sim 1$--10\,TeV), like 

\begin{itemize}
\item $SU(3)^{I+II}_H \times U(1)^{I+II}_H$
\item $SU(3)^{II}_H$ 
\item $U(1)_{H}^n$, where $n=1,2,3$ etc.
\end{itemize}

For conserving Supersymmetry we must fulfill the all equations,
$D_p^a=F_j=0$, including all breaking groups and all 
under consideration for  our case the VEVs of the scalar fields:
($\Phi^H_{1,2,4}$, $\Psi^H_{1,2,4}$, $\phi_1$, $\phi_3$, $\sigma_{1,2,3}$)

\begin{itemize}
\item[\bf 1.]
$D_{SU(3)_I}^8$
\[
({\phi}_1^2 - \bar{\phi}_1^2)=
({\Psi^H_{2}}{}^2 + {\bar{\Psi}^H_2}{}^2)+
({\Phi^H_{2}}{}^2 + {\bar{\Phi}^H_2}{}^2)
\]
\item[\bf 2.]
$D_{SU(3)_{II}}^8$
\[
(\bar{\phi}_3^2 - {\phi}_3^2)=
({\Psi^H_2}{}^2 +{\bar{\Psi}^H_2}{}^2)+
({\Phi^H_2}{}^2 +{\bar{\Phi}^H_2}{}^2)+
2({\Phi^H_1}{}^2 +{\Phi^H_2}{}^2+{\Phi^H_4}{}^2)
\]
\item[\bf 3.]
\[
N(\bar{\sigma}_3^2 -{\sigma}_3^2)=
({\Phi^H_1}{}^2 + {\bar{\Phi}^H_1}{}^2)+
({\Phi^H_2}{}^2 + {\bar{\Phi}^H_2}{}^2)+
({\Phi^H_4}{}^2 + {\bar{\Phi}^H_4}{}^2)
\]
\item[\bf 4.]
\[
\begin{array}{rcl}
N({\sigma}_2^2 -\bar{\sigma}_2^2)&=&
({\Phi^H_1}{}^2 + {\bar{\Phi}^H_1}{}^2)+
({\Phi^H_2}{}^2 + {\bar{\Phi}^H_2}{}^2)+
({\Phi^H_4}{}^2 + {\bar{\Phi}^H_4}{}^2)\\[10pt]
&+&2({\Phi^H_1}{}^2 +{\Phi^H_2}{}^2+{\Phi^H_4}{}^2)
\end{array}
\]
\item[\bf 5.]
\[
\begin{array}{rcl}
({\Psi^H_1}{}^2 +{\bar{\Psi}^H_1}{}^2)&=&
({\Psi^H_2}{}^2 + {\bar{\Psi}{}^H_2}^2)+
({\Phi^H_1}{}^2 + {\bar{\Phi}{}^H_1}^2)+
({\Phi^H_2}{}^2 + {\bar{\Phi}{}^H_2}^2)\\[10pt]
&+&2({\Phi^H_1}{}^2 +{\Phi^H_2}{}^2+{\Phi^H_4}{}^2)
\end{array}
\]
\item[\bf 6.]
\[
({\Psi^H_4}{}^2 +{\bar{\Psi}^H_4}{}^2) =
({\Phi^H_2}{}^2 +{\bar{\Phi}^H_2}{}^2) -
({\Phi^H_1}{}^2 +{\bar{\Phi}^H_1}{}^2) > 0
\]
\item[\bf 7.-8.]
\[
\begin{array}{rcl}
N{\bar{\sigma}}_1^2 =N{\sigma}_1^2&=&
1/2 N(\bar{\sigma}_3^2 -{\sigma}_3^2)-
1/2N({\sigma}_2^2 -\bar{\sigma}_2^2) +N{\sigma}_3^2\\[10pt]
&=&N{\sigma}_3^2-
({\Phi^H_1}{}^2 +{\Phi^H_2}{}^2+{\Phi^H_4}{}^2) >0
\end{array}
\]
\item[\bf 9.]
\[
D_2=
({\Psi^H_1}{}^2 - {\bar{\Psi}^H_1}{}^2) +
({\Psi^H_2}{}^2 - {\bar{\Psi}^H_2}{}^2)+
({\Psi^H_4}{}^2 - {\bar{\Psi}^H_4}{}^2)=0
\]
\item[\bf 10.]
\[
D_1=
-4({\Psi^H_2}{}^2 + {\bar{\Psi}{}^H_2}^2) -
4({\Phi^H_1}{}^2 + {\Phi^H_2}{}^2)
-2\bar{\phi}_1^2 - 2{\phi}_3^2 + 2N\bar{\sigma}_2^2+D_{\rm term}
\]
\end{itemize}

As we can see from the states list for the model, the hidden group
$U(1)_1$ in this model appears to be anomalous: Tr $U(1)_1=12$.
This means that at one-loop string level there exists Fayet-Iliopoulos
D-term determined by VEV of the dilaton and it is proportional to
Tr $U(1)_1=12$.
Potentially this term could break SUSY at the high scale
and destabilize the vacua \cite{Dine'}. This could be avoided
if the potential has
D-flat direction on $U(1)_1$-charged fields which have VEVs that break
anomalous group (and may be some other groups), compensate D-term, and
restore SUSY. Those fields have to have appropriate charges on the
remaining groups in order to  cause no  SUSY breaking via their D-terms.
Note that
since the  superfields mentioned above besides the $\hat{\sigma}_2$-field
do not participate in construction
of the renormalizable superpotential $W_1+W_2+W_3$
in this scheme we have no problem with the F-flat directions.
Finally, note that in non-flipped $SU(5)\times U(1)$ model
we can give the VEVs to the fields ${\hat\sigma}_1$
for breaking $U(1)^{\rm sym}_5$. In this case for the choice of the $D$-flat
direction we also need to use VEVs of the fields ${\hat\sigma}_3$.
From the form of the ${\hat\sigma}$-depended contribution to the
superpotential $W_2$  (\ref{eq34}) it follows that the field
of the fourth generation ${\hat\Psi}_{(1,1;1,1)}$ (the fourth neutrino
in non-flipped scheme)  obtains a heavy mass.
The more careful consideration of the neutrino mass matrix we will do
in our further papers.

\begin{table}
\caption{The possible 4-point ${\rm R}^4$ forms to superpotential
from the full gauge invariance.}
\label{table1f}
\begin{center}
{\footnotesize
\begin{tabular}{|c|cccc|}\hline
- & $\Psi_1(1,3,1,1)$ & $\Psi_2(\bar 5,3,1,1)$ &
    $\Psi_2(\bar 5,3,1,1)$ & $\Psi_3(10,1,1,1)$ \\
- & $\Psi_1(1,3,1,1)$ & $\Psi_2(\bar 5,3,1,1)$ &
    $\Psi_5(\bar 5,1,1,1)$ & $\Psi_6(10,3,1,1)$ \\
- & $\Psi_1(1,3,1,1)$ & $\Psi_5(\bar 5,1,1,1)$ &
    $\phi_2(5,1,1,1)_{+1,+3}$ & $\phi_1(1,\bar 3,1,1)_{-1,-3}$ \\
+ & $\Psi_1(1,3,1,1)$ & $\Phi^H_2(1,\bar 3,1,1)_{+1,-3}$ &
    $\sigma_1(1,1,1,1)_{-1,-4}$ & $\sigma_3(1,1,1,1)_{+3,+4}$ \\
- & $\Psi_2(\bar 5,3,1,1)$ & $\Psi_2(\bar 5,3,1,1)$ &
    $\Psi_4(1,1,1,1)$ & $\Psi_6(10,3,1,1)$ \\
- & $\Psi_2(\bar 5,3,1,1)$ & $\Psi_3(10,1,1,1)$ &
    $\Psi_6(10,3,1,1)$ & $\Psi_6(10,3,1,1)$ \\
- & $\Psi_2(\bar 5,3,1,1)$ & $\Psi_4(1,1,1,1)$ &
    $\phi_2(5,1,1,1)_{+1,+3}$ & $\phi_1(1,\bar 3,1,1)_{-1,-3}$ \\
- & $\Psi_2(\bar 5,3,1,1)$ & $\Psi_6(10,3,1,1)$ &
    $\phi_1(1,3,1,1)_{-1,+3}$ & $\phi_2(\bar 5,1,1,1)_{+1,-3}$ \\
- & $\Psi_3(10,1,1,1)$ & $\Psi_6(10,3,1,1)$ &
    $\phi_2(5,1,1,1)_{+1,+3}$ & $\phi_1(1,\bar 3,1,1)_{-1,-3}$ \\
- & $\Psi_5(\bar 5,1,1,1)$ & $\Psi_6(10,3,1,1)$ &
    $\Psi_6(10,3,1,1)$ & $\Psi_6(10,3,1,1)$ \\
- & $\Psi^H_1(1,1,1,3)_{-1,+2}$ & $\Psi^H_2(1,\bar 3,1,1)_{-1,+2}$ &
    $\Psi^H_3(1,\bar 3,1,3)_{+1,-2}$ & $\Psi^H_3(1,\bar 3,1,3)_{+1,-2}$ \\
- & $\Psi^H_1(1,1,1,3)_{-1,+2}$ & $\Psi^H_3(1,\bar 3,1,3)_{+1,+2}$ &
    $\Psi^H_2(1,\bar 3,1,1)_{-1,-2}$ & $\Psi^H_3(1,\bar 3,1,3)_{+1,-2}$ \\
- & $\Psi^H_1(1,1,1,3)_{-1,+2}$ & $\Psi^H_2(1,\bar 3,1,1)_{-1,-2}$ &
    $\Phi^H_3(1,\bar 3,1,\bar 3)_{+1,+3}$ & $\Phi^H_2(1,\bar 3,1,1)_{+1,-3}$ \\
- & $\Psi^H_1(1,1,1,3)_{-1,+2}$ & $\Psi^H_3(1,\bar 3,1,3)_{+1,-2}$ &
    $\Phi^H_3(1,\bar 3,1,\bar 3)_{+1,+3}$ &
    $\Phi^H_3(1,\bar 3,1,\bar 3)_{-1,-3}$ \\
- & $\Psi^H_2(1,\bar 3,1,1)_{-1,+2}$ & $\Psi^H_3(1,\bar 3,1,3)_{+1,+2}$ &
    $\Psi^H_1(1,1,1,3)_{-1,-2}$ & $\Psi^H_3(1,\bar 3,1,3)_{+1,-2}$ \\
- & $\Psi^H_2(1,\bar 3,1,1)_{-1,+2}$ & $\Psi^H_1(1,1,1,3)_{-1,-2}$ &
    $\Phi^H_3(1,\bar 3,1,\bar 3)_{+1,+3}$ & $\Phi^H_2(1,\bar 3,1,1)_{+1,-3}$ \\
- & $\Psi^H_2(1,\bar 3,1,1)_{-1,+2}$ & $\Psi^H_2(1,\bar 3,1,1)_{-1,-2}$ &
    $\Phi^H_4(1,1,1,1)_{+1,+3}$ & $\Phi^H_2(1,\bar 3,1,1)_{+1,-3}$ \\
- & $\Psi^H_2(1,\bar 3,1,1)_{-1,+2}$ & $\Psi^H_3(1,\bar 3,1,3)_{+1,-2}$ &
    $\Phi^H_1(1,1,1,\bar 3)_{-1,+3}$ & $\Phi^H_2(1,\bar 3,1,1)_{+1,-3}$ \\
- & $\Psi^H_2(1,\bar 3,1,1)_{-1,+2}$ & $\Psi^H_3(1,\bar 3,1,3)_{+1,-2}$ &
    $\Phi^H_2(1,\bar 3,1,1)_{-1,+3}$ & $\Phi^H_1(1,1,1,\bar 3)_{+1,-3}$ \\
- & $\Psi^H_2(1,\bar 3,1,1)_{-1,+2}$ & $\Psi^H_3(1,\bar 3,1,3)_{+1,-2}$ &
    $\Phi^H_3(1,\bar 3,1,\bar 3)_{+1,+3}$ & $\Phi^H_4(1,1,1,1)_{-1,-3}$ \\
- & $\Psi^H_2(1,\bar 3,1,1)_{-1,+2}$ & $\Psi^H_3(1,\bar 3,1,3)_{+1,-2}$ &
    $\Phi^H_4(1,1,1,1)_{+1,+3}$ & $\Phi^H_3(1,\bar 3,1,\bar 3)_{-1,-3}$ \\
- & $\Psi^H_2(1,\bar 3,1,1)_{-1,+2}$ & $\Psi^H_4(1,1,1,1)_{+1,-2}$ &
    $\Phi^H_2(1,\bar 3,1,1)_{-1,+3}$ & $\Phi^H_2(1,\bar 3,1,1)_{+1,-3}$ \\
- & $\Psi^H_3(1,\bar 3,1,3)_{+1,+2}$ & $\Psi^H_3(1,\bar 3,1,3)_{+1,+2}$ &
    $\Psi^H_1(1,1,1,3)_{-1,-2}$ & $\Psi^H_2(1,\bar 3,1,1)_{-1,-2}$ \\
- & $\Psi^H_3(1,\bar 3,1,3)_{+1,+2}$ & $\Psi^H_1(1,1,1,3)_{-1,-2}$ &
    $\Phi^H_3(1,\bar 3,1,\bar 3)_{+1,+3}$ &
    $\Phi^H_3(1,\bar 3,1,\bar 3)_{-1,-3}$ \\
- & $\Psi^H_3(1,\bar 3,1,3)_{+1,+2}$ & $\Psi^H_2(1,\bar 3,1,1)_{-1,-2}$ &
    $\Phi^H_1(1,1,1,\bar 3)_{-1,+3}$ & $\Phi^H_2(1,\bar 3,1,1)_{+1,-3}$ \\
- & $\Psi^H_3(1,\bar 3,1,3)_{+1,+2}$ & $\Psi^H_2(1,\bar 3,1,1)_{-1,-2}$ &
    $\Phi^H_2(1,\bar 3,1,1)_{-1,+3}$ & $\Phi^H_1(1,1,1,\bar 3)_{+1,-3}$ \\
- & $\Psi^H_3(1,\bar 3,1,3)_{+1,+2}$ & $\Psi^H_2(1,\bar 3,1,1)_{-1,-2}$ &
    $\Phi^H_3(1,\bar 3,1,\bar 3)_{+1,+3}$ & $\Phi^H_4(1,1,1,1)_{-1,-3}$ \\
- & $\Psi^H_3(1,\bar 3,1,3)_{+1,+2}$ & $\Psi^H_2(1,\bar 3,1,1)_{-1,-2}$ &
    $\Phi^H_4(1,1,1,1)_{+1,+3}$ & $\Phi^H_3(1,\bar 3,1,\bar 3)_{-1,-3}$ \\
- & $\Psi^H_3(1,\bar 3,1,3)_{+1,+2}$ & $\Psi^H_3(1,\bar 3,1,3)_{+1,-2}$ &
    $\Phi^H_1(1,1,1,\bar 3)_{-1,+3}$ & $\Phi^H_3(1,\bar 3,1,\bar 3)_{-1,-3}$ \\
+ & $\Psi^H_3(1,\bar 3,1,3)_{+1,+2}$ & $\Psi^H_3(1,\bar 3,1,3)_{+1,-2}$ &
    $\phi_3(1,1,1,3)_{-1,+3}$ & $\phi_1(1,\bar 3,1,1)_{-1,-3}$ \\
- & $\Psi^H_3(1,\bar 3,1,3)_{+1,+2}$ & $\Psi^H_4(1,1,1,1)_{+1,-2}$ &
    $\Phi^H_2(1,\bar 3,1,1)_{-1,+3}$ & $\Phi^H_3(1,\bar 3,1,\bar 3)_{-1,-3}$ \\
+ & $\Psi^H_3(1,\bar 3,1,3)_{+1,+2}$ & $\Psi^H_4(1,1,1,1)_{+1,-2}$ &
    $\phi_1(1,3,1,1)_{-1,+3}$ & $\phi_3(1,1,1,\bar 3)_{-1,-3}$ \\
- & $\Psi^H_4(1,1,1,1)_{+1,+2}$ & $\Psi^H_2(1,\bar 3,1,1)_{-1,-2}$ &
    $\Phi^H_2(1,\bar 3,1,1)_{-1,+3}$ & $\Phi^H_2(1,\bar 3,1,1)_{+1,-3}$ \\
- & $\Psi^H_4(1,1,1,1)_{+1,+2}$ & $\Psi^H_3(1,\bar 3,1,3)_{+1,-2}$ &
    $\Phi^H_2(1,\bar 3,1,1)_{-1,+3}$ & $\Phi^H_3(1,\bar 3,1,\bar 3)_{-1,-3}$ \\
+ & $\Psi^H_4(1,1,1,1)_{+1,+2}$ & $\Psi^H_3(1,\bar 3,1,3)_{+1,-2}$ &
    $\phi_1(1,3,1,1)_{-1,+3}$ & $\phi_3(1,1,1,\bar 3)_{-1,-3}$ \\
- & $\Phi^H_1(1,1,1,\bar 3)_{-1,+3}$ & $\Phi^H_3(1,\bar 3,1,\bar 3)_{+1,+3}$ &
    $\Phi^H_2(1,\bar 3,1,1)_{+1,-3}$ & $\Phi^H_3(1,\bar 3,1,\bar 3)_{-1,-3}$ \\
- & $\Phi^H_2(1,\bar 3,1,1)_{-1,+3}$ & $\Phi^H_3(1,\bar 3,1,\bar 3)_{+1,+3}$ &
    $\Phi^H_1(1,1,1,\bar 3)_{+1,-3}$ & $\Phi^H_3(1,\bar 3,1,\bar 3)_{-1,-3}$ \\
- & $\Phi^H_3(1,\bar 3,1,\bar 3)_{+1,+3}$ &
    $\Phi^H_3(1,\bar 3,1,\bar 3)_{+1,+3}$ &
    $\Phi^H_3(1,\bar 3,1,\bar 3)_{-1,-3}$ & $\Phi^H_4(1,1,1,1)_{-1,-3}$ \\
- & $\Phi^H_3(1,\bar 3,1,\bar 3)_{+1,+3}$ & $\Phi^H_4(1,1,1,1)_{+1,+3}$ &
    $\Phi^H_3(1,\bar 3,1,\bar 3)_{-1,-3}$ &
    $\Phi^H_3(1,\bar 3,1,\bar 3)_{-1,-3}$ \\
+ & $\Phi^H_3(1,\bar 3,1,\bar 3)_{+1,+3}$ & $\Phi^H_1(1,1,1,\bar 3)_{+1,-3}$ &
    $\phi_1(1,3,1,1)_{-1,+3}$ & $\phi_3(1,1,1,\bar 3)_{-1,-3}$ \\
+ & $\Phi^H_3(1,\bar 3,1,\bar 3)_{+1,+3}$ & $\Phi^H_2(1,\bar 3,1,1)_{+1,-3}$ &
    $\phi_3(1,1,1,3)_{-1,+3}$ & $\phi_1(1,\bar 3,1,1)_{-1,-3}$ \\
- & $\phi_1(1,3,1,1)_{-1,+3}$ & $\phi_2(5,1,1,1)_{+1,+3}$ &
    $\phi_1(1,\bar 3,1,1)_{-1,-3}$ & $\phi_2(\bar 5,1,1,1)_{+1,-3}$ \\
- & $\phi_1(1,3,1,1)_{-1,+3}$ & $\phi_4(1,1,5,1)_{+1,+3}$ &
    $\phi_1(1,\bar 3,1,1)_{-1,-3}$ & $\phi_4(1,1,\bar 5,1)_{+1,-3}$ \\
+ & $\phi_1(1,3,1,1)_{-1,+3}$ & $\phi_1(1,\bar 3,1,1)_{-1,-3}$ &
    $\sigma_2(1,1,1,1)_{+1,-4}$ & $\sigma_2(1,1,1,1)_{+1,+4}$ \\
- & $\phi_2(5,1,1,1)_{+1,+3}$ & $\phi_3(1,1,1,3)_{-1,+3}$ &
    $\phi_2(\bar 5,1,1,1)_{+1,-3}$ & $\phi_3(1,1,1,\bar 3)_{-1,-3}$ \\
- & $\phi_3(1,1,1,3)_{-1,+3}$ & $\phi_4(1,1,5,1)_{+1,+3}$ &
    $\phi_3(1,1,1,\bar 3)_{-1,-3}$ & $\phi_4(1,1,\bar 5,1)_{+1,-3}$ \\
+ & $\phi_3(1,1,1,3)_{-1,+3}$ & $\phi_3(1,1,1,\bar 3)_{-1,-3}$ &
    $\sigma_2(1,1,1,1)_{+1,-4}$ & $\sigma_2(1,1,1,1)_{+1,+4}$ \\
- & $\sigma_1(1,1,1,1)_{+1,-4}$ & $\sigma_1(1,1,1,1)_{+1,-4}$ &
    $\sigma_1(1,1,1,1)_{-1,-4}$ & $\sigma_1(1,1,1,1)_{-1,-4}$ \\
+ & $\sigma_1(1,1,1,1)_{+1,-4}$ & $\sigma_3(1,1,1,1)_{-3,+4}$ &
    $\sigma_1(1,1,1,1)_{-1,-4}$ & $\sigma_3(1,1,1,1)_{+3,+4}$ \\
- & $\sigma_3(1,1,1,1)_{-3,+4}$ & $\sigma_3(1,1,1,1)_{-3,+4}$ &
    $\sigma_3(1,1,1,1)_{+3,+4}$ & $\sigma_3(1,1,1,1)_{+3,+4}$ \\
\hline
\end{tabular}%
}%
\end{center}%
\end{table}

\begin{table}
\caption{The possible 5-point ${\rm R}^4{\rm NS}$
forms to superpotential
from the full gauge invariance.}
\label{table2f}
\begin{center}
{\footnotesize
\begin{tabular}{|c|ccccc|}\hline
- & $\Phi_3(5,1,5,1)$ & $\Psi_1(1,3,1,1)$ &
    $\Psi_2(\bar 5,3,1,1)$ & $\phi_1(1,3,1,1)_{-1,+3}$ &
    $\phi_4(1,1,\bar 5,1)_{+1,-3}$ \\
- & $\Phi_3(5,1,5,1)$ & $\Psi_6(10,3,1,1)$ &
    $\Psi_6(10,3,1,1)$ & $\phi_1(1,3,1,1)_{-1,+3}$ &
    $\phi_4(1,1,\bar 5,1)_{+1,-3}$ \\
- & $\Phi_4(1,3,1,3)$ & $\Psi_1(1,3,1,1)$ &
    $\Psi_2(\bar 5,3,1,1)$ & $\phi_2(5,1,1,1)_{+1,+3}$ &
    $\phi_3(1,1,1,\bar 3)_{-1,-3}$ \\
- & $\Phi_4(1,3,1,3)$ & $\Psi_6(10,3,1,1)$ &
    $\Psi_6(10,3,1,1)$ & $\phi_2(5,1,1,1)_{+1,+3}$ &
    $\phi_3(1,1,1,\bar 3)_{-1,-3}$ \\
+ & $\Phi_5(5,1,1,3)$ & $\Psi_1(1,3,1,1)$ &
    $\Psi_5(\bar 5,1,1,1)$ & $\Psi^H_1(1,1,1,3)_{-1,+2}$ &
    $\Psi^H_3(1,\bar 3,1,3)_{+1,-2}$ \\
+ & $\Phi_5(5,1,1,3)$ & $\Psi_1(1,3,1,1)$ &
    $\Psi_5(\bar 5,1,1,1)$ & $\Psi^H_3(1,\bar 3,1,3)_{+1,+2}$ &
    $\Psi^H_1(1,1,1,3)_{-1,-2}$ \\
+ & $\Phi_5(5,1,1,3)$ & $\Psi_1(1,3,1,1)$ &
    $\Psi_5(\bar 5,1,1,1)$ & $\Phi^H_1(1,1,1,\bar 3)_{-1,+3}$ &
    $\Phi^H_2(1,\bar 3,1,1)_{+1,-3}$ \\
+ & $\Phi_5(5,1,1,3)$ & $\Psi_1(1,3,1,1)$ &
    $\Psi_5(\bar 5,1,1,1)$ & $\Phi^H_2(1,\bar 3,1,1)_{-1,+3}$ &
    $\Phi^H_1(1,1,1,\bar 3)_{+1,-3}$ \\
+ & $\Phi_5(5,1,1,3)$ & $\Psi_1(1,3,1,1)$ &
    $\Psi_5(\bar 5,1,1,1)$ & $\Phi^H_3(1,\bar 3,1,\bar 3)_{+1,+3}$ &
    $\Phi^H_4(1,1,1,1)_{-1,-3}$ \\
+ & $\Phi_5(5,1,1,3)$ & $\Psi_1(1,3,1,1)$ &
    $\Psi_5(\bar 5,1,1,1)$ & $\Phi^H_4(1,1,1,1)_{+1,+3}$ &
    $\Phi^H_3(1,\bar 3,1,\bar 3)_{-1,-3}$ \\
+ & $\Phi_5(5,1,1,3)$ & $\Psi_2(\bar 5,3,1,1)$ &
    $\Psi_4(1,1,1,1)$ & $\Psi^H_1(1,1,1,3)_{-1,+2}$ &
    $\Psi^H_3(1,\bar 3,1,3)_{+1,-2}$ \\
+ & $\Phi_5(5,1,1,3)$ & $\Psi_2(\bar 5,3,1,1)$ &
    $\Psi_4(1,1,1,1)$ & $\Psi^H_3(1,\bar 3,1,3)_{+1,+2}$ &
    $\Psi^H_1(1,1,1,3)_{-1,-2}$ \\
+ & $\Phi_5(5,1,1,3)$ & $\Psi_2(\bar 5,3,1,1)$ &
    $\Psi_4(1,1,1,1)$ & $\Phi^H_1(1,1,1,\bar 3)_{-1,+3}$ &
    $\Phi^H_2(1,\bar 3,1,1)_{+1,-3}$ \\
+ & $\Phi_5(5,1,1,3)$ & $\Psi_2(\bar 5,3,1,1)$ &
    $\Psi_4(1,1,1,1)$ & $\Phi^H_2(1,\bar 3,1,1)_{-1,+3}$ &
    $\Phi^H_1(1,1,1,\bar 3)_{+1,-3}$ \\
+ & $\Phi_5(5,1,1,3)$ & $\Psi_2(\bar 5,3,1,1)$ &
    $\Psi_4(1,1,1,1)$ & $\Phi^H_3(1,\bar 3,1,\bar 3)_{+1,+3}$ &
    $\Phi^H_4(1,1,1,1)_{-1,-3}$ \\
+ & $\Phi_5(5,1,1,3)$ & $\Psi_2(\bar 5,3,1,1)$ &
    $\Psi_4(1,1,1,1)$ & $\Phi^H_4(1,1,1,1)_{+1,+3}$ &
    $\Phi^H_3(1,\bar 3,1,\bar 3)_{-1,-3}$ \\
+ & $\Phi_5(5,1,1,3)$ & $\Psi_3(10,1,1,1)$ &
    $\Psi_6(10,3,1,1)$ & $\Psi^H_1(1,1,1,3)_{-1,+2}$ &
    $\Psi^H_3(1,\bar 3,1,3)_{+1,-2}$ \\
+ & $\Phi_5(5,1,1,3)$ & $\Psi_3(10,1,1,1)$ &
    $\Psi_6(10,3,1,1)$ & $\Psi^H_3(1,\bar 3,1,3)_{+1,+2}$ &
    $\Psi^H_1(1,1,1,3)_{-1,-2}$ \\
+ & $\Phi_5(5,1,1,3)$ & $\Psi_3(10,1,1,1)$ &
    $\Psi_6(10,3,1,1)$ & $\Phi^H_1(1,1,1,\bar 3)_{-1,+3}$ &
    $\Phi^H_2(1,\bar 3,1,1)_{+1,-3}$ \\
+ & $\Phi_5(5,1,1,3)$ & $\Psi_3(10,1,1,1)$ &
    $\Psi_6(10,3,1,1)$ & $\Phi^H_2(1,\bar 3,1,1)_{-1,+3}$ &
    $\Phi^H_1(1,1,1,\bar 3)_{+1,-3}$ \\
+ & $\Phi_5(5,1,1,3)$ & $\Psi_3(10,1,1,1)$ &
    $\Psi_6(10,3,1,1)$ & $\Phi^H_3(1,\bar 3,1,\bar 3)_{+1,+3}$ &
    $\Phi^H_4(1,1,1,1)_{-1,-3}$ \\
+ & $\Phi_5(5,1,1,3)$ & $\Psi_3(10,1,1,1)$ &
    $\Psi_6(10,3,1,1)$ & $\Phi^H_4(1,1,1,1)_{+1,+3}$ &
    $\Phi^H_3(1,\bar 3,1,\bar 3)_{-1,-3}$ \\
+ & $\Phi_5(5,1,1,3)$ & $\Psi_5(\bar 5,1,1,1)$ &
    $\Phi^H_1(1,1,1,\bar 3)_{-1,+3}$ & $\sigma_1(1,1,1,1)_{+1,-4}$ &
    $\sigma_3(1,1,1,1)_{-3,+4}$ \\
+ & $\Phi_5(5,1,1,3)$ & $\Psi^H_1(1,1,1,3)_{-1,+2}$ &
    $\Psi^H_3(1,\bar 3,1,3)_{+1,-2}$ & $\phi_1(1,3,1,1)_{-1,+3}$ &
    $\phi_2(\bar 5,1,1,1)_{+1,-3}$ \\
+ & $\Phi_5(5,1,1,3)$ & $\Psi^H_3(1,\bar 3,1,3)_{+1,+2}$ &
    $\Psi^H_1(1,1,1,3)_{-1,-2}$ & $\phi_1(1,3,1,1)_{-1,+3}$ &
    $\phi_2(\bar 5,1,1,1)_{+1,-3}$ \\
+ & $\Phi_5(5,1,1,3)$ & $\Phi^H_1(1,1,1,\bar 3)_{-1,+3}$ &
    $\Phi^H_2(1,\bar 3,1,1)_{+1,-3}$ & $\phi_1(1,3,1,1)_{-1,+3}$ &
    $\phi_2(\bar 5,1,1,1)_{+1,-3}$ \\
+ & $\Phi_5(5,1,1,3)$ & $\Phi^H_2(1,\bar 3,1,1)_{-1,+3}$ &
    $\Phi^H_1(1,1,1,\bar 3)_{+1,-3}$ & $\phi_1(1,3,1,1)_{-1,+3}$ &
    $\phi_2(\bar 5,1,1,1)_{+1,-3}$ \\
+ & $\Phi_5(5,1,1,3)$ & $\Phi^H_3(1,\bar 3,1,\bar 3)_{+1,+3}$ &
    $\Phi^H_4(1,1,1,1)_{-1,-3}$ & $\phi_1(1,3,1,1)_{-1,+3}$ &
    $\phi_2(\bar 5,1,1,1)_{+1,-3}$ \\
+ & $\Phi_5(5,1,1,3)$ & $\Phi^H_4(1,1,1,1)_{+1,+3}$ &
    $\Phi^H_3(1,\bar 3,1,\bar 3)_{-1,-3}$ & $\phi_1(1,3,1,1)_{-1,+3}$ &
    $\phi_2(\bar 5,1,1,1)_{+1,-3}$ \\
+ & $\Phi_6(1,3,5,1)$ & $\Psi^H_1(1,1,1,3)_{-1,+2}$ &
    $\Psi^H_3(1,\bar 3,1,3)_{+1,-2}$ & $\phi_3(1,1,1,3)_{-1,+3}$ &
    $\phi_4(1,1,\bar 5,1)_{+1,-3}$ \\
+ & $\Phi_6(1,3,5,1)$ & $\Psi^H_3(1,\bar 3,1,3)_{+1,+2}$ &
    $\Psi^H_1(1,1,1,3)_{-1,-2}$ & $\phi_3(1,1,1,3)_{-1,+3}$ &
    $\phi_4(1,1,\bar 5,1)_{+1,-3}$ \\
+ & $\Phi_6(1,3,5,1)$ & $\Phi^H_1(1,1,1,\bar 3)_{-1,+3}$ &
    $\Phi^H_2(1,\bar 3,1,1)_{+1,-3}$ & $\phi_3(1,1,1,3)_{-1,+3}$ &
    $\phi_4(1,1,\bar 5,1)_{+1,-3}$ \\
+ & $\Phi_6(1,3,5,1)$ & $\Phi^H_2(1,\bar 3,1,1)_{-1,+3}$ &
    $\Phi^H_1(1,1,1,\bar 3)_{+1,-3}$ & $\phi_3(1,1,1,3)_{-1,+3}$ &
    $\phi_4(1,1,\bar 5,1)_{+1,-3}$ \\
+ & $\Phi_6(1,3,5,1)$ & $\Phi^H_3(1,\bar 3,1,\bar 3)_{+1,+3}$ &
    $\Phi^H_4(1,1,1,1)_{-1,-3}$ & $\phi_3(1,1,1,3)_{-1,+3}$ &
    $\phi_4(1,1,\bar 5,1)_{+1,-3}$ \\
+ & $\Phi_6(1,3,5,1)$ & $\Phi^H_4(1,1,1,1)_{+1,+3}$ &
    $\Phi^H_3(1,\bar 3,1,\bar 3)_{-1,-3}$ & $\phi_3(1,1,1,3)_{-1,+3}$ &
    $\phi_4(1,1,\bar 5,1)_{+1,-3}$ \\
- & $\bar\Phi_3(\bar 5,1,\bar 5,1)$ & $\Psi_2(\bar 5,3,1,1)$ &
    $\Psi_3(10,1,1,1)$ & $\phi_4(1,1,5,1)_{+1,+3}$ &
    $\phi_1(1,\bar 3,1,1)_{-1,-3}$ \\
- & $\bar\Phi_3(\bar 5,1,\bar 5,1)$ & $\Psi_5(\bar 5,1,1,1)$ &
    $\Psi_6(10,3,1,1)$ & $\phi_4(1,1,5,1)_{+1,+3}$ &
    $\phi_1(1,\bar 3,1,1)_{-1,-3}$ \\
- & $\bar\Phi_4(1,\bar 3,1,\bar 3)$ & $\Psi_2(\bar 5,3,1,1)$ &
    $\Psi_3(10,1,1,1)$ & $\phi_3(1,1,1,3)_{-1,+3}$ &
    $\phi_2(\bar 5,1,1,1)_{+1,-3}$ \\
- & $\bar\Phi_4(1,\bar 3,1,\bar 3)$ & $\Psi_5(\bar 5,1,1,1)$ &
    $\Psi_6(10,3,1,1)$ & $\phi_3(1,1,1,3)_{-1,+3}$ &
    $\phi_2(\bar 5,1,1,1)_{+1,-3}$ \\
+ & $\bar\Phi_5(\bar 5,1,1,\bar 3)$ & $\Psi_2(\bar 5,3,1,1)$ &
    $\Psi_6(10,3,1,1)$ & $\Psi^H_2(1,\bar 3,1,1)_{-1,+2}$ &
    $\Psi^H_3(1,\bar 3,1,3)_{+1,-2}$ \\
+ & $\bar\Phi_5(\bar 5,1,1,\bar 3)$ & $\Psi_2(\bar 5,3,1,1)$ &
    $\Psi_6(10,3,1,1)$ & $\Psi^H_3(1,\bar 3,1,3)_{+1,+2}$ &
    $\Psi^H_2(1,\bar 3,1,1)_{-1,-2}$ \\
+ & $\bar\Phi_5(\bar 5,1,1,\bar 3)$ & $\Psi_2(\bar 5,3,1,1)$ &
    $\Psi_6(10,3,1,1)$ & $\Phi^H_3(1,\bar 3,1,\bar 3)_{+1,+3}$ &
    $\Phi^H_3(1,\bar 3,1,\bar 3)_{-1,-3}$ \\
+ & $\bar\Phi_5(\bar 5,1,1,\bar 3)$ & $\Psi_3(10,1,1,1)$ &
    $\Psi_5(\bar 5,1,1,1)$ & $\Psi^H_1(1,1,1,3)_{-1,+2}$ &
    $\Psi^H_4(1,1,1,1)_{+1,-2}$ \\
+ & $\bar\Phi_5(\bar 5,1,1,\bar 3)$ & $\Psi_3(10,1,1,1)$ &
    $\Psi_5(\bar 5,1,1,1)$ & $\Psi^H_4(1,1,1,1)_{+1,+2}$ &
    $\Psi^H_1(1,1,1,3)_{-1,-2}$ \\
+ & $\bar\Phi_5(\bar 5,1,1,\bar 3)$ & $\Psi_3(10,1,1,1)$ &
    $\Psi_5(\bar 5,1,1,1)$ & $\Phi^H_1(1,1,1,\bar 3)_{-1,+3}$ &
    $\Phi^H_1(1,1,1,\bar 3)_{+1,-3}$ \\
+ & $\bar\Phi_5(\bar 5,1,1,\bar 3)$ & $\Psi^H_2(1,\bar 3,1,1)_{-1,+2}$ &
    $\Psi^H_3(1,\bar 3,1,3)_{+1,-2}$ & $\phi_2(5,1,1,1)_{+1,+3}$ &
    $\phi_1(1,\bar 3,1,1)_{-1,-3}$ \\
+ & $\bar\Phi_5(\bar 5,1,1,\bar 3)$ & $\Psi^H_3(1,\bar 3,1,3)_{+1,+2}$ &
    $\Psi^H_2(1,\bar 3,1,1)_{-1,-2}$ & $\phi_2(5,1,1,1)_{+1,+3}$ &
    $\phi_1(1,\bar 3,1,1)_{-1,-3}$ \\
+ & $\bar\Phi_5(\bar 5,1,1,\bar 3)$ & $\Phi^H_3(1,\bar 3,1,\bar 3)_{+1,+3}$ &
    $\Phi^H_3(1,\bar 3,1,\bar 3)_{-1,-3}$ & $\phi_2(5,1,1,1)_{+1,+3}$ &
    $\phi_1(1,\bar 3,1,1)_{-1,-3}$ \\
+ & $\bar\Phi_6(1,\bar 3,\bar 5,1)$ & $\Psi^H_2(1,\bar 3,1,1)_{-1,+2}$ &
    $\Psi^H_3(1,\bar 3,1,3)_{+1,-2}$ & $\phi_4(1,1,5,1)_{+1,+3}$ &
    $\phi_3(1,1,1,\bar 3)_{-1,-3}$ \\
+ & $\bar\Phi_6(1,\bar 3,\bar 5,1)$ & $\Psi^H_3(1,\bar 3,1,3)_{+1,+2}$ &
    $\Psi^H_2(1,\bar 3,1,1)_{-1,-2}$ & $\phi_4(1,1,5,1)_{+1,+3}$ &
    $\phi_3(1,1,1,\bar 3)_{-1,-3}$ \\
+ & $\bar\Phi_6(1,\bar 3,\bar 5,1)$ & $\Phi^H_3(1,\bar 3,1,\bar 3)_{+1,+3}$ &
    $\Phi^H_3(1,\bar 3,1,\bar 3)_{-1,-3}$ & $\phi_4(1,1,5,1)_{+1,+3}$ &
    $\phi_3(1,1,1,\bar 3)_{-1,-3}$ \\
\hline
\end{tabular}%
}%
\end{center}%
\end{table}

\begin{table}
\caption{The 6-point ${\rm R}^6$ forms to superpotential
from the full gauge invariance and all internal global symmetries.}
\label{table3f}
\begin{center}
{\footnotesize
\begin{tabular}{|ccc|}\hline
$\Psi_1(1,3,1,1)$ & $\Psi_2(\bar 5,3,1,1)$ &
$\Psi^H_2(1,\bar 3,1,1)_{-1,+2}$ \\
$\Psi^H_3(1,\bar 3,1,3)_{+1,-2}$ & $\phi_2(5,1,1,1)_{+1,+3}$ &
$\phi_3(1,1,1,\bar 3)_{-1,-3}$ \\ \hline
$\Psi_1(1,3,1,1)$ & $\Psi_2(\bar 5,3,1,1)$ &
$\Psi^H_3(1,\bar 3,1,3)_{+1,+2}$ \\
$\Psi^H_2(1,\bar 3,1,1)_{-1,-2}$ & $\phi_2(5,1,1,1)_{+1,+3}$ &
$\phi_3(1,1,1,\bar 3)_{-1,-3}$ \\ \hline
$\Psi_1(1,3,1,1)$ & $\Psi_2(\bar 5,3,1,1)$ &
$\Phi^H_3(1,\bar 3,1,\bar 3)_{+1,+3}$ \\
$\Phi^H_3(1,\bar 3,1,\bar 3)_{-1,-3}$ & $\phi_2(5,1,1,1)_{+1,+3}$ &
$\phi_3(1,1,1,\bar 3)_{-1,-3}$ \\ \hline
$\Psi_2(\bar 5,3,1,1)$ & $\Psi_3(10,1,1,1)$ &
$\Psi^H_1(1,1,1,3)_{-1,+2}$ \\
$\Psi^H_3(1,\bar 3,1,3)_{+1,-2}$ & $\phi_3(1,1,1,3)_{-1,+3}$ &
$\phi_2(\bar 5,1,1,1)_{+1,-3}$ \\ \hline
$\Psi_2(\bar 5,3,1,1)$ & $\Psi_3(10,1,1,1)$ &
$\Psi^H_3(1,\bar 3,1,3)_{+1,+2}$ \\
$\Psi^H_1(1,1,1,3)_{-1,-2}$ & $\phi_3(1,1,1,3)_{-1,+3}$ &
$\phi_2(\bar 5,1,1,1)_{+1,-3}$ \\ \hline
$\Psi_2(\bar 5,3,1,1)$ & $\Psi_3(10,1,1,1)$ &
$\Phi^H_1(1,1,1,\bar 3)_{-1,+3}$ \\
$\Phi^H_2(1,\bar 3,1,1)_{+1,-3}$ & $\phi_3(1,1,1,3)_{-1,+3}$ &
$\phi_2(\bar 5,1,1,1)_{+1,-3}$ \\ \hline
$\Psi_2(\bar 5,3,1,1)$ & $\Psi_3(10,1,1,1)$ &
$\Phi^H_2(1,\bar 3,1,1)_{-1,+3}$ \\
$\Phi^H_1(1,1,1,\bar 3)_{+1,-3}$ & $\phi_3(1,1,1,3)_{-1,+3}$ &
$\phi_2(\bar 5,1,1,1)_{+1,-3}$ \\ \hline
$\Psi_2(\bar 5,3,1,1)$ & $\Psi_3(10,1,1,1)$ &
$\Phi^H_3(1,\bar 3,1,\bar 3)_{+1,+3}$ \\
$\Phi^H_4(1,1,1,1)_{-1,-3}$ & $\phi_3(1,1,1,3)_{-1,+3}$ &
$\phi_2(\bar 5,1,1,1)_{+1,-3}$ \\ \hline
$\Psi_2(\bar 5,3,1,1)$ & $\Psi_3(10,1,1,1)$ &
$\Phi^H_4(1,1,1,1)_{+1,+3}$ \\
$\Phi^H_3(1,\bar 3,1,\bar 3)_{-1,-3}$ & $\phi_3(1,1,1,3)_{-1,+3}$ &
$\phi_2(\bar 5,1,1,1)_{+1,-3}$ \\ \hline
$\Psi_3(10,1,1,1)$ & $\Psi_3(10,1,1,1)$ &
$\Psi^H_1(1,1,1,3)_{-1,+2}$ \\
$\Psi^H_4(1,1,1,1)_{+1,-2}$ & $\phi_2(5,1,1,1)_{+1,+3}$ &
$\phi_3(1,1,1,\bar 3)_{-1,-3}$ \\ \hline
$\Psi_3(10,1,1,1)$ & $\Psi_3(10,1,1,1)$ &
$\Psi^H_4(1,1,1,1)_{+1,+2}$ \\
$\Psi^H_1(1,1,1,3)_{-1,-2}$ & $\phi_2(5,1,1,1)_{+1,+3}$ &
$\phi_3(1,1,1,\bar 3)_{-1,-3}$ \\ \hline
$\Psi_3(10,1,1,1)$ & $\Psi_3(10,1,1,1)$ &
$\Phi^H_1(1,1,1,\bar 3)_{-1,+3}$ \\
$\Phi^H_1(1,1,1,\bar 3)_{+1,-3}$ & $\phi_2(5,1,1,1)_{+1,+3}$ &
$\phi_3(1,1,1,\bar 3)_{-1,-3}$ \\ \hline
$\Psi_4(1,1,1,1)$ & $\Psi_5(\bar 5,1,1,1)$ &
$\Psi^H_1(1,1,1,3)_{-1,+2}$ \\
$\Psi^H_4(1,1,1,1)_{+1,-2}$ & $\phi_2(5,1,1,1)_{+1,+3}$ &
$\phi_3(1,1,1,\bar 3)_{-1,-3}$ \\ \hline
$\Psi_4(1,1,1,1)$ & $\Psi_5(\bar 5,1,1,1)$ &
$\Psi^H_4(1,1,1,1)_{+1,+2}$ \\
$\Psi^H_1(1,1,1,3)_{-1,-2}$ & $\phi_2(5,1,1,1)_{+1,+3}$ &
$\phi_3(1,1,1,\bar 3)_{-1,-3}$ \\ \hline
$\Psi_4(1,1,1,1)$ & $\Psi_5(\bar 5,1,1,1)$ &
$\Phi^H_1(1,1,1,\bar 3)_{-1,+3}$ \\
$\Phi^H_1(1,1,1,\bar 3)_{+1,-3}$ & $\phi_2(5,1,1,1)_{+1,+3}$ &
$\phi_3(1,1,1,\bar 3)_{-1,-3}$ \\ \hline
$\Psi_4(1,1,1,1)$ & $\Phi^H_4(1,1,1,1)_{-1,-3}$ &
$\phi_2(5,1,1,1)_{+1,+3}$ \\
$\phi_2(\bar 5,1,1,1)_{+1,-3}$ & $\sigma_1(1,1,1,1)_{-1,-4}$ &
$\sigma_3(1,1,1,1)_{+3,+4}$ \\ \hline
$\Psi_4(1,1,1,1)$ & $\Phi^H_4(1,1,1,1)_{-1,-3}$ &
$\phi_4(1,1,5,1)_{+1,+3}$ \\
$\phi_4(1,1,\bar 5,1)_{+1,-3}$ & $\sigma_1(1,1,1,1)_{-1,-4}$ &
$\sigma_3(1,1,1,1)_{+3,+4}$ \\ \hline
$\Psi_5(\bar 5,1,1,1)$ & $\Psi_6(10,3,1,1)$ &
$\Psi^H_1(1,1,1,3)_{-1,+2}$ \\
$\Psi^H_3(1,\bar 3,1,3)_{+1,-2}$ & $\phi_3(1,1,1,3)_{-1,+3}$ &
$\phi_2(\bar 5,1,1,1)_{+1,-3}$ \\ \hline
$\Psi_5(\bar 5,1,1,1)$ & $\Psi_6(10,3,1,1)$ &
$\Psi^H_3(1,\bar 3,1,3)_{+1,+2}$ \\
$\Psi^H_1(1,1,1,3)_{-1,-2}$ & $\phi_3(1,1,1,3)_{-1,+3}$ &
$\phi_2(\bar 5,1,1,1)_{+1,-3}$ \\ \hline
$\Psi_5(\bar 5,1,1,1)$ & $\Psi_6(10,3,1,1)$ &
$\Phi^H_1(1,1,1,\bar 3)_{-1,+3}$ \\
$\Phi^H_2(1,\bar 3,1,1)_{+1,-3}$ & $\phi_3(1,1,1,3)_{-1,+3}$ &
$\phi_2(\bar 5,1,1,1)_{+1,-3}$ \\ \hline
$\Psi_5(\bar 5,1,1,1)$ & $\Psi_6(10,3,1,1)$ &
$\Phi^H_2(1,\bar 3,1,1)_{-1,+3}$ \\
$\Phi^H_1(1,1,1,\bar 3)_{+1,-3}$ & $\phi_3(1,1,1,3)_{-1,+3}$ &
$\phi_2(\bar 5,1,1,1)_{+1,-3}$ \\ \hline
$\Psi_5(\bar 5,1,1,1)$ & $\Psi_6(10,3,1,1)$ &
$\Phi^H_3(1,\bar 3,1,\bar 3)_{+1,+3}$ \\
$\Phi^H_4(1,1,1,1)_{-1,-3}$ & $\phi_3(1,1,1,3)_{-1,+3}$ &
$\phi_2(\bar 5,1,1,1)_{+1,-3}$ \\ \hline
$\Psi_5(\bar 5,1,1,1)$ & $\Psi_6(10,3,1,1)$ &
$\Phi^H_4(1,1,1,1)_{+1,+3}$ \\
$\Phi^H_3(1,\bar 3,1,\bar 3)_{-1,-3}$ & $\phi_3(1,1,1,3)_{-1,+3}$ &
$\phi_2(\bar 5,1,1,1)_{+1,-3}$ \\ \hline
$\Psi_6(10,3,1,1)$ & $\Psi_6(10,3,1,1)$ &
$\Psi^H_2(1,\bar 3,1,1)_{-1,+2}$ \\
$\Psi^H_3(1,\bar 3,1,3)_{+1,-2}$ & $\phi_2(5,1,1,1)_{+1,+3}$ &
$\phi_3(1,1,1,\bar 3)_{-1,-3}$ \\ \hline
$\Psi_6(10,3,1,1)$ & $\Psi_6(10,3,1,1)$ &
$\Psi^H_3(1,\bar 3,1,3)_{+1,+2}$ \\
$\Psi^H_2(1,\bar 3,1,1)_{-1,-2}$ & $\phi_2(5,1,1,1)_{+1,+3}$ &
$\phi_3(1,1,1,\bar 3)_{-1,-3}$ \\ \hline
$\Psi_6(10,3,1,1)$ & $\Psi_6(10,3,1,1)$ &
$\Phi^H_3(1,\bar 3,1,\bar 3)_{+1,+3}$ \\
$\Phi^H_3(1,\bar 3,1,\bar 3)_{-1,-3}$ & $\phi_2(5,1,1,1)_{+1,+3}$ &
$\phi_3(1,1,1,\bar 3)_{-1,-3}$ \\
\hline
\end{tabular}%
}%
\end{center}
\end{table}

\section{The Paths of Unification with the $G\times G$ Gauge Groups}
In this chapter we explicitly investigate the paths of the unification
in the  GUST with gauge symmetry
$ G\times G = [SU(5) \times U(1)\times (SU(3) \times U(1))_H]^{\otimes 2}$.
We show that the GUSTs with the $G\times G$ gauge group allow to make
the scale of unification to be consistent with
the string scale $M_{SU} \sim g_{string}\cdot 5 \cdot 10^{17} GeV$.
 The main goal of this chapter is to solve the problem of discrepancy between
$[SU(5)\times U(1) \times G_H]^{\otimes 2}$,
along with horizontal gauge symmetry $G_H=U(3)$.
Note that when we place some reasonable phenomenological constraints
namely when we want to obtain a model with the  gauge group  mentioned above
without matter doubling and with necessary Higgs fields then modular
invariance restrict the possibilities for the spectrum highly enough.
Actually one can build only a few classes of such a model with different
structure of the hidden gauge group (rank 6).
In particular, the attempts to build a model that conforms to
our condition using different set of boundary conditions for world-sheet
fermions in principle lead to the same spectrum. Thus we consider
the model presented  worthy of the detailed study.

There are no so many GUSTs describing the observable sector of Standard
Model. They are well known: the $SU(3)\times SU(2)\times U(1)^n\times G_{hid}$
gauge group, the Pati-Salam ($SU(4)\times SU(2)\times SU(2)\times U(1)
\times G_{hid}$) gauge group,
the flipped $SU(5) \times U(1)\times G_{hid}$ gauge group
and $SO(10)\times G_{hid}$ gauge group \cite{20i,20ii}.
For the heterotic 10-dimensional  string the groups $E_8\times E_8$ and
$spin(32)/Z_2$ are characteristic. Hence it is interesting to consider GUSTs
in four dimension based on its various rank 8 and 16 subgroup \cite{25i}.
appear in consideration \cite {25i,26i}.
Moreover for the observable gauge symmetry one can consider the diagonal
subgroups ${G'}^{\rm sym}$ of the rank 16 group $G \times G \subset SO(16)
\times SO(16)$ or $\subset E(8) \times E(8)$.
In the paper \cite {25i} we have considered the possible 
ways of breaking the "string" gauge subgroups
$\subset E_8\times E_8$ down to low energy supersymmetric model that includes
Standard Model group and horizontal factor $SU(3)$.
The constraints of horizontal model parameters followed from this approach
allow the existence of the interesting flavour-changing physics in the
TeV region. Also these models gives rise to a rather natural way of
the superweak-like CP-violation\cite {9i}.
All this leads us naturally to considering possible forms for horizontal
symmetry
$G_H$, and $G_H$ quantum number assignments for quarks (anti-quarks) and
leptons (anti-leptons) which can be realized within GUST's framework.

We consider two possibilities of the breaking of the primordial
$[U(5) \times G_H]^{\otimes 2}$ gauge symmetry with two variants of
$Q_{em}$ charge  quantization correspondingly.
For the various chains of gauge symmetry breaking in flipped and non-flipped
 cases of the $SU(5)$ model we carry out the RGE analysis of the behaviour of
the gauge coupling constants taking into account the  possible intermediate
thresholds (the additional Higgs doublets, color triplets,
SUSY threshold, massive fourth generation)
 and the threshold effects due to the  massive string states.
We show that only in
non-flipped case of $[U(5) \times G_H]^{\otimes 2}$ GUST it is possible
to make the unification scale of $g_{1,2,3}$-
coupling constants in supersymmetric standard model, $M_G$, to be
consistent  with the string
scale unification, $M_{GUST} = M_{SU}$ and
obtain estimation of the string coupling constant $g_{str}=O(1)$.
As an additional benefit,
the values of the $g_{str}$ and $M_{SU}$ allow us to estimate
the horizontal coupling constant $g_{3H}$ on the scale of $\sim$1 TeV.
The important prediction of the GUSTs class considered is the existence
of the 4th heavy ($> 200$ GeV) generation that is singlet under
$SU(3)_H$.

\subsection{The Ways of the
$[SU(5)\times U(1)\times SU(3)\times U(1)]^{\otimes 2}$
Gauge Symmetry Breaking. } \label{sbsec43}
Further we shortly discuss the problem of
gauge symmetry breaking in Model~1.  The most
important point  is that the Higgs fields
$(\underline{10}_{1/2}+\underline{\bar{10}}_{-1/2})$ do not appear.
However there exists some
possibilities to break the GUST group $[(U(5)\times U(3))^I]^{\times 2}$
down to the symmetric subgroups using the following VEVs of the Higgs fields
$(\underline 5,\underline 1;\underline 5,\underline 1)_{(-1,0;-1,0)}$:
\begin{description}
\itemsep=0cm
\item[a)] $\qquad\qquad \langle (\underline 5,\underline 5)\rangle_0 =
a\cdot \mbox{diag}(1,1,1,1,1)$
\item[b)] $\qquad\qquad \langle (\underline 5,\underline 5)\rangle_0 =
 \mbox{diag}(x,x,x,y,y)$,
\item[c)] $\qquad\qquad \langle (\underline 5,\underline 5)\rangle_0 =
a \cdot \mbox{diag}(0,0,0,1,1)$,
\item[d)] $\qquad\qquad \langle (\underline 5,\underline 5)\rangle_0 =
a \cdot \mbox{diag}(1,1,1,0,0)$.
\end{description}
With the VEV a) the GUST group $[(U(5)\times U(3))^I]^{\otimes 2}$
 breaks to symmetric group:
$$ \mbox{\bf a)}\qquad\qquad U(5)^I\times U(5)^{II}
\rightarrow {U(5)}^{\rm sym} \rightarrow \ldots$$
With such a breaking tensor Higgs fields transform
under the $(SU(5)\times U(1))^{\rm sym}\times G_H$ group in the
following way:
\begin{eqnarray}\label{eq19}
(\underline 5,\underline 1;\underline 5,\underline 1)_{(-1,0;-1,0)}
{}~&\rightarrow~~& {(\underline {24},\underline 1) }_{(0,0)}~~
+~~(\underline 1,\underline 1) _{(0,0)}.
\end{eqnarray}

The diagonal VEVs of the Higgs fields break the GUST with $G \times G$ down
to the "skew"-symmetric group with the generators $\triangle^{\rm sym}$ of
the form:
\begin{equation}\label{eq210}
\triangle^{\rm sym} (t) = -t^* \times 1 + 1\times t,
\end{equation}
Adjoint  representations which appear on the $rhs$ of (\ref{eq19}) can be used
for
further breaking of the symmetric group. This can lead to the final physical
symmetry
$$ \qquad SU(3^c)\times SU(2_{EW})\times U(1)_5 \times U(1)^{\rm sym}
\times G_H^{'}$$
with the low-energy gauge symmetry of the quark--lepton generations with an
additional $U(1)^{\rm sym}$--factor. As we already discussed above
the form of the $G_H^{'}$ depends on the way of the U(1)
anomaly group cancelations and the complete breaking of this group is
realized by the VEVs of the ${\hat \Phi}_{(1,3;1,3)}$ Higgs fields
and/or ${\hat \Phi}_{(1,1;5,{\bar 3})}$,
${\hat \Phi}_{(5,{\bar 3};1,1)}$ Higgs fields.

Note, that when we use the VEVs b),c),d) there also
exist another interesting ways 
of breaking the $G^I \times G^{II}$ gauge symmetry
down to
\begin{description}
\itemsep=0cm
\leftmargin=5mm
\rightmargin=0mm
\item[b)] $\quad SU(3^c) \times
 SU(2)_{EW} \times U(1)_5 \times U(1)^{\rm sym} \times G_H^{'}
 \rightarrow \ldots$,
\item[c)] $\quad  SU(3^c) \times SU(2)^I_{EW}
\times SU(2)^{II}_{EW} \times U(1)_5\times U(1)^{\rm sym}
\times G_H^{'} \rightarrow \ldots $ ,
\item[d)] $\quad SU(3^c)_I \times SU(3^c)_{II} \times SU(2)_{EW} \times U(1)_5
\times U(1)^{\rm sym} \times G_H^{'} \rightarrow \ldots$ .
\end{description}
 We could consider the GUST construction involving
$[SO(10)  \times G_H]^{\otimes 2}\times G_{hidden}$
 as the  gauge group \cite{25i}.
In that model the only Higgs fields appeared are $(10,10)$ of
$SO(10) \times SO(10) $ and the hidden group $G_{hidden}=U(1)\times SO(6)$
 is anomaly free.
As an illustration we would like to remark that for the
$SO(10) \times SO(10) \times G_H \times G_H$ GUST we can consider similarly
the following VEVs of the Higgs fields $(10,10)$:
\begin{description}
\itemsep=0cm
\item[a')]$\quad \langle (10,10)\rangle_0 =
           a\cdot \mbox{diag} (1,1,1,1,1,1,1,1,1,1)$,
\item[b')]$\quad \langle (10,10)\rangle_0 =
           a \cdot \mbox{diag} (1,1,1,1,1,1,0,0,0,0)$,
\item[c')]$\quad \langle (10,10)\rangle_0 =
           a \cdot \mbox{diag} (1,1,1,1,1,1,x,x,x,x)$.
\end{description}
These cases
lead  correspondingly to the following chains of
$[SO(10)]^{\otimes 2}$ breaking ways:
\begin{description}
\itemsep=0cm
\item[a')]$\quad [SO(10)]^{\rm sym}$,
\item[b')]$\quad SU(4)\times SU(2)_{I1}\times SU(2)_{I2}\times
 SU(2)_{II1} \times SU(2)_{II2}$,
\item[c')]$\quad SU(4)\times SU(2)_{I}\times SU(2)_{II}$.
\end{description}

\subsection{The GUT and String Unification Scales.}

Indeed the estimates on the $M_{H_0}$ scale depend on the
value of the family gauge coupling.
String theories imply a natural unification of the gauge
and gravitational
couplings, $g_i$ and $G_N$ respectively. For example, it turns out that
these couplings unify
at tree level to form one coupling constant $g_{string}$ \cite{Ginsparg'}:
\begin{eqnarray}
8\pi \frac{G_N}{\alpha'}=g_ik_i=g^2_{str}.
\end{eqnarray}
\medbreak\par\noindent
Here $\alpha'$ is the Regge slope, the coupling constants $g_i$
correspond to the gauge group $G_i$ with the Kac-Moody levels $k_i$.
In string theory the scale of unification is fixed by the Planck scale
$M_{Pl}\approx 10^{19}$GeV.
In one-loop string calculations
the value of the unification scale could be divided into moduli independent
part and a part that depends on the VEVs of the moduli fields. The latter
part considered as the correction to the former and obviously it is
different for the various models. Like in the gauge fields theory
this correction is called the string threshold correction of the massive
string states. Moduli independent contribution depends on the renormalizing
scheme used,
so in $\overline{DR}$ renormalization scheme the scale of string
unification is shifted to \cite{Kaplunovsky'}a :
\begin{equation}
M_{SU} = \frac{e^{(1-\gamma)/2} 3^{-3/4}}{4\pi} g_{str} M_{Pl}
 \approx 5\,g_{str} \times {10}^{17} GeV
\label{Mstring}
\end{equation}

There  exists the most important difference between the unification
scales of gauge coupling
unification in string theory and  in field theory. In field theory this scale
is determined via extrapolation of data within the Supersymmetric
$SU(3) \times SU(2) \times U(1)$ Standard Model using
 the renormalization group equations (RGE) for the gauge couplings
$${\alpha}_{3}^{-1}(M_Z) = 8.93\ ,$$
$${\alpha}_{2}^{-1}= ({\alpha}_{em}/
{\sin}^{2}{\theta}_{W})^{-1}|_{M_Z}=29.609\ ,$$
$${\alpha}_{1}^{-1}= (5{\alpha}_{em}/ 3{\cos}^{2}
{\theta}_{W})^{-1}|_{M_Z}=58.975\ .$$
The factor 5/3 in the definition of ${\alpha}_1$ has been included for the
normalization at the unification scale $M_{G}$.

The one loop   renormalization group equations for these gauge couplings
 are given by

\begin{equation}
\frac {d{\alpha}_{i}}{d \ln \mu}= \frac{{\alpha}_{i}^2}{2\pi} b_i
\label{RGEi}
\end{equation}

 Beta functions coefficients
for the $SU(3)\times SU(2) \times U(1)$ coupling constants
in SUSY models are given by the following:

\begin{eqnarray}
b_3&=& -9 +2N_g + 0\cdot N_h + \frac{1}{2}\cdot N_3,\\
b_2&=& -6 +2N_g + \frac{1}{2}\cdot N_h +  0\cdot N_3,\\
b_1&=&\,0 +2N_g + \frac{3}{10}\cdot N_h + \frac{1}{5}\cdot N_3,
\label{bi}
\end{eqnarray}
where $N_g$ is the number of generation and $N_h$ is the number of Higgs
doublets and we also include some possible intermediate thresholds for heavy
Higgs doublets $(N_h - 2)$ and color triplets, $N_3$,
which exist in the spectrum of the model and can be take into account for
RGE.

The RGE are integrated from $M_Z$-mass to the unification scale $M_{G}$.
In the presence of various intermediate scale, $M_I$, $I=1,2,3,\ldots$,
the RGE are given by:

\begin{eqnarray}
{\alpha}_i^{-1}(M_Z)=
{\alpha}_i^{-1}(M_{G}) + \frac{b_i}{2\pi}\ln{\frac{M_{G}}{M_Z}}
-\frac{b_{i1}}{2\pi}\ln{\frac{M_1}{M_Z}}
-\frac{b_{i2}}{2\pi}\ln{\frac{M_2}{M_Z}}-\ldots
\end{eqnarray}
where $b_{iI}$ are the additional corresponding contributions of the new
thresholds to the beta functions. At the Z-mass
scale we have:

\begin{eqnarray}
{\sin}^2{\theta}_{W}(M_Z) &\approx & 0.2315 \pm 0.001\ ,\nonumber\\
M_Z&=&91.161\pm 0.031 GeV, \label{experiment}
\end{eqnarray}

Note, that for flipped models
\begin{equation}
\sin^2 \theta_W(M_{G}) =\frac{15k^2}{16k^2 +24}
\end{equation}
where $k^2=g_1^2/g_5^2$.
In the SO(10) limit (or for non-flipped case) we have
$k^2=1$ and $\sin^2 \theta_W = 3/8$.

The string unification scale could be contrasted with
MSSM, $SU(3^c)\times SU(2) \times U(1)$
naive unification scale,
\begin{equation}
M_{G} \approx 2\times {10}^{16} GeV
\end{equation}
obtained by running the SM particles and their SUSY-partners to high energies.

\subsection{The String Thresholds Corrections.}

One of the first way to explain the difference between these two mass
scales, $M_{SU}$ and $M_{MSSM}=M_G$, was the  attempts to take into account
the  string thresholds corrections of the massive string states
\cite {Kaplunovsky'}:
\begin{equation}
\frac{1}{g_i^2(\mu)} = k_i\frac{1}{g_{string}^2}
+2b_i\ln(\frac{\mu}{ M_{SU}}) + {\tilde\Delta}_{G_i}.
\end{equation}
where the index $i$ runs over gauge coupling and  $\mu$ is
some phenomenological scale such as $M_Z$ or $M_{G}$.
The coefficients $k_i$ are the Kac-Moody levels (e.g.
for SU(5) $k_2=k_3=1,k_1=5/3$).
The quantities $ {\tilde\Delta}_{G_i}$ represent
the heavy thresholds corrections, which are the corrections arising
from the infinite towers of massive string states.
Although these states have the Planck mass scale, there are infinite
number of them, so they together could have the considerable effect.
In general the full string thresholds corrections are  of the form
\begin{equation}
{\tilde {\Delta}}_i={\Delta}_i +k_i Y,
\end{equation}
where Y is independent of the gauge
group factor. Moreover, the low energy predictions for
${\sin}^2{\theta}_W(M_Z)$ and ${\alpha}_3(M_Z)$ depend only on the differences
$
({\tilde{\Delta}}_i -{\tilde{\Delta}}_j)=(\Delta_i -\Delta_j)
$
for the different gauge groups.

However the $Y$ factor makes influence on the estimation of the value of
$g_{str}$ if we base on low energy gauge constants and use RGE. Note that
in general $g_{str}$ is defined by VEV of the dilaton moduli field but
because of degeneracy of the classic potential of the moduli fields
we do not know the $g_{str}$ a priori. If we could have the value of the
$g_{str}$ then the its coincidence with our estimates will show the
correctness of our model.

It is supposed that the value of the $Y$ factor is small enough
\cite{KalaraLopez} so we neglect it in our calculations of $g_{str}$
via RGE.

In the  GUSTs examples considered the threshold corrections
of the massive string states are not large enough to
compensate the difference between the scales of unification of
the string and the MSSM (GUT) models.
Later we will discuss the possible effects of them in $G \times G$ models,
for example in Model 1.

In our calculations for Model 1
we will follow the way suggested in \cite{Kaplunovsky'}c,\cite {Dienes'}.
According to it
we have to calculate an integral of the modified partition function
over the fundamental domain $\Gamma$ of modular group.
$$
\delta\Delta=\int_\Gamma\frac{\mbox{d}^2\tau}{\tau_2}
\left(|\eta(\tau)|^{-4}\hat{Z}(\tau)-b_G\right)
$$
where $b_G$ is the beta function coefficient,
$\hat Z(\tau)$ is modified partition function.  Modified in this
context means that the charge operators $Q$ are inserted in the trace
in partition function in the following way:
$$ \hat
Z(\tau)=-\mbox{Tr}Q^2_sQ^2_Iq^Hq^{\bar H}, $$
where $Q_s$ is helicity operator and
$Q_i$ is a generator of a gauge group.  We are interesting in particular
in the difference between groups $SU(5)_I$ and $U(1)_I$ in Model~1.
For this case charge polynomial is $5Q_1Q_2$  (see \cite{Kaplunovsky'}c).
Rewritten via well known theta
function, the modified function $\hat Z(\tau)$ for our case reads:
$$ \hat{Z}(\tau)=-\frac5{512}\sum_{{\bf a,b}}c\left({\bf a\atop b}\right)
\eta^{-1}\theta'\left( a_\Psi\atop b_\Psi\right)\bar\eta^{-2}
 \hat{\bar\theta}^2\left( a_I\atop b_I\right)
 \prod\eta^{-1}\theta\left(a_j\atop b_j\right)
 \prod\bar\eta^{-1}\bar\theta\left(a_k\atop b_k\right)
 $$
where  512 is normalizing factor
and products calculated over all fermions excluding fermions
 which $Q$ operators apply to. Namely $\theta'$ and $\hat{\bar\theta}$
 denote action of helicity and gauge group operators respectively.
 The sum is taken over all pairs of boundary condition vectors that
 appear on the model.

According to \cite{Dienes'} we expand the resulting expression with
$\theta$-functions via $q$ and $\bar q$ in order to achieve appropriate
precision.

The final results for Model 1 are as follows ($\delta$ denotes the difference
between corresponding quantities for $U(1)_I$ and $SU(5)_I$).
$\delta b=26.875; \qquad \delta\Delta=5.97 $
Given this relative threshold corrections we can compute its effect on
string unification scale $M_{SU}$.
We find that the correction unification scale is:
\begin{equation}\label{Mstrcor}
M_{SU}^{corr}=M_{SU} \,\exp\frac{({\Delta}_5-{\Delta}_1)}{2(b_5-b_1)}
\approx g_{str.}\cdot 5.6\cdot 10^{17} GeV\ .
\end{equation}
\subsection{The Paths of Unification in Flipped and Non-Flipped
GUST Models.}
However there are some  ways to explain the difference between
scales of string ($M_{SU}$) and ordinary ($M_{G}=M_{SU}$) unifications
(without additional intermediate exotic vector matter fields
that does not fit into $5$ or $\bar 5$ representations of SU(5).
\cite {Dienes'})
Perhaps the most natural  way is related to  the $G^I \times G^{II}$
String GUT.
If one uses the breaking scheme $G^I\times G^{II}\,\rightarrow G^{\rm sym}$
( where $G^{I,II}=U(5)\times U(3)_H \subset E_8$ ) on the $M_{\rm sym}$-scale,
then unification scale $M_{MSSM}= M_{G} \sim 2\cdot 10^{16}\,$GeV is
the scale of breaking
the $G_{\rm sym}$ group, and string unification do supply the equality
of coupling constant $G\times G$ on the string scale
$M_{SU}\sim g \cdot 5\cdot10^{17}\,$GeV.
Otherwise, we can have an addition scale of the symmetry breaking
$M_{\rm sym} > M_{G}$.
In any case on the scale of breaking
$U(5)^I\times U(5)^{II}\,\rightarrow U(5)^{\rm sym}$
the gauge coupling constants satisfy the equation
\begin{equation}
({\alpha}^{Sym})^{-1}= ({\alpha}^I)^{-1} + ({\alpha}^{II})^{-1} \ .
\label{symscale}
\end{equation}
Thus in this scheme the knowledge of scales $M_{SU}$ and $M_{Sym}$ gives
us a principal possibility to trace the evolution of coupling
constant of the original  group
$SU(5)^I \times U(1)^I \times SU(5)^{II} \times U(1)^{II}$  through
the $SU(5)^{Sym} \times U(1)^{Sym}$ to the low energies
and estimate the values of all coupling constants including
the horizontal gauge constant $g_{3H}$.

The coincidence of $\sin^2 \theta_W$ and ${\alpha}_3$ with experiment will
show how realistic this model is.

The evolution of the gauge constant from the string constant $g_{str}$
to the scale of $M_G$ is described by the equation:
\begin{equation}
({\alpha}_{5,1}^{Sym})^{-1}(M_{G})
=2({\alpha}_{5,1}^{Str})^{-1}(M_{SU}) +
\frac{(b_{5,1}^{I}+ b_{5,1}^{II})}{2 \pi}\ln(M_{SU}/M_{Sym})
+\frac{b_{5,1}^{Sym}}{2 \pi}\ln(M_{Sym}/M_{G})
\label{RGES}
\end{equation}
where ${\alpha}_{5}^{Str}(M_{SU})=g_{str}^2/4 \pi$
and $g_{str}$ is the string coupling.
\bigskip
Now we can get the relation between $g_{str}=g$ and $M_{Sym}$ from RGE's
for gauge running constants $g_5^{Sym}=g_5$, $g_5^{I}$ and , $g_5^{II}$
on the $M_{G}-M_{SU}$ scale.
For example in  Model 1 for the breaking scheme a) one can get:
\begin{equation} \label{b5values}
b_5^{Sym}=12,\,\,\,
b_5^{I}=5,\,\,\,
b_5^{II}=-3
\end{equation}
and
\begin{equation} \label{b1values}
b_1^{Sym}=27,\,\,\,
b_1^{I}=\frac{105}{4},\,\,\,
b_1^{II}=\frac{73}{4}.
\end{equation}

Let us try to make the behaviours of
these coupling constants consistent above and below $M_{G}$ scale.
Note that while taking the renormgroup analysis of the $g_{1,2,3}$
behaviour below the $M_G$ scale in our model we need to take into
consideration only $3+1$ generations and the Higgs fields.
The remaining fields are either heavy (see superpotential
and anomalous D-term suppression discussion) or sterile
under the standard observable gauge group.
There are two possibilities to embed quark-lepton matter in
$SU(5)\times U(1)$ of SO(10) multiplets,
$\underline {1}_{5/2} $,
$ \underline {\bar 5}_{-3/2} $, $\underline {10}_{1/2}$.
For the electromagnetic charge we get:
\begin{eqnarray}
\label{eq023}
Q_{em}&=& Q^{II} - Q^{I} = {\bar T}_3 + \frac{1}{2}{\bar y},\nonumber\\
{\bar T}_3&=&T^{II}-T^{I};\,\,\,\,\quad
{\bar y}= y^{II} - y^{I}.
\end{eqnarray}
where
\begin{equation}
\label {23iii}
\frac{1}{2}{y^{I,II}}
=\alpha {T^{I,II}}_{5z} + \beta {Y^{I,II}}_5\ , \quad
{T^{I,II}}_{5z} = \mbox{diag} (-1/3,-1/3,-1/3,1/2,1/2).
\end{equation}
In usual non-flipped Georgi-Glashow  SU(5)
model the $Q_{e.m.}$ is expressed
via SU(5) generators only: $\alpha= 1;\,\,\, \beta= 0$. For flipped
$SU(5)\times U(1)$ we have
$\alpha=- \frac{1}{5};\,\,\, \beta= \frac{2}{5}.$

Note, that this charge quantization does not lead to exotic
states with fractional electromagnetic charges (e.g. $Q_{em} =\pm 1/2,
\pm 1/6$)\cite{27i,25i}. Also in non-flipped
$SU(5) \times U(1)$ gauge symmetry breaking
scheme  there are  no  SU(3) color triplets and SU(2) doublets with
exotic hypercharges.

In our  model   we can
use the Higgs doublets from
$(5,1;1,1) + ({\bar 5},1;1,1)$,
(the fields $\hat{\phi}_2+\bar{\hat{\phi}}_2$ from sector 5)
for breaking the SM symmetry and low energy $U(1)_5^{\rm sym}$-symmetry.
Below the $M_{G}$  scale in non-horizontal sector
the evolution of gauge coupling constants is described by equations
\begin{eqnarray}
&&\alpha^{-1}_S(\mu )=\alpha^{-1}_5(M_{G} ) + \frac{b_3}{2\pi}
\ln (M_{G}/\mu) \\
&&\alpha^{-1}(\mu )\sin^2\theta_W =\alpha^{-1}_5(M_{G} ) +
\frac{b_2}{2 \pi}\ln (M_{G}/\mu) \\
&&\frac{15k^2}{k^2+24}{\alpha}^{-1}(\mu )\cos^2\theta_W =
{\alpha}_5^{-1}(M_{G}) +\frac{{\bar b}_1}{2 \pi}\ln (M_{G}/\mu)  ,
\end{eqnarray}
where for $N_g=4$ generations and for the minimal set of Higgs fields we have:
$$b_3=-1\ ,\quad b_2=3\ ,\quad
{\bar b}_1=\frac{25k^2}{k^2+24}\cdot b_1 |_{4\ gen.}
=\frac{15k^2}{(k^2+24)}\frac{43}{3} .$$

From these equations and from the experimental data we can find for $N_g=3,4$,
respectively:
\begin{eqnarray}
M_{G}&=&1.2\cdot 10^{16}\mbox{GeV}\ ,\qquad \alpha_5^{-1}=24.4,
\qquad k^2=0.98\ ,\nonumber\\
M_{G}&=&1.17\cdot 10^{16}\mbox{GeV}\ ,\qquad \alpha_5^{-1}=14.1,
\qquad k^2=0.97.
\end{eqnarray}

Here we assume that additional Higgs doublets and triplets appeared
in the theory are heavy ($>M_G$).

From the other hand for the Model 1 and with the mass of the fourth
generation  sufficiently heavy to be invisible but less than $M_G$
the equations
for ${\alpha}_{5,1}^{I.II.Sym}$ for all $M_{\rm sym}$-scale in the range,
$M_G <M_{\rm sym}< M_{SU}$,
give the contradicting value for $k^2$ that is considerably less than 1.
For example, for $M_{Sym}=1.6 \cdot10^{17}GeV$, we get:
\begin{eqnarray}
 M_{SU} = 9.6\cdot10^{17}GeV,\,\,\,
g_{str}=1.7\,\,\rightarrow k^2=0.44.
\end{eqnarray}

In the non-flipped case in Model 1  we have an additional neutral singlet
$\hat{\sigma}^1$ field, which could be used for breaking $U(1)^{Sym}$ group
(of  $U(1)_5^I \times U(1)_5^{II}$) at any high scale, independently on
the $M_{Sym}$ scale, where the
$$SU(5)^I\times SU(5)^{II}\longrightarrow SU(5)^{Sym}$$
 Therefore we have no
constraints on $k^2$ parameter in the range from string scale down to $M_G$.
As a result in  non-flipped case of the
$G^I \times G^{II}$ ($G=SU(5) \times U(1)\times G_H$) GUST the string
unification scale, $M_{SU}$, can be consistent with the
$M_{G}$ ($M_{MSSM}$) scale, i.e using low energy values of the
$g_{1,2,3}$- coupling constants and their RGE (\ref {RGEi}, \ref {bi},
\ref {symscale}, \ref {RGES})
 we get for the GUST scale $M_{SU}$  the expression
(\ref {Mstrcor})
with the corresponding value of the string coupling constant.

In this case while $M_{\rm sym}$ changes in the range from
$M_G=1.17\cdot 10^{16}$ GeV to $10^{18}$ GeV we could
expect the string constant and string scale (\ref{Mstrcor}) to be in the range
$$g\sim (1.40\div 2.13)\ ,
\quad M_{SU}\sim (0.79\div1.20) \times 10^{18} GeV\ . $$

It is interesting to estimate the value of horizontal coupling constant.
The analysis of RG-equations allows to
state that the horizontal coupling constant $g_{3H}$ does not exceed
the electro-weak one $g_2$.

In Model~1 after cancelation of the U(1) anomaly by VEVs
of the fields ${\hat\phi}_1$ or ${\hat\phi}_3$
corresponding $SU(3)^{II}$ or $SU(3)^{I}$ horizontal gauge
symmetry group survives.

Using RG equations for the running constant $g_{3H}^{I,II}$
and the value of the string coupling constant $g_{str}$ at $M_{SU}$
we can estimate a value of
the horizontal coupling constant at low energies.
For Model~1 we have
$$
b_{3H}^{I}= 21\ ,\qquad
b_{3H}^{II}= 13\ ,$$
and
we find from RGE for $g_{3H}$ that
$$ g^I_H\sim 0.3\ ,\qquad g^{II}_H\sim 0.4\ .$$
Below we consider in details the gauge symmetry breaking by VEV's
b), c), d) applied to Model 1 both in flipped and non-flipped cases.
We assume that some additional Higgs doublets and triplets originating
from the group $G^{II}$ (see sector 1 in table \ref{tabl3}) could
be lighter than $M_G$. The fact that this Higgs fields were initially
(i.e. before breaking $G^I\times G^{II}\rightarrow G$) related to
the group $G^{II}$ excludes their dangerous interaction with matter
that could lead to proton decay.

Also we investigate the dependency on the fourth generation mass $M_4$
and take into account the SUSY threshold.

In general the RGE with thresholds between $M_Z$ and $M_{G}$ are as follows
($k^2\equiv 1$ for non-flipped $SU(5)$):
\begin{eqnarray}\label{rgeeq}
\alpha_3^{-1}&=&{\alpha_5^{-1}} +\frac{1}{2\pi}\ (z + b_3 \ln M_{G}
-b_3^0 \ln M_Z)
\nonumber \\
\alpha_2^{-1}&=&{\alpha_5^{-1}} +\frac{1}{2\pi}\ (y + b_2 \ln
M_{G}-b_2^0 \ln M_Z) \\
\frac{25k^2}{k^2+24}\ \alpha_1^{-1}&=& \alpha_5^{-1}
+\frac{1}{2\pi}\ (x + {\bar b}_1 \ln M_{G}-{\bar b}_1^0 \ln M_Z)\nonumber
\end{eqnarray}
${\bar b}_1,\ b_2, \ b_3$ denote beta function coefficients for the
corresponding coupling constants that take into account all fields below
$M_{G}$ scale.
Similarly ${\bar b}^0_1,\ b^0_2,\ b^0_3$ are beta function coefficients that
take all fields of the MSSM excluding superpartners.
We also introduce thresholds factors $x,\  y,\ z$ that depend on various
thresholds $M_I$:
$$ x,y,z= -\sum_I \Delta b^I_{1,2,3}\ \ln M_I $$
In particular we are going to consider SUSY threshold $M_{SUSY}$,
4th generation masses  $M_4$ and effective masses of addition doublets
and triplets $M_{2,3}$.

In this scheme the representations like
$({\underline{5}, \underline{3}})$ of the Model 1
break into
the equal number of vector-like triplets and doublets
  ( $({\underline{3},\underline{1}})$ and
$({\underline{1},\underline{2}})$ under the $SU(3)\times SU(2)$ group).
We consider the case
when the fields in one $({\underline{5}, \underline{3}})$
representation  with the masses below
$M_{G}$ are from the second group $U(5) \times U(3)$.
(In this case we have no problems with the dimension four, five,
six operators of proton decay in Higgs sector.)
 Hence in addition to the MSSM Higgs  vector-like
doublet we have 2 doublets and 3 triplets.

Below are the values of the  $b$ coefficients for our case:
$$
{\bar b}_1=\frac{275k^2}{24+k^2}\ ,\ b_2=5\ ,\ b_3=2\ ,\
{\bar b}^0_1=\frac{105k^2}{24+k^2}\ ,\ b^0_2=-3\ ,\ b^0_3=-7\ .
$$

 Considering  flipped $SU(5)$ case we have to pay attention to the
consistency of the value of $k^2$ derived from (\ref{rgeeq}) and from
RGE of  the string coupling $g_{str}$ above the $M_{G}$ scale.
We use $b$
  coefficients from (\ref{b5values}, \ref{b1values}).

From the system (\ref{rgeeq}) we can get a set of appropriate masses in the
range of $M_Z-M_{G}$ and values of $\alpha_5,\ k^2$ and $g^I_5/g^{II}_5$
as well. But then we should apply RGE between the string scale
and $M_G$ scale
to check out whether
this values are consistent. This equations give us $k^2 < 1$.
 Our calculations show
that with  $k^2\leq 1$ for flipped $SU(5)$ in the Model 1
 one cannot get appropriate values for
$M_{SUSY},\, M_{2,3,4},\,M_{G}$ (i.e. that are within range
$M_{SUSY}-M_{G}$) that are consistent with string RGE.

For non-flipped case we apparently obtain the demand that
constants $\alpha_{1,2,3}$ converge to one point (that is equivalent
to $k^2\equiv 1$) which is consistent with RGE in the framework
of the MSSM-like models.

For this case we consider the b)  breaking way of
$ [SU(5) \times U(1)]^{\otimes 2}$. We can consider the cases c) and d)
as a limits ($x \ll y$ and $x \gg y$). As it follows from our analysis
there exists a range of parameters values (threshold masses) that make
system (\ref{rgeeq}) consistent and we have an appropriate hierarchy
of the scales \cite{PL}.

The maximum value of $M_G$ one can obtain in this case is
$M_G\sim 1.3\cdot 10^{16}$ GeV.

The horizontal gauge constant on the scale of 1 TeV for first
or second $SU(3)_H$ group (depending on which of them will survive
after suppression of the $U(1)$ anomaly) appears to be of the order

$${g_{3H}^I}^2 \biggl( O(1\mbox{TeV})\biggr)\approx 0.10\div 0.11\ , \quad
{g_{3H}^{II}}^2 \biggl( O(1\mbox{TeV})\biggr)\approx 0.15\div 0.17\ .
$$

The calculations for our model for different breaking chains
show that for evaluation of intensity
of a processes with a gauge horizontal bosons at low energies
we can use inequality
${{\alpha}_{3H}(\mu)}\,\leq \,{{\alpha}_2(\mu)}\ .$

\section{Appendix A. Rules for Constructing Consistent String
 Models out of Free World-Sheet Fermions.}

 The partition function of the theory is a sum over terms
 corresponding to world-sheets of different genus $g$.
 For consistency of the theory we must require that partition
 function to be invariant under modular transformation, which
 is reparametrizations not continuously connected to the identity.
 For this we must sum over the different possible boundary conditions
 for the world-sheet fermions with appropriate weights \cite{SW}.

 If the fermions are parallel transported around a nontrivial loop
 of a genus-$g$ world-sheet $M_g$, they must transform into themselves:
 \begin{equation}
 \chi^I\longrightarrow L_g(\alpha )^I_J\chi^J \label{tr}
 \end{equation}
 and similar for the right-moving fermions.
 The only constraints on $L_g(\alpha )$ and $R_g(\alpha )$ are that
 it be orthogonal matrix representation of $\pi_1 (M_g)$ to leave
 the energy-momentum current invariant and supercharge (\ref{sch})
 invariant up to a sign. It means that
 \begin{eqnarray}
 && \psi^{\mu}\longrightarrow -\delta_{\alpha}\psi^{\mu}\ ,\qquad
 \delta_{\alpha}=\pm 1\ ,\label{bc1} \\
 && L^I_{gI'}L^J_{gJ'}L^K_{gK'} f_{IJK}=-\delta_{\alpha}f_{I'J'K'}
 \label{bc2}
 \end{eqnarray}
 and consequently $-\delta_{\alpha}L_g(\alpha )$ is an automorphism
 of the Lie algebra of $G$.

 Farther, the following restrictions on $L_g(\alpha )$ and
 $R_g(\alpha )$ are imposed:

 $(a)$ $L_g(\alpha )$ and $R_g(\alpha )$ are abelian matrix
 representations of $\pi_1(M_g)$. Thus all of the $L_g(\alpha )$
 and all of the $R_g(\alpha )$ can be simultaneously diagonalized in
 some basis.

 $(b)$ There is commutativity between the boundary conditions on
 surfaces of different genus.

 When all of the $L(\alpha )$ and $R(\alpha )$ have been
 simultaneously diagonalized the transformations like (\ref{tr})
 can be written as
 \begin{equation}
 f\longrightarrow -\exp (i\pi\alpha_f)f\ .
 \end{equation}
 Here and in eqs. (\ref{bc1}), (\ref{bc2}) the minus signs
 are conventional.

 The boundary conditions (\ref{tr}), (\ref{bc1}) are specified
 in this basis by a vector of phases
 \begin{equation}
 \alpha=[\alpha(f_1^L),\cdots,\alpha(f_k^L)\ |\
 \alpha(f_1^R),\cdots,\alpha(f_l^R)]\ .
 \end{equation}
 For complex fermions and $d=4$, $k=10$ and $l=22$.
 The phases in this formula are reduced mod(2) and are chosen
 to be in the interval $(-1,+1]$.

 Modular transformations mix spin-structures amongst one another
 within a surface of a given genus. Thus, requiring the modular
 invariance of the partition function imposes constraints
 on the coefficients ${\cal C}
 \left[
 \begin{array}{c}
 {\alpha}_1\ \cdots\ \alpha_g \\
 {\beta}_1\ \cdots\ \beta_g
 \end{array}\right]$
 (weights in the partition function sum, for example see eq.(\ref{Z})
 which in turn impose constraints on what spin-structures are allowed
 in a consistent theory. In according to the assumptions $(a)$ and $(b)$
 these coefficients factorize:
 \begin{equation}
 {\cal C}
 \left[
 \begin{array}{c}
 {\alpha}_1\ \cdots\ \alpha_g \\
 {\beta}_1\ \cdots\ \beta_g
 \end{array}\right]={\cal C}
 \left[
 \begin{array}{c}
 {\alpha}_1 \\
 {\beta}_1
 \end{array}\right]{\cal C}
 \left[
 \begin{array}{c}
 {\alpha}_2 \\
 {\beta}_2
 \end{array}\right] \cdots {\cal C}
 \left[
 \begin{array}{c}
 {\alpha}_g \\
 {\beta}_g
 \end{array}\right]
 \end{equation}
 The requirement of modular invariance of the partition function
 thus gives rise to constraints on the one-loop coefficients
 ${\cal C}$ and hence on the possible spin structures
 $(\alpha , \beta)$ on the torus.

 For rational phases $\alpha(f)$ (we consider only this case)
 the possible boundary conditions $\alpha$ comprise a finite additive
 group $\Xi=\sum_{i=1}^k\oplus Z_{N_i}$ which is generated by
 a basis $(b_1,\cdots ,b_k)$, where $N_i$ is the smallest integer
 for which $N_i b_i =0\,\mbox{mod}(2)$.
 A multiplication of two vectors from $\Xi$ is defined by
 \begin{equation}
 \alpha\cdot\beta=(\alpha^i_L\beta^i_L-\alpha^j_R\beta^j_R)_{\rm complex}
 +1/2\,(\alpha^k_L\beta^k_L-\alpha^l_R\beta^l_R)_{real}\ .
 \end{equation}
 \medskip
 The basis satisfies the following conditions derived in \cite{19i}:\\
 \medskip
 (A1) The basis $(b_1,\cdots ,b_k)$ is chosen to be canonical:
 $$\sum m_i b_i=0 \Longleftrightarrow m_i=0\,\mbox{mod}(N_i)
 \quad\forall i\ .$$
 Then an arbitrary vector $\alpha$ from $\Xi$ is a linear
 combination $\alpha=\sum a_i b_i$.\\
 \medskip
 (A2) The vector $b_1$ satisfies $1/2\,N_1 b_1={\bf 1}$.
 This is clearly satisfied by $b_1={\bf 1}$.\\
 \medskip
 (A3) $N_{ij}b_i\cdot b_j=0\,\mbox{mod}(4)$ where $N_{ij}$ is the
 least common multiple of $N_i$ and $N_j$.\\
 \medskip
 (A4) $N_i b_i^2=0\,\mbox{mod}(4)$; however, if $N_i$ is even we
 must have $N_i b_i^2=0\,\mbox{mod}(8)$.\\
 \medskip
 (A5) The number of real fermions that are simultaneously periodic
 under any three boundary conditions $b_i$, $b_j$, $b_k$
 is even, where $i$, $j$, $k$ are not necessarily distinct.
 This implies that the number of periodic real fermions in any
 $b_i$ be even.\\
 \medskip
 (A6) The boundary condition matrix corresponding to each $b_i$
 is an automorphism of the Lie algebra that defines the supercharge
 (\ref{sch}). All such automorphisms must commute with one another,
 since they must simultaneously diagonalizable.
 \medskip

 For each group of boundary conditions $\Xi$ there are a number
 of consistent choices for coefficients ${\cal C}[\cdots ]$,
 which determine from requirement of invariant under modular
 transformation. The number of such theories corresponds to the
 number of different choices of ${\cal C}
 \left[
 \begin{array}{c}
 b_i \\ b_j
 \end{array}\right]$.
 This set must satisfy equations:\\
 \medskip
 (B1)$\qquad\qquad\qquad {\cal C}
 \left[
 \begin{array}{c}
 b_i \\ b_j
 \end{array}\right] =\delta_{b_i} e^{2\pi in/N_j}
 =\delta_{b_j} e^{i\pi(b_i\cdot b_j)/2} e^{2\pi im/N_i}\ .$\\
 \medskip
 (B2)$\qquad\qquad\qquad {\cal C}
 \left[
 \begin{array}{c}
 b_1 \\ b_1
 \end{array}\right] =\pm e^{i\pi b_1^2/4}\ .$\\
 \medskip
 The values of ${\cal C}
 \left[
 \begin{array}{c}
 \alpha \\ \beta
 \end{array}\right]$ for arbitrary $\alpha,\,\beta\in\Xi$ can be
 obtained by means of the following rules:\\
 \medskip
 (B3)$\qquad\qquad\qquad {\cal C}
 \left[
 \begin{array}{c}
 \alpha \\ \alpha
 \end{array}\right] =e^{i\pi (\alpha\cdot\alpha +1\cdot 1)/4}{\cal C}
 {\left[
 \begin{array}{c}
 \alpha \\ b_1
 \end{array}\right]}^{N_1/2}\ .$\\
 \medskip
 (B4)$\qquad\qquad\qquad {\cal C}
 \left[
 \begin{array}{c}
 \alpha \\ \beta
 \end{array}\right] =e^{i\pi (\alpha\cdot\beta )/2} {\cal C}
 {\left[
 \begin{array}{c}
 \beta \\ \alpha
 \end{array}\right]}^{\ast}\ .$\\
 \medskip
 (B5)$\qquad\qquad\qquad {\cal C}
 \left[
 \begin{array}{c}
 \alpha \\ \beta +\gamma
 \end{array}\right] =\delta_{\alpha}{\cal C}
 \left[
 \begin{array}{c}
 \alpha \\ \beta
 \end{array}\right] {\cal C}
 \left[
 \begin{array}{c}
 \alpha \\ \gamma
 \end{array}\right]\ .$
 \medskip

 The relative normalization of all the ${\cal C}[\cdots ]$ is fixed
 in these expressions conventionally to be ${\cal C}
 \left[
 \begin{array}{c}
 0 \\ 0
 \end{array}\right] \equiv 1\ .$
 \medskip

 For each $\alpha\in\Xi$ there is a corresponding Hilbert space of
 string states ${\cal H}_{\alpha}$ that potentially contribute to the
 one-loop partition function. If we write
 $\alpha =(\alpha_L\,|\,\alpha_R)$, then the states in
 ${\cal H}_{\alpha}$ are those that satisfy the Virasoro condition:
 \begin{equation}
 M_L^2=-c_L +1/8\,\alpha_L\cdot\alpha_L+\sum_{L-mov.}({\rm frequencies})
 =-c_R +1/8\,\alpha_R\cdot\alpha_R +\sum_{R-mov.}({\rm freq.})=M_R^2\,.
 \end{equation}
 Here $c_L=1/2$ and $c_R=1$ in the heterotic case.
 In ${\cal H}_{\alpha}$ sector the fermion $f$ $(f^{\ast})$ has
 oscillator frequencies
 \begin{equation}
 \frac{1\pm\alpha (f)}{2} +\mbox{integer}\ .
 \end{equation}

 The only states $|s{\rangle }$ in ${\cal H}_{\alpha}$ that contribute to the
 partition function are those that satisfy the generalized GSO conditions
 \begin{equation}
 \left\{ e^{i\pi (b_i\cdot F_{\alpha})}-\delta_{\alpha} {\cal C}
 {\left[
 \begin{array}{c}
 \alpha \\ b_i
 \end{array}\right]}^{\ast}\right\} |s{\rangle }=0
 \end{equation}
 for all $b_i$. Where $F_{\alpha}(f)$ is a fermion number operator.
 If $\alpha$ contains periodic fermions then $|0{\rangle }_{\alpha}$ is
 degenerate, transforming as a representation of an $SO(2n)$
 Clifford algebra.

\end{document}